\documentclass[12pt]{article} 
\usepackage{amssymb,amsmath,amsfonts,theorem,subfigure} 
\usepackage{graphicx}
\usepackage{epic,curves}
\usepackage{multirow}
\usepackage{array}

\usepackage{cite}
\usepackage[usenames]{color}
\usepackage{pstricks}
\usepackage[backref=false]{hyperref}
\hypersetup{
colorlinks=true,
citecolor=red,
linkcolor=darkblue
}
\definecolor{darkblue}{rgb}{0,0,.8}
\definecolor{red}{rgb}{1,0,0}

\newcommand{\nc}{\newcommand}
\nc{\be}{\begin{equation}}
\nc{\ee}{\end{equation}}
\nc{\Ib}{\mbox{\boldmath $I$}}
\nc{\Jb}{\mbox{\boldmath $J$}}
\nc{\Tb}{\mbox{\boldmath $T$}}

\newcommand{\lw}{\psset{linewidth=0.5pt}}
\newcommand{\unlw}{\psset{linewidth=1pt}}
\newrgbcolor{myc}{1 0 0}
\newrgbcolor{myc2}{0 0 1}
\newrgbcolor{myc3}{1 0 1}
\newrgbcolor{mycthree}{1 0 1}
\newrgbcolor{lightermyc}{1 0.2 0.2}
\newrgbcolor{lightermyc2}{0.5 0.5 1}
\newrgbcolor{darkgreen}{0., 0.733333, 0.0621042}
\newrgbcolor{lightergreen}{0.625 0.9 0.648289}
\newrgbcolor{orange}{0.933333, 0.569436, 0}
\newrgbcolor{trialcolor1}{0.933333, 0.569436, 0}

\definecolor{lightblue}{rgb}{.61,.61,1}
\definecolor{midblue}{rgb}{.7,.7,1}
 \definecolor{lightlightblue}{rgb}{.85,.85,1}

\def\facegrid#1#2{
\psframe[fillstyle=solid,fillcolor=lightlightblue,linewidth=0pt]#1#2
\psgrid[gridlabels=0pt,subgriddiv=1]#1#2}
\def\loopa{
\psframe[linewidth=.25pt](0,0)(1,1)
\psarc[linewidth=1.5pt,linecolor=blue](1,0){.5}{90}{180}
\psarc[linewidth=1.5pt,linecolor=blue](0,1){.5}{-90}{0}
}
\def\loopb{
\psframe[linewidth=.25pt](0,0)(1,1)
\psarc[linewidth=1.5pt,linecolor=blue](0,0){.5}{0}{90}
\psarc[linewidth=1.5pt,linecolor=blue](1,1){.5}{180}{270}
}

\theorembodyfont{\itshape} 
\theoremheaderfont{\scshape}
\theoremstyle{plain}  
\newtheorem{Lemme}{Lemma}[section]

\newtheorem{Proposition}[Lemme]{Proposition}

\numberwithin{equation}{section}

\definecolor{lightlightblue}{rgb}{.85,.85,1}

\def\facegrid#1#2{
\psframe[fillstyle=solid,fillcolor=lightlightblue,linewidth=0pt]#1#2
\psgrid[gridlabels=0pt,subgriddiv=1]#1#2}
\def\loopa{
\psframe[linewidth=.25pt](0,0)(1,1)
\psarc[linewidth=1.5pt,linecolor=blue](1,0){.5}{90}{180}
\psarc[linewidth=1.5pt,linecolor=blue](0,1){.5}{-90}{0}
}
\def\loopb{
\psframe[linewidth=.25pt](0,0)(1,1)
\psarc[linewidth=1.5pt,linecolor=blue](0,0){.5}{0}{90}
\psarc[linewidth=1.5pt,linecolor=blue](1,1){.5}{180}{270}
}

\begin{document} 

\title{\mbox{}\vspace{-.5in}\bf Modular invariant partition function \\[37pt] of critical dense polymers\vspace{-.4in}}

\author{}
\date{}
\maketitle 

\begin{center}
{\vspace{-5mm}\Large Alexi Morin-Duchesne$^\ast$, Paul A. Pearce$^\dagger$, J{\o}rgen Rasmussen$^\ast$}
\\[.5cm]
{\em {}$^\ast$School of Mathematics and Physics, University of Queensland}\\
{\em St Lucia, Brisbane, Queensland 4072, Australia}
\\[.2cm]
{\em {}$^\dagger$Department of Mathematics and Statistics, University of Melbourne}\\
{\em Parkville, Victoria 3010, Australia}
\\[.4cm]
{\tt a.morinduchesne\,@\,uq.edu.au}
\quad
{\tt p.pearce\,@\,ms.unimelb.edu.au}
\quad
{\tt j.rasmussen\,@\,uq.edu.au}
\end{center}

\begin{abstract}
A lattice model of critical dense polymers is solved exactly for arbitrary system size on the torus.
More generally, an infinite family of lattice loop models is 
 studied on the torus and related to the corresponding Fortuin-Kasteleyn random cluster models.
Starting with a cylinder, the commuting periodic single-row transfer matrices are built from the periodic Temperley-Lieb algebra extended by the shift operators $\Omega^{\pm1}$. In this enlarged algebra, the non-contractible loop fugacity is $\alpha$ and the contractible loop fugacity is $\beta$. 
The torus is formed by gluing the top and bottom of the cylinder. This gives rise to a variety of non-contractible loops winding around the torus. 
Because of their nonlocal nature, the standard matrix trace does not produce the proper geometric torus. Instead, we introduce a modified matrix trace for this purpose.
This is achieved by using a representation of the enlarged periodic Temperley-Lieb algebra with a parameter $v$ that keeps track of the winding of defects on the cylinder. The transfer matrix representatives and their eigenvalues thus depend on $v$. The modified trace is constructed as a linear functional on planar connectivity diagrams in terms of matrix traces $\mbox{Tr}_d$ (with a fixed number of defects $d$) and Chebyshev polynomials of the first kind. 
For critical dense polymers, where $\beta=0$,
the transfer matrix eigenvalues are obtained by solving a functional equation in the form of an inversion identity.
The solution depends on $d$ and is subject to selection rules which 
we prove.
Simplifications occur if all non-contractible loop fugacities are set to $\alpha=2$ in which case the traces are evaluated at $v=1$. In the continuum scaling limit, the corresponding conformal torus partition function obtained from finite-size corrections agrees with the known modular invariant partition function of symplectic fermions.
\end{abstract}

\newpage
\tableofcontents

\newpage
\section{Introduction}

Solvable critical dense polymers ${\cal LM}(1,2)$ is the first member of the Yang-Baxter integrable family ${\cal LM}(p,p')$ of logarithmic minimal models~\cite{PRZ2006}. Algebraically, the logarithmic minimal models are
described by the planar Temperley-Lieb (TL) algebra~\cite{TL,Jones} with loop fugacity $\beta=2\cos\lambda$ and crossing parameter $\lambda=\frac{(p'-p)\pi}{p'}$.
For solvable critical dense polymers, $\lambda=\frac{\pi}{2}$ so the loop fugacity vanishes, $\beta=0$. 
This implies that closed contractible loops are not allowed so the loop segments form long polymer segments.
An elementary face of the square lattice is assigned a statistical face weight 
according to the configuration of the face. The two 
possible configurations with their associated weights are combined into a single face operator as
\psset{unit=0.9cm}
\setlength{\unitlength}{0.9cm}
\begin{eqnarray}
\begin{pspicture}[shift=-.45](-.5,-.1)(1.25,1.1)
\facegrid{(0,0)}{(1,1)}
\psarc[linewidth=.5pt](0,0){.15}{0}{90}
\rput(.5,.5){\small $u$}
\end{pspicture}
=\ \sin(\lambda-u)\!\!
\begin{pspicture}[shift=-.45](-.5,-.1)(1.25,1.1)
\facegrid{(0,0)}{(1,1)}
\put(0,0){\loopa}
\end{pspicture}
\ +\ \sin u\!\!
\begin{pspicture}[shift=-.45](-.5,-.1)(1.25,1.1)
\facegrid{(0,0)}{(1,1)}
\put(0,0){\loopb}
\end{pspicture}\;
=\!\!\begin{pspicture}[shift=-.45](-.5,-.1)(1.25,1.1)
\facegrid{(0,0)}{(1,1)}
\psarc[linewidth=.5pt](1,0){.15}{90}{180}
\rput(.5,.5){\small $\lambda\!-\!u$}
\end{pspicture}
\label{u}
\end{eqnarray}
where $u$ is the spectral parameter and the lower-left corner has been marked to fix the orientation of 
the square. By the crossing symmetry, rotating the face by 90 degrees changes the spectral parameter from $u$ to $\lambda-u$. The polymer segments begin and end at {\em nodes} at the midpoints of the edges of the face. 

Because of the nonlocal degrees of freedom, in the form of long polymer segments, the topology of the lattice has profound effects on the properties of dense polymers. Critical dense polymers has been solved exactly for arbitrary finite sizes on the strip~\cite{PR2007,AMD2011,PRV2013} and the cylinder~\cite{PRV2010}. Among other results, these studies have firmly established the central charge $c=-2$ and an infinitely extended Kac table of conformal dimensions $\Delta_{r,s}$, $r,s=1,2,3\ldots$ given by the usual Kac formula. The integrals of motion and Baxter $Q$-operators have also been studied~\cite{Nigro2009}. 
Recently, generalized order parameters were calculated~\cite{OffCrit} for an exactly solvable $\varphi_{1,3}$ off-critical perturbation of dense polymers. For $(r,s)$ satisfying $(2r-s)^2<8$, the conformal dimensions $\Delta_{r,s}=\beta_{r,s}$ precisely agree with the critical exponents $\beta_{r,s}$ associated with these generalized order parameters. 

Perhaps, the only other Yang-Baxter integrable model that has been studied in such detail is the Ising or free fermion model. The glaring anomaly is that the lattice model of critical dense polymers has not yet been solved exactly
on the torus. 
The fundamental obstacle in applying transfer matrix techniques is that, because of the nonlocal degrees of freedom and winding, the naive matrix trace of a cylinder transfer matrix
does not produce the proper geometric torus~\cite{PRV2010}. 
In this paper, we use link representations that depend on a winding parameter $v$
and construct a modified trace, initially as a linear functional on planar connectivity diagrams, and ultimately in terms of matrix traces $\mbox{Tr}_d$ (with a fixed number of defects $d$). 
These representations, which we refer to as {\em twist representations} and denote by $\omega_d$, 
were first studied in~\cite{MS1993,GL1998} and more recently in~\cite{AMDYSAinprep}.

Modified matrix traces have appeared before as a way to relate Potts models and loop transfer matrices~\cite{RJ2006,RJ2007}. 
In~\cite{RJ2006}, the partition function of the Potts model on the strip is expressed in terms of representations labeled by a number $\ell$ of bridges, isomorphic to link state representations with $\ell$ defects. The coefficients of the corresponding modified matrix trace are given by Chebyshev polynomials of the 
second kind. On the torus, this construction fails, but an alternative construction was subsequently obtained by the same 
authors~\cite{RJ2007} using a larger class of representations
involving the cyclic group $C_\ell$. The coefficients of the corresponding modified matrix trace are given by Chebyshev polynomials of the first kind.

Here we use a different construction based on the twist representations $\omega_d$, and to the best of our knowledge, the ensuing modified matrix trace is new. 
Furthermore, the fugacities of non-contractible loops considered in~\cite{RJ2006, RJ2007} are very specific, while the ones we consider are general. 
In principle, for critical dense polymers, this allows us to calculate the torus partition function exactly for arbitrary finite system sizes and arbitrary non-contractible loop fugacities by solving an inversion identity in the form of a functional equation for the transfer matrix eigenvalues in sectors with $d$ defects. 

In general, the double-row transfer matrices on the strip~\cite{PRZ2006,AMDYSA2011} and the periodic single-row transfer matrix on the cylinder~\cite{PRV2010,AMDYSA2013} exhibit nontrivial Jordan cells. In this paper, we are interested in partition functions so we only study the eigenvalues, conformal spectra and associated characters, and do not consider the structure and indecomposability of the associated Virasoro representations. 

Using the modified trace, and calculating finite-size corrections, allows us to obtain the torus partition function in the continuum scaling limit. From general principles of conformal invariance~\cite{CardyModular}, this is expected to be modular invariant. Indeed, we find that the torus partition function is Coulombic~\cite{FSZcoulomb1987}, modular invariant and in agreement with~\cite{SaleurSUSY}. 
This modular invariant partition function coincides with that of the triplet model~\cite{GabKausch96} and symplectic fermions on a $\mathbb{Z}_2$ orbifold~\cite{Kausch00}. Both of these models are described by a 
logarithmic Conformal Field Theory (CFT)~\cite{Gurarie} with central charge $c=-2$~\cite{Kausch95,GabKausch99,KW01,BF02,GaberdielRunkelBdy,GaberdielRunkelBdyToBulk,Runkel12}. 

A striking feature of our derivation of the modular invariant is that, contrary to most results on loop models and the associated conformal field theories, it is entirely rigorous. In particular, we present a mathematical proof of the selection rules determining the eigenvalues (and their degeneracies) of the transfer matrix. This proof is based on a relation between the loop model and an XXZ Hamiltonian, an idea previously used in~\cite{AMD2011} to construct a similar proof of the selection rules applicable to the loop model defined on the strip. A proof of the selection rules conjectured in~\cite{PRV2010} is provided in the process. 

The layout of this paper is as follows. In Section~\ref{sec:LoopsFKandTL}, 
we define a general loop model on the torus and explain its relation to the Fortuin-Kasteleyn (FK) random cluster model~\cite{FK}. 
We also introduce the enlarged periodic Temperley-Lieb (TL) algebra. Lastly, we introduce link states, the twist representation on these link states, which uses the {\em winding parameter} $v$ to keep track of windings, and our new modified trace which we use to close the cylinder to the geometric torus. In Section~\ref{sec:CDP}, we focus on critical dense polymers. We use the twist representation in sectors with $d$ defects and the known transfer matrix inversion identity~\cite{PRV2010} to write down functional equations involving $v$ for the eigenvalues. These equations are solved for the eigenvalues subject to certain selection rules which we prove in Appendix~\ref{app:a}. Our modified trace is then applied to build the finitized torus partition function. In the continuum scaling limit, this becomes a simple Coulombic partition function and so is modular invariant. 
Loop models and critical dense polymers on helical tori~\cite{OkabeEtAl1999,LiawEtAl2006,IzmHu2007} are considered in Appendix~\ref{app:Twist}. 
Finally, Section~\ref{sec:Conclusion} contains some concluding remarks.

%
\section{Loops, FK clusters and the periodic TL algebra}
\label{sec:LoopsFKandTL}
%

\subsection{Loop model}
\label{sec:loopmodel}

Configurations of the loop model of our interest are drawn on a square lattice of $M \times N$ tiles in the plane. Every tile is decorated with one of the two diagrams 
$\begin{pspicture}(0,0.1)(0.5,0.5)
\psset{unit=0.5}
\psline[linewidth=0.5pt]{-}(0,0)(0,1)(1,1)(1,0)(0,0)
\psset{linecolor=myc2}
\psset{linewidth=1.7pt}
\psarc(0,0){0.5}{0}{90}\psarc(1,1){0.5}{180}{270}
\end{pspicture}
$ and
$\begin{pspicture}(0,0.1)(0.6,0.5)
\psset{unit=0.5}
\psline[linewidth=0.5pt]{-}(0,0)(0,1)(1,1)(1,0)(0,0)
\psset{linecolor=darkgreen}
\psset{linewidth=1.7pt}
\psarc(1,0){0.5}{90}{180}\psarc(0,1){0.5}{270}{0}
\end{pspicture}
$, as indicated in Figure~\ref{fig:loopconfig}. 
The boundary conditions are taken to be periodic in both the vertical and horizontal directions. The resulting diagram is a graph of non-intersecting closed curves (loops) on an $M\times N$ torus (of trivial helicity, see Appendix~\ref{app:Twist}). 
Because of the toroidal geometry, the loops can have different homologies. Contractible loops, which can be continuously deformed to a point, are said to be {\it homotopic to a point}, or of homotopy $\{0\}$. Loops can also wind around the torus in a nontrivial manner; $a$ times in the horizontal direction and $b$ times in the vertical direction. The constraint that the loops do not intersect imposes that the greatest common divisor of $a$ and $b$, $a \wedge b$, is $1$ (with $a \wedge 0 = 0 \wedge a \equiv a$). 

As a curve moves upward to wind around the torus vertically ($b$ times), it can wind around the horizontal direction ($a$ times) by either going to the right or the left. To establish the difference in homotopy between these two cases, we fix the convention that $a$ is positive if the curve winds toward the right, and negative if it winds toward the left. The ensuing loop will be said to have homotopy $\{a,b\}$, with $a \in \mathbb Z$, $b \in \mathbb N_0$ and $a \wedge b =1$. For example, the configuration in Figure~\ref{fig:loopconfig} has two loops of homotopy $\{1,1\}$. 

\begin{figure}[ht]
\begin{center}
\psset{unit=0.75}
\begin{pspicture}(0,-0.5)(6,4)
\psset{linewidth=1pt}
\psline[linewidth=0.5pt]{-}(0,0)(0,4)
\psline[linewidth=0.5pt]{-}(1,0)(1,4)
\psline[linewidth=0.5pt]{-}(2,0)(2,4)
\psline[linewidth=0.5pt]{-}(3,0)(3,4)
\psline[linewidth=0.5pt]{-}(4,0)(4,4)
\psline[linewidth=0.5pt]{-}(5,0)(5,4)
\psline[linewidth=0.5pt]{-}(6,0)(6,4)
\psline[linewidth=0.5pt]{-}(0,0)(6,0)
\psline[linewidth=0.5pt]{-}(0,1)(6,1)
\psline[linewidth=0.5pt]{-}(0,2)(6,2)
\psline[linewidth=0.5pt]{-}(0,3)(6,3)
\psline[linewidth=0.5pt]{-}(0,4)(6,4)
\psline{<->}(0.1,-0.3)(5.9,-0.3)
\psline{<->}(-0.3, 0.1)(-0.3, 3.9)
\rput(3,-0.7){$N$}
\rput(-0.7,2){$M$}
\psset{linecolor=myc2}
\psset{linewidth=1.7pt}
\psarc(2,0){0.5}{0}{90}\psarc(3,1){0.5}{180}{270}
\psarc(3,0){0.5}{0}{90}\psarc(4,1){0.5}{180}{270}
\psarc(5,0){0.5}{0}{90}\psarc(6,1){0.5}{180}{270}
\psarc(0,1){0.5}{0}{90}\psarc(1,2){0.5}{180}{270}
\psarc(0,2){0.5}{0}{90}\psarc(1,3){0.5}{180}{270}
\psarc(3,1){0.5}{0}{90}\psarc(4,2){0.5}{180}{270}
\psarc(3,2){0.5}{0}{90}\psarc(4,3){0.5}{180}{270}
\psarc(4,3){0.5}{0}{90}\psarc(5,4){0.5}{180}{270}
\psset{linecolor=darkgreen}
\psarc(1,0){0.5}{90}{180}\psarc(0,1){0.5}{270}{0}
\psarc(2,0){0.5}{90}{180}\psarc(1,1){0.5}{270}{0}
\psarc(5,0){0.5}{90}{180}\psarc(4,1){0.5}{270}{0}
\psarc(2,1){0.5}{90}{180}\psarc(1,2){0.5}{270}{0}
\psarc(3,1){0.5}{90}{180}\psarc(2,2){0.5}{270}{0}
\psarc(5,1){0.5}{90}{180}\psarc(4,2){0.5}{270}{0}
\psarc(6,1){0.5}{90}{180}\psarc(5,2){0.5}{270}{0}
\psarc(2,2){0.5}{90}{180}\psarc(1,3){0.5}{270}{0}
\psarc(3,2){0.5}{90}{180}\psarc(2,3){0.5}{270}{0}
\psarc(5,2){0.5}{90}{180}\psarc(4,3){0.5}{270}{0}
\psarc(6,2){0.5}{90}{180}\psarc(5,3){0.5}{270}{0}
\psarc(1,3){0.5}{90}{180}\psarc(0,4){0.5}{270}{0}
\psarc(2,3){0.5}{90}{180}\psarc(1,4){0.5}{270}{0}
\psarc(3,3){0.5}{90}{180}\psarc(2,4){0.5}{270}{0}
\psarc(4,3){0.5}{90}{180}\psarc(3,4){0.5}{270}{0}
\psarc(6,3){0.5}{90}{180}\psarc(5,4){0.5}{270}{0}
\end{pspicture}
\caption{A configuration of the loop model  with weight $W_L(\sigma) = \beta^4 \alpha_{1,1}^2 \sin^{8}\!u \sin^{16} (\lambda - u)$.}
\label{fig:loopconfig}
\end{center}
\end{figure}
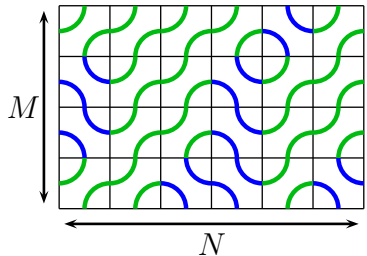

In a given configuration, there can be more than one loop that winds nontrivially 
around the torus, but because they are non-intersecting, these non-contractible loops must all have the {\em same} homotopy.  A configuration is therefore classified by the homotopy of its loops that wind around the torus. If a configuration has no loop winding around the torus, it has homotopy $\{0\}$.
 
To every configuration $\sigma$, we associate a weight 
\be 
W_L(\sigma) = \beta^{n_0} \Big(\!\prod_{a\wedge b=1}\!\alpha_{a,b}^{n_{a,b}}\Big) p_1^{n(
\begin{pspicture}(0,0.05)(0.25,0.25)
\psset{unit=0.25}
\psline[linewidth=0.5pt]{-}(0,0)(0,1)(1,1)(1,0)(0,0)
\psset{linecolor=darkgreen}
\psset{linewidth=1.0pt}
\psarc(1,0){0.5}{90}{180}\psarc(0,1){0.5}{270}{0}\end{pspicture}
)}p_2^{n(
\begin{pspicture}(0,0.05)(0.25,0.25)
\psset{unit=0.25}
\psline[linewidth=0.5pt]{-}(0,0)(0,1)(1,1)(1,0)(0,0)
\psset{linecolor=myc2}
\psset{linewidth=1.0pt}
\psarc(0,0){0.5}{0}{90}\psarc(1,1){0.5}{180}{270}
\end{pspicture}
)} 
\ee
where  
$n(\begin{pspicture}(0,0.05)(0.3,0.25)
\psset{unit=0.3}
\psline[linewidth=0.5pt]{-}(0,0)(0,1)(1,1)(1,0)(0,0)
\psset{linecolor=darkgreen}
\psset{linewidth=1.0pt}
\psarc(1,0){0.5}{90}{180}\psarc(0,1){0.5}{270}{0}
\end{pspicture}
)$ and 
$n(
\begin{pspicture}(0,0.05)(0.3,0.25)
\psset{unit=0.3}
\psline[linewidth=0.5pt]{-}(0,0)(0,1)(1,1)(1,0)(0,0)
\psset{linecolor=myc2}
\psset{linewidth=1.0pt}
\psarc(0,0){0.5}{0}{90}\psarc(1,1){0.5}{180}{270}
\end{pspicture}
)$
are the number of times the tiles  
$\begin{pspicture}(0,0.1)(0.55,0.5)
\psset{unit=0.5}
\psline[linewidth=0.5pt]{-}(0,0)(0,1)(1,1)(1,0)(0,0)
\psset{linecolor=darkgreen}
\psset{linewidth=1.7pt}
\psarc(1,0){0.5}{90}{180}\psarc(0,1){0.5}{270}{0}
\end{pspicture}
$ 
and 
$\begin{pspicture}(0,0.1)(0.55,0.5)
\psset{unit=0.5}
\psline[linewidth=0.5pt]{-}(0,0)(0,1)(1,1)(1,0)(0,0)
\psset{linecolor=myc2}
\psset{linewidth=1.7pt}
\psarc(0,0){0.5}{0}{90}\psarc(1,1){0.5}{180}{270}
\end{pspicture}
$
appear in the configuration~$\sigma$, $n_0$ is the number of loops of homotopy $\{0\}$ and $n_{a,b}$ the number of loops of homotopy $\{a,b\}$, while $\prod_{a \wedge b = 1}$ is a product over $a \in \mathbb Z$ and $b \in \mathbb N_0$ with the constraint $a \wedge b = 1$. Note that at most one $n_{a,b}$ is nonzero, so the product will contain at most one term $\alpha_{a,b}^{n_{a,b}}$. If no loop winds around the cylinder, then the product is replaced by $1$. We refer to $\beta$ and $\alpha_{a,b}$ as contractible and non-contractible loop fugacities, respectively, while $p_1$ and $p_2$ are the weights of the face configurations 
$\begin{pspicture}(0,0.1)(0.55,0.5)
\psset{unit=0.5}
\psline[linewidth=0.5pt]{-}(0,0)(0,1)(1,1)(1,0)(0,0)
\psset{linecolor=darkgreen}
\psset{linewidth=1.7pt}
\psarc(1,0){0.5}{90}{180}\psarc(0,1){0.5}{270}{0}
\end{pspicture}
$ 
and 
$\begin{pspicture}(0,0.1)(0.55,0.5)
\psset{unit=0.5}
\psline[linewidth=0.5pt]{-}(0,0)(0,1)(1,1)(1,0)(0,0)
\psset{linecolor=myc2}
\psset{linewidth=1.7pt}
\psarc(0,0){0.5}{0}{90}\psarc(1,1){0.5}{180}{270}
\end{pspicture}
$, 
respectively. 

The {\em partition function} is defined as the sum of the weights of the $2^{MN}$ configurations,
\be 
Z_L = \sum_{\sigma} W_L(\sigma).
\ee
From the discussion above, the sum in $Z_L$ can be split between the different possible homotopies of the configurations, and we write $Z_L(h)$ for the partition function obtained from restricting the sum over $\sigma$ to configurations with homotopy $h$. We thus have
\be 
Z_L = Z_L(\{0\}) + \sum_{a \wedge b = 1} Z_L(\{a,b\}),
\ee
where $\sum_{a \wedge b = 1}$ is a sum over $a \in \mathbb Z$ and $b \in \mathbb N_0$ with the constraint $a \wedge b = 1$. By construction, for $M$ and $N$ finite, $Z_L(\{a,b\})$ is zero if $|a|>M$ or $b>N$.

The partition function depends on many free parameters: $\beta$, $\alpha_{a,b}$, $p_1$ and $p_2$. Since 
$n(\begin{pspicture}(0,0.05)(0.3,0.25)
\psset{unit=0.3}
\psline[linewidth=0.5pt]{-}(0,0)(0,1)(1,1)(1,0)(0,0)
\psset{linecolor=darkgreen}
\psset{linewidth=1.0pt}
\psarc(1,0){0.5}{90}{180}\psarc(0,1){0.5}{270}{0}
\end{pspicture}
) + n(
\begin{pspicture}(0,0.05)(0.3,0.25)
\psset{unit=0.3}
\psline[linewidth=0.5pt]{-}(0,0)(0,1)(1,1)(1,0)(0,0)
\psset{linecolor=myc2}
\psset{linewidth=1.0pt}
\psarc(0,0){0.5}{0}{90}\psarc(1,1){0.5}{180}{270}
\end{pspicture}
) = MN$, up to a $\sigma$-independent constant, $W_L(\sigma)$ depends only on the ratio of $p_1$ and $p_2$. In the following, we parameterize the weights $p_1,p_2$ and the contractible loop fugacity $\beta$ as 
\be
 p_1 = \sin(\lambda - u),\qquad p_2 = \sin u,\qquad \beta = 2\cos \lambda,
\ee 
where $\lambda$ and $u$ are the crossing and spectral parameters, respectively.

\subsection{Relating the loop and FK cluster models}
\label{sec:FKmodel}

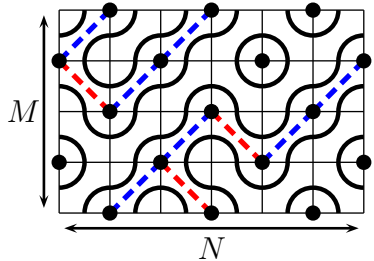
\begin{figure}[ht]
\begin{center}
\psset{unit=0.75}
\begin{pspicture}(0,-0.5)(6,4)
\psset{linewidth=1pt}
\psline[linewidth=0.5pt]{-}(0,0)(0,4)
\psline[linewidth=0.5pt]{-}(1,0)(1,4)
\psline[linewidth=0.5pt]{-}(2,0)(2,4)
\psline[linewidth=0.5pt]{-}(3,0)(3,4)
\psline[linewidth=0.5pt]{-}(4,0)(4,4)
\psline[linewidth=0.5pt]{-}(5,0)(5,4)
\psline[linewidth=0.5pt]{-}(6,0)(6,4)
\psline[linewidth=0.5pt]{-}(0,0)(6,0)
\psline[linewidth=0.5pt]{-}(0,1)(6,1)
\psline[linewidth=0.5pt]{-}(0,2)(6,2)
\psline[linewidth=0.5pt]{-}(0,3)(6,3)
\psline[linewidth=0.5pt]{-}(0,4)(6,4)
\psline{<->}(0.1,-0.3)(5.9,-0.3)
\psline{<->}(-0.3, 0.1)(-0.3, 3.9)
\rput(3,-0.7){$N$}
\rput(-0.7,2){$M$}
\psline[linecolor=myc,linestyle=dashed,linewidth=2pt]{-}(0,3)(1,2)
\psline[linecolor=myc,linestyle=dashed,linewidth=2pt]{-}(2,1)(3,0)
\psline[linecolor=myc,linestyle=dashed,linewidth=2pt]{-}(3,2)(4,1)
\psline[linecolor=myc2,linestyle=dashed,linewidth=2pt]{-}(1,2)(2,3)
\psline[linecolor=myc2,linestyle=dashed,linewidth=2pt]{-}(1,0)(2,1)
\psline[linecolor=myc2,linestyle=dashed,linewidth=2pt]{-}(2,1)(3,2)
\psline[linecolor=myc2,linestyle=dashed,linewidth=2pt]{-}(2,3)(3,4)
\psline[linecolor=myc2,linestyle=dashed,linewidth=2pt]{-}(0,3)(1,4)
\psline[linecolor=myc2,linestyle=dashed,linewidth=2pt]{-}(4,1)(5,2)
\psline[linecolor=myc2,linestyle=dashed,linewidth=2pt]{-}(5,2)(6,3)
\psset{linewidth=1.7pt}%
\psarc(2,0){0.5}{0}{90}\psarc(3,1){0.5}{180}{270}
\psarc(3,0){0.5}{0}{90}\psarc(4,1){0.5}{180}{270}
\psarc(5,0){0.5}{0}{90}\psarc(6,1){0.5}{180}{270}
\psarc(0,1){0.5}{0}{90}\psarc(1,2){0.5}{180}{270}
\psarc(0,2){0.5}{0}{90}\psarc(1,3){0.5}{180}{270}
\psarc(3,1){0.5}{0}{90}\psarc(4,2){0.5}{180}{270}
\psarc(3,2){0.5}{0}{90}\psarc(4,3){0.5}{180}{270}
\psarc(4,3){0.5}{0}{90}\psarc(5,4){0.5}{180}{270}
\psarc(1,0){0.5}{90}{180}\psarc(0,1){0.5}{270}{0}
\psarc(2,0){0.5}{90}{180}\psarc(1,1){0.5}{270}{0}
\psarc(5,0){0.5}{90}{180}\psarc(4,1){0.5}{270}{0}
\psarc(2,1){0.5}{90}{180}\psarc(1,2){0.5}{270}{0}
\psarc(3,1){0.5}{90}{180}\psarc(2,2){0.5}{270}{0}
\psarc(5,1){0.5}{90}{180}\psarc(4,2){0.5}{270}{0}
\psarc(6,1){0.5}{90}{180}\psarc(5,2){0.5}{270}{0}
\psarc(2,2){0.5}{90}{180}\psarc(1,3){0.5}{270}{0}
\psarc(3,2){0.5}{90}{180}\psarc(2,3){0.5}{270}{0}
\psarc(5,2){0.5}{90}{180}\psarc(4,3){0.5}{270}{0}
\psarc(6,2){0.5}{90}{180}\psarc(5,3){0.5}{270}{0}
\psarc(1,3){0.5}{90}{180}\psarc(0,4){0.5}{270}{0}
\psarc(2,3){0.5}{90}{180}\psarc(1,4){0.5}{270}{0}
\psarc(3,3){0.5}{90}{180}\psarc(2,4){0.5}{270}{0}
\psarc(4,3){0.5}{90}{180}\psarc(3,4){0.5}{270}{0}
\psarc(6,3){0.5}{90}{180}\psarc(5,4){0.5}{270}{0}
\pscircle[fillstyle=solid,linecolor=black,fillcolor=black](1,0){0.15}
\pscircle[fillstyle=solid,linecolor=black,fillcolor=black](3,0){0.15}
\pscircle[fillstyle=solid,linecolor=black,fillcolor=black](5,0){0.15}
\pscircle[fillstyle=solid,linecolor=black,fillcolor=black](0,1){0.15}
\pscircle[fillstyle=solid,linecolor=black,fillcolor=black](2,1){0.15}
\pscircle[fillstyle=solid,linecolor=black,fillcolor=black](4,1){0.15}
\pscircle[fillstyle=solid,linecolor=black,fillcolor=black](6,1){0.15}
\pscircle[fillstyle=solid,linecolor=black,fillcolor=black](1,2){0.15}
\pscircle[fillstyle=solid,linecolor=black,fillcolor=black](3,2){0.15}
\pscircle[fillstyle=solid,linecolor=black,fillcolor=black](5,2){0.15}
\pscircle[fillstyle=solid,linecolor=black,fillcolor=black](0,3){0.15}
\pscircle[fillstyle=solid,linecolor=black,fillcolor=black](2,3){0.15}
\pscircle[fillstyle=solid,linecolor=black,fillcolor=black](4,3){0.15}
\pscircle[fillstyle=solid,linecolor=black,fillcolor=black](6,3){0.15}
\pscircle[fillstyle=solid,linecolor=black,fillcolor=black](1,4){0.15}
\pscircle[fillstyle=solid,linecolor=black,fillcolor=black](3,4){0.15}
\pscircle[fillstyle=solid,linecolor=black,fillcolor=black](5,4){0.15}
\end{pspicture} 
\caption{An FK configuration with weight $W_{FK}(\sigma) = Q^3 w_{1,1} v_1^{7} v_2^{3}$.
}
\label{fig:FKconfig}
\end{center}
\end{figure}

The Fortuin-Kasteleyn (FK) random cluster model~\cite{FK} can also be defined on the toroidal geometry of the loop model above. We label the lower-left corners of the tiles by their positions $(x,y)$ on $\mathbb Z^2$, with $(0,0)$ in the lower-left corner of the lattice. A distinguished sublattice is added to the lattice: Sites, identified by solid dots in Figure~\ref{fig:FKconfig}, appear in the positions $(x,y)$ if $x + y$ is odd. The integers $M$ and $N$ are chosen even to ensure periodicity in both directions. 
As before, a configuration consists of a choice of either 
$\begin{pspicture}(0,0.1)(0.55,0.5)
\psset{unit=0.5}
\psline[linewidth=0.5pt]{-}(0,0)(0,1)(1,1)(1,0)(0,0)
\psset{linecolor=black}
\psset{linewidth=1.7pt}
\psarc(0,0){0.5}{0}{90}\psarc(1,1){0.5}{180}{270}
\end{pspicture}
$ or 
$\begin{pspicture}(0,0.1)(0.55,0.5)
\psset{unit=0.5}
\psline[linewidth=0.5pt]{-}(0,0)(0,1)(1,1)(1,0)(0,0)
\psset{linecolor=black}
\psset{linewidth=1.7pt}
\psarc(1,0){0.5}{90}{180}\psarc(0,1){0.5}{270}{0}
\end{pspicture}
$ for each of the tiles. Dashed lines are then added between neighbouring sites of the distinguished sublattice if the configuration allows one to draw a bond without intersecting the loop segments.
Blue lines 
\mbox{(\! 
$
\begin{pspicture}(0,0.1)(0.5,0.5)
\psset{unit=0.4}
\psline[linecolor=myc2,linestyle=dashed, dash=2.5pt 1.5pt, linewidth=1.2pt]{-}(0,0)(1,1)
\pscircle[fillstyle=solid,linecolor=black,fillcolor=black](0,0){0.15}
\pscircle[fillstyle=solid,linecolor=black,fillcolor=black](1,1){0.15}
\end{pspicture}$)
}\!\!
are for bonds connecting upper-right and lower-left corners of the faces ({\it type $1$ bonds}), while red lines 
\mbox{(\! 
$
\begin{pspicture}(0,0.1)(0.5,0.5)
\psset{unit=0.4}
\psline[linecolor=myc,linestyle=dashed, dash=2.5pt 1.5pt, linewidth=1.2pt]{-}(1,0)(0,1)
\pscircle[fillstyle=solid,linecolor=black,fillcolor=black](1,0){0.15}
\pscircle[fillstyle=solid,linecolor=black,fillcolor=black](0,1){0.15}
\end{pspicture}$)
}\!\!
are for bonds connecting upper-left and lower-right corners of the faces ({\it type $2$ bonds}). For a given configuration, we denote by $N_{B_1}$ and $N_{B_2}$ the numbers of type $1$ and type $2$ bonds.

The curves formed by concatenating the loop segments of the tiles separate the set of sites into clusters consisting of sites connected by bonds. 
The definition of a cluster site is therefore intimately related to the notion of the interior of a closed loop, here defined to be occupied by sites. For $M$ or $N$ odd, the periodicity of the lattice renders this definition ill-defined, which, as already announced, prompts us to only consider FK clusters on lattices for which both $M$ and $N$ are even.

\begin{figure}[ht]
\begin{center}
\psset{unit=0.75}
\begin{pspicture}(0,-0.5)(6,4)
\psset{linewidth=1pt}
\psline[linewidth=0.5pt]{-}(0,0)(0,4)
\psline[linewidth=0.5pt]{-}(1,0)(1,4)
\psline[linewidth=0.5pt]{-}(2,0)(2,4)
\psline[linewidth=0.5pt]{-}(3,0)(3,4)
\psline[linewidth=0.5pt]{-}(4,0)(4,4)
\psline[linewidth=0.5pt]{-}(5,0)(5,4)
\psline[linewidth=0.5pt]{-}(6,0)(6,4)
\psline[linewidth=0.5pt]{-}(0,0)(6,0)
\psline[linewidth=0.5pt]{-}(0,1)(6,1)
\psline[linewidth=0.5pt]{-}(0,2)(6,2)
\psline[linewidth=0.5pt]{-}(0,3)(6,3)
\psline[linewidth=0.5pt]{-}(0,4)(6,4)
\psline[linewidth=2pt, linecolor=myc3]{-}(0,0)(1,0)(1,4)(0,4)(0,0)
\psline[linecolor=myc,linestyle=dashed,linewidth=2pt]{-}(0,3)(1,2)
\psline[linecolor=myc,linestyle=dashed,linewidth=2pt]{-}(2,3)(3,2)
\psline[linecolor=myc,linestyle=dashed,linewidth=2pt]{-}(2,1)(3,0)
\psline[linecolor=myc,linestyle=dashed,linewidth=2pt]{-}(3,2)(4,1)
\psline[linecolor=myc2,linestyle=dashed,linewidth=2pt]{-}(1,2)(2,3)
\psline[linecolor=myc2,linestyle=dashed,linewidth=2pt]{-}(3,0)(4,1)
\psline[linecolor=myc2,linestyle=dashed,linewidth=2pt]{-}(5,0)(6,1)
\psline[linecolor=myc2,linestyle=dashed,linewidth=2pt]{-}(2,3)(3,4)
\psline[linecolor=myc2,linestyle=dashed,linewidth=2pt]{-}(0,3)(1,4)
\psline[linecolor=myc2,linestyle=dashed,linewidth=2pt]{-}(4,1)(5,2)
\psline[linecolor=myc2,linestyle=dashed,linewidth=2pt]{-}(5,2)(6,3)
\psset{linewidth=1.7pt}
\psarc(1,0){0.5}{0}{90}\psarc(2,1){0.5}{180}{270}
\psarc(2,0){0.5}{0}{90}\psarc(3,1){0.5}{180}{270}
\psarc(0,1){0.5}{0}{90}\psarc(1,2){0.5}{180}{270}
\psarc(0,2){0.5}{0}{90}\psarc(1,3){0.5}{180}{270}
\psarc(2,2){0.5}{0}{90}\psarc(3,3){0.5}{180}{270}
\psarc(3,1){0.5}{0}{90}\psarc(4,2){0.5}{180}{270}
\psarc(3,2){0.5}{0}{90}\psarc(4,3){0.5}{180}{270}
\psarc(4,3){0.5}{0}{90}\psarc(5,4){0.5}{180}{270}
\psarc(2,1){0.5}{0}{90}\psarc(3,2){0.5}{180}{270}
\psarc(1,0){0.5}{90}{180}\psarc(0,1){0.5}{270}{0}
\psarc(4,0){0.5}{90}{180}\psarc(3,1){0.5}{270}{0}
\psarc(5,0){0.5}{90}{180}\psarc(4,1){0.5}{270}{0}
\psarc(6,0){0.5}{90}{180}\psarc(5,1){0.5}{270}{0}
\psarc(2,1){0.5}{90}{180}\psarc(1,2){0.5}{270}{0}
\psarc(5,1){0.5}{90}{180}\psarc(4,2){0.5}{270}{0}
\psarc(6,1){0.5}{90}{180}\psarc(5,2){0.5}{270}{0}
\psarc(2,2){0.5}{90}{180}\psarc(1,3){0.5}{270}{0}
\psarc(5,2){0.5}{90}{180}\psarc(4,3){0.5}{270}{0}
\psarc(6,2){0.5}{90}{180}\psarc(5,3){0.5}{270}{0}
\psarc(1,3){0.5}{90}{180}\psarc(0,4){0.5}{270}{0}
\psarc(2,3){0.5}{90}{180}\psarc(1,4){0.5}{270}{0}
\psarc(3,3){0.5}{90}{180}\psarc(2,4){0.5}{270}{0}
\psarc(4,3){0.5}{90}{180}\psarc(3,4){0.5}{270}{0}
\psarc(6,3){0.5}{90}{180}\psarc(5,4){0.5}{270}{0}
\pscircle[fillstyle=solid,linecolor=black,fillcolor=black](1,0){0.15}
\pscircle[fillstyle=solid,linecolor=black,fillcolor=black](3,0){0.15}
\pscircle[fillstyle=solid,linecolor=black,fillcolor=black](5,0){0.15}
\pscircle[fillstyle=solid,linecolor=black,fillcolor=black](0,1){0.15}
\pscircle[fillstyle=solid,linecolor=black,fillcolor=black](2,1){0.15}
\pscircle[fillstyle=solid,linecolor=black,fillcolor=black](4,1){0.15}
\pscircle[fillstyle=solid,linecolor=black,fillcolor=black](6,1){0.15}
\pscircle[fillstyle=solid,linecolor=black,fillcolor=black](1,2){0.15}
\pscircle[fillstyle=solid,linecolor=black,fillcolor=black](3,2){0.15}
\pscircle[fillstyle=solid,linecolor=black,fillcolor=black](5,2){0.15}
\pscircle[fillstyle=solid,linecolor=black,fillcolor=black](0,3){0.15}
\pscircle[fillstyle=solid,linecolor=black,fillcolor=black](2,3){0.15}
\pscircle[fillstyle=solid,linecolor=black,fillcolor=black](4,3){0.15}
\pscircle[fillstyle=solid,linecolor=black,fillcolor=black](6,3){0.15}
\pscircle[fillstyle=solid,linecolor=black,fillcolor=black](1,4){0.15}
\pscircle[fillstyle=solid,linecolor=black,fillcolor=black](3,4){0.15}
\pscircle[fillstyle=solid,linecolor=black,fillcolor=black](5,4){0.15}
\end{pspicture}
\quad  \quad \quad
\begin{pspicture}(0,-0.5)(6,4)
\rput(-0.9,2){$\leftrightarrow$}\;
\psset{linewidth=1pt}
\psline[linewidth=0.5pt]{-}(0,0)(0,4)
\psline[linewidth=0.5pt]{-}(1,0)(1,4)
\psline[linewidth=0.5pt]{-}(2,0)(2,4)
\psline[linewidth=0.5pt]{-}(3,0)(3,4)
\psline[linewidth=0.5pt]{-}(4,0)(4,4)
\psline[linewidth=0.5pt]{-}(5,0)(5,4)
\psline[linewidth=0.5pt]{-}(6,0)(6,4)
\psline[linewidth=0.5pt]{-}(0,0)(6,0)
\psline[linewidth=0.5pt]{-}(0,1)(6,1)
\psline[linewidth=0.5pt]{-}(0,2)(6,2)
\psline[linewidth=0.5pt]{-}(0,3)(6,3)
\psline[linewidth=0.5pt]{-}(0,4)(6,4)
\psline[linewidth=2pt, linecolor=myc3]{-}(1,0)(2,0)(2,4)(1,4)(1,0)
\psline[linecolor=myc,linestyle=dashed,linewidth=2pt]{-}(2,1)(1,2)
\psline[linecolor=myc,linestyle=dashed,linewidth=2pt]{-}(4,3)(5,2)
\psline[linecolor=myc,linestyle=dashed,linewidth=2pt]{-}(5,4)(6,3)
\psline[linecolor=myc,linestyle=dashed,linewidth=2pt]{-}(2,1)(3,0)
\psline[linecolor=myc,linestyle=dashed,linewidth=2pt]{-}(3,2)(4,1)
\psline[linecolor=myc2,linestyle=dashed,linewidth=2pt]{-}(1,2)(0,1)
\psline[linecolor=myc2,linestyle=dashed,linewidth=2pt]{-}(1,0)(2,1)
\psline[linecolor=myc2,linestyle=dashed,linewidth=2pt]{-}(5,0)(6,1)
\psline[linecolor=myc2,linestyle=dashed,linewidth=2pt]{-}(2,1)(3,2)
\psline[linecolor=myc2,linestyle=dashed,linewidth=2pt]{-}(2,3)(3,4)
\psline[linecolor=myc2,linestyle=dashed,linewidth=2pt]{-}(4,3)(5,4)
\psline[linecolor=myc2,linestyle=dashed,linewidth=2pt]{-}(0,3)(1,4)
\psline[linecolor=myc2,linestyle=dashed,linewidth=2pt]{-}(5,2)(6,3)
\psset{linewidth=1.7pt}
\psarc(2,0){0.5}{0}{90}\psarc(3,1){0.5}{180}{270}
\psarc(3,0){0.5}{0}{90}\psarc(4,1){0.5}{180}{270}
\psarc(1,1){0.5}{0}{90}\psarc(2,2){0.5}{180}{270}
\psarc(1,2){0.5}{0}{90}\psarc(2,3){0.5}{180}{270}
\psarc(3,2){0.5}{0}{90}\psarc(4,3){0.5}{180}{270}
\psarc(4,1){0.5}{0}{90}\psarc(5,2){0.5}{180}{270}
\psarc(4,2){0.5}{0}{90}\psarc(5,3){0.5}{180}{270}
\psarc(5,3){0.5}{0}{90}\psarc(6,4){0.5}{180}{270}
\psarc(3,1){0.5}{0}{90}\psarc(4,2){0.5}{180}{270}
\psarc(2,0){0.5}{90}{180}\psarc(1,1){0.5}{270}{0}
\psarc(5,0){0.5}{90}{180}\psarc(4,1){0.5}{270}{0}
\psarc(6,0){0.5}{90}{180}\psarc(0,1){0.5}{270}{0}
\psarc(1,0){0.5}{90}{180}\psarc(5,1){0.5}{270}{0}
\psarc(3,1){0.5}{90}{180}\psarc(2,2){0.5}{270}{0}
\psarc(6,1){0.5}{90}{180}\psarc(0,2){0.5}{270}{0}
\psarc(1,1){0.5}{90}{180}\psarc(5,2){0.5}{270}{0}
\psarc(3,2){0.5}{90}{180}\psarc(2,3){0.5}{270}{0}
\psarc(6,2){0.5}{90}{180}\psarc(0,3){0.5}{270}{0}
\psarc(1,2){0.5}{90}{180}\psarc(5,3){0.5}{270}{0}
\psarc(2,3){0.5}{90}{180}\psarc(1,4){0.5}{270}{0}
\psarc(3,3){0.5}{90}{180}\psarc(2,4){0.5}{270}{0}
\psarc(4,3){0.5}{90}{180}\psarc(3,4){0.5}{270}{0}
\psarc(5,3){0.5}{90}{180}\psarc(4,4){0.5}{270}{0}
\psarc(1,3){0.5}{90}{180}\psarc(0,4){0.5}{270}{0}
\pscircle[fillstyle=solid,linecolor=black,fillcolor=black](1,0){0.15}
\pscircle[fillstyle=solid,linecolor=black,fillcolor=black](3,0){0.15}
\pscircle[fillstyle=solid,linecolor=black,fillcolor=black](5,0){0.15}
\pscircle[fillstyle=solid,linecolor=black,fillcolor=black](0,1){0.15}
\pscircle[fillstyle=solid,linecolor=black,fillcolor=black](2,1){0.15}
\pscircle[fillstyle=solid,linecolor=black,fillcolor=black](4,1){0.15}
\pscircle[fillstyle=solid,linecolor=black,fillcolor=black](6,1){0.15}
\pscircle[fillstyle=solid,linecolor=black,fillcolor=black](1,2){0.15}
\pscircle[fillstyle=solid,linecolor=black,fillcolor=black](3,2){0.15}
\pscircle[fillstyle=solid,linecolor=black,fillcolor=black](5,2){0.15}
\pscircle[fillstyle=solid,linecolor=black,fillcolor=black](0,3){0.15}
\pscircle[fillstyle=solid,linecolor=black,fillcolor=black](2,3){0.15}
\pscircle[fillstyle=solid,linecolor=black,fillcolor=black](4,3){0.15}
\pscircle[fillstyle=solid,linecolor=black,fillcolor=black](6,3){0.15}
\pscircle[fillstyle=solid,linecolor=black,fillcolor=black](1,4){0.15}
\pscircle[fillstyle=solid,linecolor=black,fillcolor=black](3,4){0.15}
\pscircle[fillstyle=solid,linecolor=black,fillcolor=black](5,4){0.15}
\end{pspicture}
\caption{An illustration of the bijection between a cross-topology configuration and a configuration with homotopy $\{0\}$. The second configuration of tiles is a shift towards the right of the first.
}
\label{fig:FKbij}
\end{center}
\end{figure}
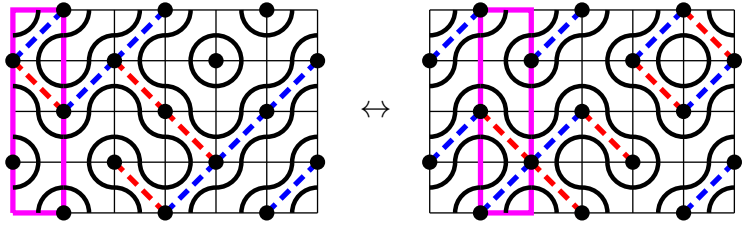

As for loops, homology properties are well-defined for these clusters, some of which are contractible and said to have cluster homotopy $\{0\}$. The contour of such a cluster is a union of loops with homotopy $\{0\}$. Other clusters have homotopy $\{a,b\}$ and wind around the cylinder $a$ times in the horizontal direction and $b$ times in the vertical direction (with $a \wedge b = 1$). The boundary of such a cluster contains two loops of homotopy $\{a,b\}$. Finally, unlike loops, a cluster of sites can wind around the torus in another nontrivial manner: by wrapping it in {\em both} directions. Such a cluster appears in the first configuration in Figure~\ref{fig:FKbij}. Such clusters are said to have {\it cross-topology}, or homotopy $\mathbb Z^2$. Only a single cross-topology cluster can appear in a given configuration, and it cannot coexist with any cluster with homotopy $\{a,b\}$. The boundary of a cluster with cross-topology consists of loops with homotopy $\{0\}$. By construction, FK configurations are characterized by the homotopy of their clusters that wind the torus nontrivially.

The FK weight of a configuration is  
\be 
W_{FK}(\sigma) = Q^{N_0} \Big(\!\prod_{a\wedge b=1}\!w_{a,b}^{N_{a,b}}\Big) w_{+}^{N_+} v_1^{N_{B_1}}v_2^{N_{B_2}}
\ee
where $Q$, $w_{a,b}$, $w_+$, $v_1$ and $v_2$ are all free parameters, while $N_0$, $N_{a,b}$, $N_+$ are the respective numbers of clusters homotopic to a point, of homotopy $\{a,b\}$, and with cross-topology. Only $N_+$ or one $N_{a,b}$ can be nonzero, and $N_+ \in \{0,1\}$. The partition function $Z_{FK}$ is the sum of $W_{FK}(\sigma)$ over all possible configurations and can be separated into contributions coming from each homotopy group. 
We thus write $\sigma(h)$ for the set of configurations $\sigma$ with homotopy $h$, and
$\sum_{\sigma(h)}$ for the sum over configurations with homotopy $h$. 

Our next objective is to show how to calculate $Z_{FK}$ at the critical point from $Z_L$, the partition function of the loop model. To proceed, we note that there exists an Euler relation between the cluster numbers $N_0$ and $N_{a,b}$, the total number of bonds $N_B = N_{B_1} + N_{B_2}$, the number of sites $N_s = NM/2$ and the number of loops in the diagram, $\#(\sigma) = n_0 + n_{a,b}$:
\be
N_0 + N_{a,b}=   \tfrac12 \big(\#(\sigma) + N_s - N_B \big).
\label{eq:euler}
\ee
This can be verified by first checking that it holds for the configuration $\sigma_0$ with zero bonds, where $N_0 = \#(\sigma_0) = NM/2$ and $N_{a,b}$ is zero. By adding bonds, one either reduces the number of clusters, adds cycles in clusters, creates $\{a,b\}$ clusters or creates a $\mathbb Z^2$ cluster, and equation \eqref{eq:euler} remains satisfied in all cases. 
We note that the number of cross-topology clusters, $N_+$, does not appear in the relation \eqref{eq:euler}.

With these observations, we can write the FK partition function as
\begin{align}
Z_{FK} &= \Big(\!\sum_{\sigma(\{0\})} + \sum_{a \wedge b = 1} \sum_{\sigma(\{a,b\})}\!\Big) Q^{N_0+N_{a,b}}\left(\frac{w_{a,b}}{Q}\right)^{N_{a,b}} v_1^{N_{B_1}}v_2^{N_{B_2}}+ w_+ \sum_{\sigma(\mathbb Z^2)} Q^{N_0} v_1^{N_{B_1}}v_2^{N_{B_2}}
\nonumber\\
&= \Big(\!\sum_{\sigma(\{0\})} + \sum_{a \wedge b = 1} \sum_{\sigma(\{a,b\})}\!\Big) Q^{ \frac12 ( \#(\sigma) + N_s)}\left(\frac{w_{a,b}}{Q}\right)^{\frac12 n_{a,b}} \left(\frac{v_1}{\sqrt{Q}}\right)^{N_{B_1}}\left(\frac{v_2}{\sqrt{Q}}\right)^{N_{B_2}} 
\nonumber\\ 
& \hspace{2cm}+ w_+ \sum_{\sigma(\mathbb Z^2)} Q^{ \frac12 ( \#(\sigma) + N_s)} \left(\frac{v_1}{\sqrt{Q}}\right)^{N_{B_1}}\left(\frac{v_2}{\sqrt{Q}}\right)^{N_{B_2}}\\
& =Q^{\frac12 N_s} \Big(\!\sum_{\sigma(\{0\})} + \sum_{a \wedge b = 1} \sum_{\sigma(\{a,b\})} + \,w_+ \sum_{\sigma(\mathbb Z^2)}\!\Big) Q^{\frac12 n_0}(w_{a,b})^{\frac12 n_{a,b}} \left(\frac{v_1}{\sqrt{Q}}\right)^{N_{B_1}}\left(\frac{v_2}{\sqrt{Q}}\right)^{N_{B_2}} 
\nonumber
\end{align}
where the equation \eqref{eq:euler} and the relation $N_{a,b} = \frac12 n_{a,b}$ were used in the second equality. Furthermore, our notation uses implicitly that, for configurations of homotopy $\{0\}$, $N_{a,b}=0$ for all $a\wedge b=1$.

Two more ingredients are required to proceed further.  First, we set the parameters $v_1$ and $v_2$ to critical values~\cite{Hintermann} by imposing the relation
\be
v_1 v_2 = Q.
\ee
These values are parameterized by the crossing parameter $\lambda$ and the spectral parameter $u$ through the relations
\be
 \frac{v_1}{\sqrt{Q}} =\frac{\sqrt{Q}}{v_2} =  \frac{\sin(\lambda-u)}{\sin u}, \qquad \sqrt Q = 2 \cos \lambda.
\label{eq:critcond}
\ee

The other key element is a bijection that exists between configurations with cross-topology and configurations with homotopy $\{0\}$, obtained by shifting every tile one position to the right. An example is given in Figure~\ref{fig:FKbij}. (Equivalently, one could interchange the distinguished and non-distinguished sublattices.) The number $n_0$ is the same in the two configurations, whereas the numbers of bonds typically differ. The bonds in the first configuration are all absent in the second one, and vice versa. For $i=1,2$, let $N_{B_i}$ and $N_{B_i}'$ denote the numbers of type $i$ bonds in the configuration with cross-topology and homotopy $\{0\}$, respectively. It then follows that 
\be
N_{B_1} + N'_{B_2}= NM/2 = N'_{B_1} + N_{B_2}
\ee
and hence
\be
 N_{B_1}-N_{B_2}=N'_{B_1}-N'_{B_2}.
\ee

From the above discussion, we now find
\be 
Z_{FK} = \frac{Q^{\frac12 N_s}}{K(u)} \Big(\! (1+ w_+)\mathcal Z(\{0\}) + \sum_{a \wedge b = 1} \mathcal Z(\{a,b\})\!\Big),\qquad
 K(u) = \big(\!\sin u \sin (\lambda - u)\big)^{\frac12 NM}
\ee
where
\begin{align}
\mathcal Z(\{0\}) & =  \sum_{\sigma (\{0\})}Q^{\frac12 n_0} (\sin u)^{\frac12 NM+N_{B_2}  - N_{B_1}} \big(\!\sin (\lambda - u)\big)^{\frac12 NM+N_{B_1} - N_{B_2}}, 
\label{eq:Zzz}\\
\mathcal Z(\{a,b\}) & = \sum_{\sigma (\{a,b\})}Q^{\frac12 n_0}(w_{a,b})^{\frac12 n_{a,b}} (\sin u)^{\frac12 NM+N_{B_2} - N_{B_1}} \big(\!\sin (\lambda - u)\big)^{\frac12 NM +N_{B_1} - N_{B_2}}.
\label{eq:Zab}
\end{align}
To express these in terms of the partition function of the loop model $Z_L$, we note that the powers of $\sin u$ and $\sin(\lambda - u)$ are given by
\be
n(
\begin{pspicture}(0,0.05)(0.3,0.25)
\psset{unit=0.3}
\psline[linewidth=0.5pt]{-}(0,0)(0,1)(1,1)(1,0)(0,0)
\psset{linecolor=myc2}
\psset{linewidth=1.0pt}
\psarc(0,0){0.5}{0}{90}\psarc(1,1){0.5}{180}{270}
\end{pspicture}
) = \tfrac12 NM+N_{B_2} - N_{B_1}
,\qquad
n(\begin{pspicture}(0,0.05)(0.3,0.25)
\psset{unit=0.3}
\psline[linewidth=0.5pt]{-}(0,0)(0,1)(1,1)(1,0)(0,0)
\psset{linecolor=darkgreen}
\psset{linewidth=1.0pt}
\psarc(1,0){0.5}{90}{180}\psarc(0,1){0.5}{270}{0}
\end{pspicture}) = \tfrac12 NM+N_{B_1} - N_{B_2}.
\ee 
By setting $\beta = \sqrt Q$ and $\alpha_{a,b} = \sqrt{w_{a,b}}$, we find $Z_L(\{a,b\}) = \mathcal Z(\{a,b\})$. The sum over configurations in $Z_L(\{0\})$ of the loop model contains all configurations with only loops that are homotopic to a point, and is therefore $\sum_{\sigma(\{0\})} +\sum_{\sigma(\mathbb Z^2)}$. By setting $\beta = \sqrt{Q}$ and using the previous bijection, we find $Z_L(\{0\}) = 2 \mathcal Z(\{0\})$. Finally, setting all non-contractible loop fugacities $\alpha_{a,b}$ to $0$ yields $\left.Z_L \right|_{\alpha_{a,b} = 0} = Z_L(\{0\})$. In conclusion, we have found that the FK partition function can be written as
\be
Z_{FK} = \frac{Q^{\frac12 N_s}}{K(u)} \left( \frac{w_+ - 1}2 \displaystyle{ Z_L \Big|_{ \substack{\tiny \alpha_{a,b} =\,0 \\ \beta = \sqrt Q}}} + \displaystyle{ Z_L \Big|_{ \substack{\tiny \alpha_{a,b} = \sqrt{w_{a,b}} \\ \beta = \sqrt Q}}} \right).
\label{eq:finalZFK}
\ee

\subsection[FK model for {$Q = 0$}]{FK model with \boldmath{$Q = 0$}}
\label{sec:FKcdp}

Even though the FK and loop models both are well-defined in the $\beta \to 0$ limit, the passage from the first to the second is awkward for $\beta = 0$: The 
factor $Q^{\frac12 N_s}$ in equation \eqref{eq:finalZFK} goes to zero, the critical condition requires that either $v_1$ or $v_2$ (or both $v_1$ and $v_2$) be zero, and the parameterization (\ref{eq:critcond}) of $v_1$ and $v_2$ in terms of the variable $u$ is ill-defined. 
Here we show how to treat the case $\sqrt{Q}=\beta = 0$ properly.

In the FK model, setting $Q = 0$ is equivalent to giving a weight zero to configurations with clusters of homotopy $\{0\}$. We implement this by writing
\be 
W_{FK}(\sigma) = \delta_{N_0,0}\Big(\!\prod_{a\wedge b=1}\!w_{a,b}^{N_{a,b}}\Big) w_{+}^{N_+} v_1^{N_{B_1}}v_2^{N_{B_2}}
\ee
and accordingly
\be 
Z_{FK} =  \sum_{a \wedge b = 1} \sum_{\sigma(\{a,b\})} \delta_{N_0,0}\, w_{a,b}^{N_{a,b}} v_1^{N_{B_1}}v_2^{N_{B_2}}+ w_+ \sum_{\sigma(\mathbb Z^2)} \delta_{N_0,0}\, v_1^{N_{B_1}}v_2^{N_{B_2}}.
\ee
To link the FK model with $Q=0$ to the corresponding loop model, 
we introduce a {\em restricted FK model} by imposing the further constraint on the configurations that 
\begin{center} 
{\it the clusters contain no contractible cycles of bonds.}
\end{center} 
This step is crucial. In the original FK model, loops with homotopy $\{0\}$ occur in three scenarios:   
They can act as a boundary for cross-topology clusters or for $\{0\}$ clusters, or appear if a non-contractible cluster contains a contractible cycle of bonds. With the added no-cycle constraint, the last two scenarios are prohibited. One can therefore replace $\delta_{N_0,0}$ by a similar constraint on $n_0$, the number of loops of homotopy $\{0\}$: $n_0=0$ if $\sigma \in \sigma(\{a,b\})$ but $n_0=1$ if $\sigma \in \sigma(\mathbb Z^2)$. 
The corresponding {\em restricted} partition function is denoted by $Z'_{FK}$ and,
from the bijection between $\{0\}$ and $\mathbb Z^2$ configurations, is given by 
\begin{align}
Z'_{FK} &= w_+ \sum_{\sigma(\mathbb{Z}^2)} \delta_{n_0,1}\, v_1^{N_{B_1}}v_2^{N_{B_2}}+\sum_{a \wedge b = 1} \sum_{\sigma(\{a,b\})} \delta_{n_0,0}\, w_{a,b}^{\frac12 n_{a,b}} v_1^{N_{B_1}}v_2^{N_{B_2}}\nonumber \\
&= \lim_{\beta\to 0} \Big( \frac{w^+}{\beta} \sum_{\sigma(\mathbb \{0\})}+\sum_{a \wedge b = 1} \sum_{\sigma(\{a,b\})} \Big) \beta^{n_0} w_{a,b}^{\frac12 n_{a,b}} v_1^{N_{B_1}}v_2^{N_{B_2}},
\label{ZFK0}
\end{align}
where we again use implicitly that $n_{a,b}=0$ for all configurations of homotopy $\{0\}$.
The limit $\beta\to 0$ in (\ref{ZFK0}) is well-defined. 

To relate this restricted FK model to the loop model, we set $v_1 v_2 = 1$ and choose the parameterization 
\be 
v_1 = (v_2)^{-1}= \frac{\sin(\lambda - u)}{\sin u},\qquad \beta = 2 \cos\lambda.
\ee
Then, using equations \eqref{eq:Zzz}, \eqref{eq:Zab} and
\be
 \lim_{\beta\to0}\,Z_L\big|_{\alpha_{a,b} = 0}=0,
\ee
one finds
\begin{align}
Z'_{FK} &= \lim_{\beta\to 0} \frac1{K(u)} \Big(\frac{w_+}{\beta} \mathcal Z(\{0\})\big|_{Q = \beta^2} +  \sum_{a \wedge b = 1} Z(\{a,b\})\big|_{Q = \beta^2}\Big) \nonumber\\
& = \frac1{K(u)} \lim_{\beta\to 0} \Big(Z_L\big|_{\alpha_{a,b} = \sqrt{w_{a,b}}}+\frac{w_+ - 2 \beta}{2\beta} \,Z_L\big|_{\alpha_{a,b} = 0} \Big) \nonumber\\
& =  \frac1{K(u)} \Big( \displaystyle{Z_L \Big|_{ \substack{\tiny \alpha_{a,b} = \sqrt{w_{a,b}} \\ \beta = 0 \hspace{1.0cm}}}+\frac{w_+}2\lim_{\beta \to 0} \frac{1}{\beta} \,Z_L\big|_{\alpha_{a,b} = 0}}\Big). 
\label{eq:FK0final}
\end{align}
This final result is quite different from equation 
 \eqref{eq:finalZFK} pertaining to $\beta\neq0$. This is not surprising since the criticality condition $v_1v_2=1$ imposed for $\beta=0$ ($Q=0$) is incompatible with the limit $\beta\to0$ ($Q\to0$) of the similar condition $v_1v_2=Q$ imposed for $\beta\neq0$.

\subsection{Enlarged periodic TL algebra}

Here we discuss the family of {\em Enlarged Periodic Temperley-Lieb algebras}\, $\mathcal EPTL_N(\alpha, \beta)$ labelled by $N$ and depending on the two parameters $\alpha \equiv \alpha_{1,0}$ and $\beta$. Our discussion will show how the partition function in the loop model $Z_L$ can be calculated from the eigenvalues of the transfer matrix $\Tb(u)$ in certain link-state representations of $\mathcal EPTL_N(\alpha, \beta)$.

\paragraph{Connectivities and the \boldmath{$\mathcal EPTL_N(\alpha, \beta)$} algebra:} 
On a rectangle, $N$ equidistant {\em nodes} are drawn on both the upper and lower edges. The rectangle is viewed as the planar representation of a band around a vertical cylinder, so that the left and right edges of the rectangle are identified along the so-called {\it virtual boundary}. A connectivity is then a connection of the $2N$ nodes by non-intersecting loop segments living between the upper and lower edges, 
such that every node is connected to one and only one other node. Loop segments can go through the virtual boundary, i.e.~around the back of the cylinder. For instance, 
\be 
c_1 = 
\psset{unit=0.5}
\begin{pspicture}(0,-0.2)(8.5,0.9)
\psline[linewidth=1pt]{-}(0.5,1)(8.5,1)
\psline[linewidth=1pt,linestyle=dotted]{-}(0.5,1)(0.5,-1)
\psline[linewidth=1pt,linestyle=dotted]{-}(8.5,1)(8.5,-1)
\psline[linewidth=1pt]{-}(8.5,-1)(0.5,-1)
\psset{linewidth=1pt}
\psdots(1,1)(2,1)(3,1)(4,1)(5,1)(6,1)(7,1)(8,1)(1,-1)(2,-1)(3,-1)(4,-1)(5,-1)(6,-1)(7,-1)(8,-1)
\psset{linecolor=myc}
\psarc(3.5,-1){0.5}{0}{180}
\psarc(7.5,-1){0.5}{0}{180}
\psbezier{-}(2,-1)(2,0.2)(5,0.2)(5,-1)
\psarc(0.5,1){0.5}{-90}{0}
\psarc(4.5,1){0.5}{180}{360}
\psarc(6.5,1){0.5}{180}{360}
\psarc(8.5,1){0.5}{180}{270}
\psbezier{-}(3,1)(3,0)(1,0)(1,-1)
\psbezier{-}(2,1)(2,0.25)(0.75,0)(0.5,0)
\psbezier{-}(6,-1)(6,-0.25)(8.25,0)(8.5,0)
\end{pspicture}\
\vspace{0.3cm}
\ee
is a connectivity. The product of connectivities is defined as follows: $c_2c_1$ is the connectivity obtained by drawing $c_1$ above $c_2$, reading the connectivity between the top and bottom $N$ nodes of the ensuing connectivity diagram, and multiplying this diagram by the factor $\alpha^{n_\alpha}\beta^{n_\beta}$, where $n_\beta$ is the number of contractible loops and $n_\alpha$ the number of non-contractible loops with homotopy $\{1,0\}$ in the diagram formed by drawing $c_1$ above $c_2$. Because non-contractible loops do not appear if $N$ is odd, the parameter $\alpha$ only comes into play for $N$ even. As an illustration, we consider

\be
c_2 c_1 = 
\psset{unit=0.5}
\begin{pspicture}(0,-1.3)(8.5,-0.3)
\psline[linewidth=1pt]{-}(0.5,1)(8.5,1)
\psline[linewidth=1pt,linestyle=dotted]{-}(0.5,1)(0.5,-3)
\psline[linewidth=1pt,linestyle=dotted]{-}(8.5,1)(8.5,-3)
\psline[linewidth=1pt]{-}(8.5,-1)(0.5,-1)
\psline[linewidth=1pt]{-}(8.5,-3)(0.5,-3)
\psset{linewidth=1pt}
\psset{linecolor=myc}
\psarc(3.5,-1){0.5}{0}{180}
\psarc(7.5,-1){0.5}{0}{180}
\psbezier{-}(2,-1)(2,0.2)(5,0.2)(5,-1)
\psarc(0.5,1){0.5}{-90}{0}
\psarc(4.5,1){0.5}{180}{360}
\psarc(6.5,1){0.5}{180}{360}
\psarc(8.5,1){0.5}{180}{270}
\psbezier{-}(3,1)(3,0)(1,0)(1,-1)
\psbezier{-}(2,1)(2,0.25)(0.75,0)(0.5,0)
\psbezier{-}(6,-1)(6,-0.25)(8.25,0)(8.5,0)
\psarc(6.5,-1){0.5}{180}{0}
\psarc(3.5,-1){-0.5}{0}{180}
\psarc(8.5,-1){0.5}{180}{270}
\psarc(0.5,-1){0.5}{-90}{0}
\psarc(2.5,-3){0.5}{0}{180}
\psarc(4.5,-3){0.5}{0}{180}
\psarc(7.5,-3){0.5}{0}{180}
\psbezier{-}(2,-1)(2,-1.75)(0.75,-2)(0.5,-2)
\psbezier{-}(1,-3)(1,-1.75)(6,-1.75)(6,-3)
\psbezier{-}(5,-1)(5,-2.25)(8,-2.25)(8.5,-2.)
\psset{linecolor=black}
\psdots(1,1)(2,1)(3,1)(4,1)(5,1)(6,1)(7,1)(8,1)(1,-1)(2,-1)(3,-1)(4,-1)(5,-1)(6,-1)(7,-1)(8,-1)(1,-3)(2,-3)(3,-3)(4,-3)(5,-3)(6,-3)(7,-3)(8,-3)
\end{pspicture}\, \, = \alpha \beta 
\begin{pspicture}(-0.1,-2.3)(8.5,-0.3)
\psline[linewidth=1pt,linestyle=dotted]{-}(0.5,-1)(0.5,-3)
\psline[linewidth=1pt,linestyle=dotted]{-}(8.5,-1)(8.5,-3)
\psline[linewidth=1pt]{-}(8.5,-1)(0.5,-1)
\psline[linewidth=1pt]{-}(8.5,-3)(0.5,-3)
\psset{linewidth=1pt}
\psdots(1,-1)(2,-1)(3,-1)(4,-1)(5,-1)(6,-1)(7,-1)(8,-1)(1,-3)(2,-3)(3,-3)(4,-3)(5,-3)(6,-3)(7,-3)(8,-3)
\psset{linecolor=myc}
\psarc(2.5,-1){-0.5}{0}{180}
\psarc(4.5,-1){-0.5}{0}{180}
\psarc(6.5,-1){-0.5}{0}{180}
\psarc(8.5,-1){0.5}{180}{270}
\psarc(0.5,-1){0.5}{-90}{0}
\psarc(2.5,-3){0.5}{0}{180}
\psarc(4.5,-3){0.5}{0}{180}
\psarc(7.5,-3){0.5}{0}{180}
\psbezier{-}(1,-3)(1,-1.75)(6,-1.75)(6,-3)
\end{pspicture} 
\label{c1c2}
\ee
\vspace{0.2cm}

The algebra $\mathcal EPTL_N(\alpha, \beta)$ is the infinite-dimensional algebra formed by the linear span of the connectivities endowed with the product just defined. It is generated by the connectivities $I, e_1, e_2, \ldots, e_N, \Omega$ and $\Omega^{-1}$:
\vspace{.15cm}
\begin{equation*} I =
\psset{unit=0.5}
\begin{pspicture}(0.1,-0.3)(8.5,0.8)
\psline[linewidth=1pt]{-}(0.5,1)(8.5,1)
\psline[linewidth=1pt,linestyle=dotted]{-}(0.5,1)(0.5,-1)
\psline[linewidth=1pt,linestyle=dotted]{-}(8.5,1)(8.5,-1)
\psline[linewidth=1pt]{-}(8.5,-1)(0.5,-1)
\psset{linewidth=1pt}
\psdots(1,1)(2,1)(3,1)(4,1)(5,1)(6,1)(7,1)(8,1)(1,-1)(2,-1)(3,-1)(4,-1)(5,-1)(6,-1)(7,-1)(8,-1)
\psset{linecolor=myc}
\psline(1,1)(1,-1)
\psline(2,1)(2,-1)
\rput(4.5,0){\dots}
\psline(3,1)(3,-1)
\psline(6,1)(6,-1)
\psline(7,1)(7,-1)
\psline(8,1)(8,-1)
\end{pspicture}\qquad 
e_i =
\begin{pspicture}(0.1,-0.3)(8.5,0.8)
\psline[linewidth=1pt]{-}(0.5,1)(8.5,1)
\psline[linewidth=1pt,linestyle=dotted]{-}(0.5,1)(0.5,-1)
\psline[linewidth=1pt,linestyle=dotted]{-}(8.5,1)(8.5,-1)
\psline[linewidth=1pt]{-}(8.5,-1)(0.5,-1)
\psset{linewidth=1pt}
\psdots(1,1)(2,1)(3,1)(4,1)(5,1)(6,1)(7,1)(8,1)(1,-1)(2,-1)(3,-1)(4,-1)(5,-1)(6,-1)(7,-1)(8,-1)
\psset{linecolor=myc}
\psarc(4.5,-1){0.5}{0}{180}\psarc(4.5,1){0.5}{180}{360}
\rput(2,0){\dots}\rput(7,0){\dots}
\rput(4,-1.6){$i$}
\psline(1,1)(1,-1)
\psline(3,1)(3,-1)
\psline(6,1)(6,-1)
\psline(8,1)(8,-1)
\end{pspicture}
\qquad
e_N =
\begin{pspicture}(0.1,-0.3)(8.5,0.8)
\psline[linewidth=1pt]{-}(0.5,1)(8.5,1)
\psline[linewidth=1pt,linestyle=dotted]{-}(0.5,1)(0.5,-1)
\psline[linewidth=1pt,linestyle=dotted]{-}(8.5,1)(8.5,-1)
\psline[linewidth=1pt]{-}(8.5,-1)(0.5,-1)
\psset{linewidth=1pt}
\psdots(1,1)(2,1)(3,1)(4,1)(5,1)(6,1)(7,1)(8,1)(1,-1)(2,-1)(3,-1)(4,-1)(5,-1)(6,-1)(7,-1)(8,-1)
\psset{linecolor=myc}
\psarc(8.5,-1){0.5}{90}{180}\psarc(8.5,1){0.5}{180}{270}
\psarc(0.5,-1){0.5}{0}{90}\psarc(0.5,1){0.5}{270}{0}
\rput(4.5,0){\dots}
\psline(2,1)(2,-1)
\psline(3,1)(3,-1)
\psline(6,1)(6,-1)
\psline(7,1)(7,-1)
\end{pspicture}
 \vspace{0.9cm}
\end{equation*}
\be 
\Omega =
\psset{unit=0.5}
\begin{pspicture}(0.1,-0.3)(8.5,0.8)
\psline[linewidth=1pt]{-}(0.5,1)(8.5,1)
\psline[linewidth=1pt,linestyle=dotted]{-}(0.5,1)(0.5,-1)
\psline[linewidth=1pt,linestyle=dotted]{-}(8.5,1)(8.5,-1)
\psline[linewidth=1pt]{-}(8.5,-1)(0.5,-1)
\psset{linewidth=1pt}
\psdots(1,1)(2,1)(3,1)(4,1)(5,1)(6,1)(7,1)(8,1)(1,-1)(2,-1)(3,-1)(4,-1)(5,-1)(6,-1)(7,-1)(8,-1)
\psset{linecolor=myc}
\psbezier{-}(1,1)(1,0.25)(0.625,0)(0.5,0)
\psbezier{-}(2,1)(2,0)(1,0)(1,-1)
\psbezier{-}(3,1)(3,0)(2,0)(2,-1)
\psbezier{-}(8,1)(8,0)(7,0)(7,-1)
\psbezier{-}(7,1)(7,0)(6,0)(6,-1)
\psbezier{-}(8,-1)(8,-0.25)(8.375,0)(8.5,0)
\rput(4.5,0){\dots}
\end{pspicture}\qquad 
\Omega^{-1} =
\begin{pspicture}(0.1,-0.3)(8.5,0.8)
\psline[linewidth=1pt]{-}(0.5,1)(8.5,1)
\psline[linewidth=1pt,linestyle=dotted]{-}(0.5,1)(0.5,-1)
\psline[linewidth=1pt,linestyle=dotted]{-}(8.5,1)(8.5,-1)
\psline[linewidth=1pt]{-}(8.5,-1)(0.5,-1)
\psset{linewidth=1pt}
\psdots(1,1)(2,1)(3,1)(4,1)(5,1)(6,1)(7,1)(8,1)(1,-1)(2,-1)(3,-1)(4,-1)(5,-1)(6,-1)(7,-1)(8,-1)
\psset{linecolor=myc}
\rput(4.5,0){\dots}
\psbezier{-}(1,-1)(1,-0.25)(0.625,0)(0.5,0)
\psbezier{-}(2,-1)(2,-0)(1,0)(1,1)
\psbezier{-}(3,-1)(3,-0)(2,0)(2,1)
\psbezier{-}(8,-1)(8,-0)(7,0)(7,1)
\psbezier{-}(7,-1)(7,-0)(6,0)(6,1)
\psbezier{-}(8,1)(8,0.25)(8.375,0)(8.5,0)
\end{pspicture}
\vspace{0.5cm}
\ee
where the index $i$ on the TL generators is understood to be modulo $N$,
with $e_0\equiv e_N$.
Products of connectivities then follow from the relations
\begin{alignat}{3}
e_i^2&= \beta e_i,&\qquad & \nonumber\\
e_ie_j&=e_je_i, &&
|i-j|>1,\nonumber\\
e_ie_{i\pm 1}e_i&=e_i,&& \nonumber\\
\Omega e_i \Omega^{-1} &= e_{i-1},&&\nonumber\\
\Omega \Omega^{-1} &= \Omega^{-1} \Omega = I, 
\label{eq:EPTLN}\\
(\Omega^{\pm 1} e_N)^{N-1} &= \Omega^{\pm N} (\Omega ^{\pm 1} e_N), \nonumber\\
\Omega^{\pm N} e_N \Omega^{\mp N} &= e_N,\nonumber\\
E \Omega^{\pm 1} E &= \alpha E, && 
E = e_2e_4\ldots e_{N-2}e_N, \nonumber 
\end{alignat}
where the last equation is for $N$ even only. It is noted that $\mathcal EPTL_N(\alpha,\beta)$ is actually generated by $\Omega$, $\Omega^{-1}$ and any one of the TL generators $e_i$.
The enlarged periodic TL algebra is the quotient of the affine Temperley-Lieb algebra~\cite{MS1993,Green1998,GL1998,EG1999} by the last relation in (\ref{eq:EPTLN}). 

\paragraph{The transfer matrix:} 
The loop transfer matrix is an element of $\mathcal EPTL_N(\alpha, \beta)$ given diagrammatically by
\be 
\Tb(u) = \quad
\psset{unit=1}
\psset{linewidth=1pt}
\overbrace{
\begin{pspicture}(-0,0.375)(5,1.2)
\pspolygon[fillstyle=solid,fillcolor=lightlightblue](0,0)(5,0)(5,1)(0,1)(0,0)
\psdots(0.5,0)(1.5,0)(4.5,0)
\psdots(0.5,1)(1.5,1)(4.5,1)
\lw
\psline{-}(-0.15,0.5)(0.0,0.5)
\psline{-}(5.15,0.5)(5.0,0.5)
\unlw
\psline{-}(0,0)(1,0)(1,1)(0,1)(0,0)\psarc[linewidth=0.5pt]{-}(0,0){0.15}{0}{90}\rput(0.5,0.5){$u$}
\psline{-}(1,0)(2,0)(2,1)(1,1)(1,0)\psarc[linewidth=0.5pt]{-}(1,0){0.15}{0}{90}\rput(1.5,0.5){$u$}
\psline{-}(3,0)(3,1)\rput(2.5,0.5){$\dots\hspace{0.02cm}$}\rput(3.5,0.5){$\dots\hspace{0.02cm}$}
\psline{-}(4,0)(5,0)(5,1)(4,1)(4,0)\psarc[linewidth=0.5pt]{-}(4,0){0.15}{0}{90}\rput(4.5,0.5){$u$}
\psset{linecolor=myc}\unlw
\end{pspicture}}^N \quad
\ee
where
\be
\psset{unit=1}
\psset{linewidth=1pt}
\begin{pspicture}(-0.5,-0.1)(0.5,0.75)
\pspolygon[fillstyle=solid,fillcolor=lightlightblue](-0.5,-0.5)(0.5,-0.5)(0.5,0.5)(-0.5,0.5)(-0.5,-0.5)
\psarc[linewidth=0.5pt]{-}(-0.5,-0.5){0.15}{0}{90}
\rput(0,0){$u$}
\end{pspicture}\ =\ \sin(\lambda-u)\ \ 
\begin{pspicture}(-0.5,-0.1)(0.5,0.5)
\pspolygon[fillstyle=solid,fillcolor=lightlightblue](-0.5,-0.5)(0.5,-0.5)(0.5,0.5)(-0.5,0.5)(-0.5,-0.5)
\psset{linecolor=blue,linewidth=1.5pt}
\psarc{-}(0.5,-0.5){0.5}{90}{180}
\psarc{-}(-0.5,0.5){0.5}{270}{360}
\end{pspicture}\ +\ \sin u\ \ 
\begin{pspicture}(-0.5,-0.1)(0.5,0.5)
\pspolygon[fillstyle=solid,fillcolor=lightlightblue](-0.5,-0.5)(0.5,-0.5)(0.5,0.5)(-0.5,0.5)(-0.5,-0.5)
\psset{linecolor=blue,linewidth=1.5pt}
\psarc{-}(-0.5,-0.5){0.5}{0}{90}
\psarc{-}(0.5,0.5){0.5}{180}{270}
\end{pspicture}
\ \ =\ \ 
\begin{pspicture}(-0.5,-0.1)(0.5,0.5)
\pspolygon[fillstyle=solid,fillcolor=lightlightblue](-0.5,-0.5)(0.5,-0.5)(0.5,0.5)(-0.5,0.5)(-0.5,-0.5)
\psarc[linewidth=0.5pt]{-}(0.5,-0.5){0.15}{90}{180}
\rput(0,0){\small $\lambda\!-\!u$}
\end{pspicture}\, \, ,
\ee
\\
$\beta = 2 \cos \lambda$, and $u$ is the spectral parameter.

\paragraph{Link states and twist representations \boldmath{$\omega_d$}:} 
Let $N$ equidistant nodes be drawn on a closed horizontal curve wrapping a cylinder. A link state is then a set of non-intersecting loop and line segments, drawn on the cylinder above the closed horizontal curve, linking the nodes pairwise or attaching vertical line segments to nodes. These vertical line segments are called defects and a loop segment between a pair of nodes is not allowed to connect above a defect. The number of defects of a link state is denoted by $d$, an integer that lies in the range $0, \ldots, N$ and subject to the constraint $N-d = 0\! \mod 2$. For instance, the link state

\be
w =
\psset{unit=0.5}
\begin{pspicture}(0.1,0.9)(8.5,1.8)
\psline[linewidth=1pt]{-}(0.5,1)(8.5,1)
\psset{linewidth=1pt}
\psdots(1,1)(2,1)(3,1)(4,1)(5,1)(6,1)(7,1)(8,1)
\psset{linecolor=myc2}
\psarc(0.5,1){0.5}{0}{90}
\psarc(8.5,1){0.5}{90}{180}
\psarc(5.5,1){0.5}{0}{180}
\psbezier(4,1)(4,2.25)(7,2.25)(7,1)
\psline(2,1)(2,2)
\psline(3,1)(3,2)
\end{pspicture}
\ee
has $2$ defects. As indicated, it is convenient to depict link states on a horizontal line segment with the left and right ends identified. The set of link states with $N$ nodes and $d$ defects is denoted by $B_N^d$ and its linear span, $V_N^d$, 
has dimension
\be
 \dim V_N^d =\left(\!\! \!\begin{array}{c} N \\  \frac{N-d}{2} \end{array}\!\!\!\right).
\label{Bdim}
\ee 
The counting of link states stems from a bijection previously exploited in \cite{AMDYSAinprep} between link states with $d$ defects and spin configurations $(x_1, \dots, x_N)$ with $x_i \in \pm1$ and $\sum_i x_i = d$: If position $i$ is occupied by a defect or if it marks the beginning of a half-arc (it connects to a position $j$ by going towards the right), $x_i= +1$. Otherwise $i$ marks the end of a half-arc (it connects to a position $j$ by going towards the left) and $x_i = -1$. The number of link states with $\frac{N-d}2$ half-arcs is equal to the number of states $(x_1, \dots, x_N)$ with $\frac{N-d}2$ down arrows, and \eqref{Bdim} follows readily.

To define link-state representations of $\mathcal EPTL_N(\alpha, \beta)$, one must define an action $cw$ of connectivities on link states
that respects $c_2(c_1 w) = (c_2 c_1) w$ for all connectivities $c_1,c_2$ and link states $w$. 
The action defining the {\em twist representation} $\omega_d$~\cite{AMDYSAinprep} is the following. First, the link state $w$ is drawn above $c$, with the nodes of $w$ connected to the $N$ nodes of the top edge of $c$. If any two defects of $w$ are connected in the diagram, the result is zero. Otherwise, $cw$ is equal to the link state obtained by reading the connection of the lower $N$ nodes of $c$, multiplied by the scalar factor $ \alpha^{n_\alpha}\beta^{n_\beta} v^\Delta$ where $n_\beta$ and $n_\alpha$ are respectively the number of loops in the diagram with homotopy $\{0\}$ and $\{1,0\}$, $v$ is the {\em winding parameter} and $\Delta$ is the total winding of the defects, calculated as follows. If $cw \neq 0$, any defect of $w$ travels across the connectivity and connects with a node on the lower edge of $c$. Let the defects of $w$ be labelled by $i$, $1\leq i\leq d$. One can then calculate the distance $\Delta_i\in \mathbb{Z}$ traveled by the defect $i$, 
\be 
\Delta_i = \big[ \textrm{initial position of $i$} \big] - \big[\textrm{final position of $i$}\big],
\ee
where the nodes of the link state are in positions $1,\ldots,N$.
In order for $\Delta_i$ to measure a distance, we let the final position of $i$ exit the interval $1, \ldots, N$ if the virtual boundary is crossed. Defects traveling towards the left or right result respectively in positive and negative powers of $v$. The total winding is then the sum of the windings of all the defects, 
\be 
\Delta = \sum_i \Delta_i.
\ee
As the number of defects is conserved, this defines a representation $\omega_d$ of $\mathcal EPTL_N(\alpha, \beta)$ for every allowed value of $d$. The non-contractible loop fugacity $\alpha$ comes into play in the $\omega_0$ representation only, while the winding parameter $v$ appears in all the representations with $d>0$. Here are three examples of the action of $cw$:
\begin{align} c_1w &=
\psset{unit=0.5}
\begin{pspicture}(0.1,0.0)(8.5,1.8)
\psline[linewidth=1pt]{-}(0.5,1)(8.5,1)
\psline[linewidth=1pt,linestyle=dotted]{-}(0.5,1)(0.5,-1)
\psline[linewidth=1pt,linestyle=dotted]{-}(8.5,1)(8.5,-1)
\psline[linewidth=1pt]{-}(8.5,-1)(0.5,-1)
\psset{linewidth=1pt}
\psdots(1,1)(2,1)(3,1)(4,1)(5,1)(6,1)(7,1)(8,1)(1,-1)(2,-1)(3,-1)(4,-1)(5,-1)(6,-1)(7,-1)(8,-1)
\psset{linecolor=myc2}
\psarc(0.5,1){0.5}{0}{90}
\psarc(8.5,1){0.5}{90}{180}
\psarc(5.5,1){0.5}{0}{180}
\psbezier(4,1)(4,2.25)(7,2.25)(7,1)
\psline(2,1)(2,2)
\psline(3,1)(3,2)
\psset{linecolor=myc}
\psarc(3.5,-1){0.5}{0}{180}
\psarc(7.5,-1){0.5}{0}{180}
\psbezier{-}(2,-1)(2,0.2)(5,0.2)(5,-1)
\psarc(0.5,1){0.5}{-90}{0}
\psarc(4.5,1){0.5}{180}{360}
\psarc(6.5,1){0.5}{180}{360}
\psarc(8.5,1){0.5}{180}{270}
\psbezier{-}(3,1)(3,0)(1,0)(1,-1)
\psbezier{-}(2,1)(2,0.25)(0.75,0)(0.5,0)
\psbezier{-}(6,-1)(6,-0.25)(8.25,0)(8.5,0)
\end{pspicture}
\,\, \, \,=\beta^2 v^6
\begin{pspicture}(0.1,0.9)(8.5,1.8)
\psline[linewidth=1pt]{-}(0.5,1)(8.5,1)
\psset{linewidth=1pt}
\psdots(1,1)(2,1)(3,1)(4,1)(5,1)(6,1)(7,1)(8,1)
\psset{linecolor=myc2}
\psarc(7.5,1){0.5}{0}{180}
\psarc(3.5,1){0.5}{0}{180}
\psbezier(2,1)(2,2.25)(5,2.25)(5,1)
\psline(1,1)(1,2)
\psline(6,1)(6,2)
\end{pspicture}\\ \nonumber \\ \nonumber \\
c_2w &=
\psset{unit=0.5}
\begin{pspicture}(0.1,0.0)(8.5,1.8)
\psline[linewidth=1pt]{-}(0.5,1)(8.5,1)
\psline[linewidth=1pt,linestyle=dotted]{-}(0.5,1)(0.5,-1)
\psline[linewidth=1pt,linestyle=dotted]{-}(8.5,1)(8.5,-1)
\psline[linewidth=1pt]{-}(8.5,-1)(0.5,-1)
\psset{linewidth=1pt}
\psdots(1,1)(2,1)(3,1)(4,1)(5,1)(6,1)(7,1)(8,1)(1,-1)(2,-1)(3,-1)(4,-1)(5,-1)(6,-1)(7,-1)(8,-1)
\psset{linecolor=myc2}
\psarc(0.5,1){0.5}{0}{90}
\psarc(8.5,1){0.5}{90}{180}
\psarc(5.5,1){0.5}{0}{180}
\psbezier(4,1)(4,2.25)(7,2.25)(7,1)
\psline(2,1)(2,2)
\psline(3,1)(3,2)
\psset{linecolor=myc}
\psarc(6.5,1){0.5}{180}{0}
\psarc(3.5,1){-0.5}{0}{180}
\psarc(8.5,1){0.5}{180}{270}
\psarc(0.5,1){0.5}{-90}{0}
\psarc(2.5,-1){0.5}{0}{180}
\psarc(4.5,-1){0.5}{0}{180}
\psarc(7.5,-1){0.5}{0}{180}
\psbezier{-}(2,1)(2,0.25)(0.75,0)(0.5,0)
\psbezier{-}(1,-1)(1,0.25)(6,0.25)(6,-1)
\psbezier{-}(5,1)(5,-0.25)(8,-0.25)(8.5,-0.)
\end{pspicture}\,\,
\,\, \,=0  \\ \nonumber \\ \nonumber \\
c_2w' &=
\psset{unit=0.5}
\begin{pspicture}(0.1,0.0)(8.5,1.8)
\psline[linewidth=1pt]{-}(0.5,1)(8.5,1)
\psline[linewidth=1pt,linestyle=dotted]{-}(0.5,1)(0.5,-1)
\psline[linewidth=1pt,linestyle=dotted]{-}(8.5,1)(8.5,-1)
\psline[linewidth=1pt]{-}(8.5,-1)(0.5,-1)
\psset{linewidth=1pt}
\psdots(1,1)(2,1)(3,1)(4,1)(5,1)(6,1)(7,1)(8,1)(1,-1)(2,-1)(3,-1)(4,-1)(5,-1)(6,-1)(7,-1)(8,-1)
\psset{linecolor=myc2}
\psarc(0.5,1){0.5}{0}{90}
\psarc(8.5,1){0.5}{90}{180}
\psarc(3.5,1){0.5}{0}{180}
\psbezier(2,1)(2,2.25)(5,2.25)(5,1)
\psbezier(7,1)(7,2.2)(8.4,2.2)(8.5,2.2)
\psbezier(6,1)(6,2.7)(1,2.5)(0.5,2.2)
\psset{linecolor=myc}
\psarc(6.5,1){0.5}{180}{0}
\psarc(3.5,1){-0.5}{0}{180}
\psarc(8.5,1){0.5}{180}{270}
\psarc(0.5,1){0.5}{-90}{0}
\psarc(2.5,-1){0.5}{0}{180}
\psarc(4.5,-1){0.5}{0}{180}
\psarc(7.5,-1){0.5}{0}{180}
\psbezier{-}(2,1)(2,0.25)(0.75,0)(0.5,0)
\psbezier{-}(1,-1)(1,0.25)(6,0.25)(6,-1)
\psbezier{-}(5,1)(5,-0.25)(8,-0.25)(8.5,-0.)
\end{pspicture}
\,\, \, \,=\alpha^2\beta^2 
\begin{pspicture}(0.1,0.9)(8.5,1.8)
\psline[linewidth=1pt]{-}(0.5,1)(8.5,1)
\psset{linewidth=1pt}
\psdots(1,1)(2,1)(3,1)(4,1)(5,1)(6,1)(7,1)(8,1)
\psset{linecolor=myc2}
\psarc(2.5,1){0.5}{0}{180}
\psarc(4.5,1){0.5}{0}{180}
\psarc(7.5,1){0.5}{0}{180}
\psbezier(1,1)(1,2.375)(6,2.375)(6,1)
\end{pspicture}
\end{align}
\\[-.1cm]
In the first example, the first defect has $\Delta_1 = 4$ and the second $\Delta_2 = 2$, yielding a total winding of $\Delta = 6$. We stress that the winding parameter $v$ is a parameter of the representation $\omega_d$ and not a parameter of the lattice model.

\paragraph{Closing the cylinder to form the geometric torus:} 
Our interest in the enlarged periodic TL algebra $\mathcal EPTL_N(\alpha, \beta)$ stems from the torus loop model discussed in 
Section~\ref{sec:loopmodel}. 
Indeed, we now proceed to show how the partition function of the loop model on the torus can be expressed in terms of $\Tb^M(u)$,
where $\Tb^M(u)$ is defined as the (vertical) concatenation of $M$ copies of the transfer tangle $\Tb(u)$.

As an element of $\mathcal EPTL_N(\alpha, \beta)$, $\Tb^M(u)$ is a weighted sum over all $2^{MN}$ configurations $\sigma$ appearing in $Z_L$. Each such weight is given by an expression of the form
\be
\alpha^{n_\alpha}\beta^{n_\beta} 
(\sin u)^{n(
\begin{pspicture}(0,0.05)(0.25,0.25)
\psset{unit=0.25}
\psline[linewidth=0.5pt]{-}(0,0)(0,1)(1,1)(1,0)(0,0)
\psset{linecolor=darkgreen}
\psset{linewidth=1.0pt}
\psarc(1,0){0.5}{90}{180}\psarc(0,1){0.5}{270}{0}\end{pspicture}
)}\big(\!\sin(\lambda-u)\big)^{n(
\begin{pspicture}(0,0.05)(0.25,0.25)
\psset{unit=0.25}
\psline[linewidth=0.5pt]{-}(0,0)(0,1)(1,1)(1,0)(0,0)
\psset{linecolor=myc2}
\psset{linewidth=1.0pt}
\psarc(0,0){0.5}{0}{90}\psarc(1,1){0.5}{180}{270}
\end{pspicture}
)}
\label{eq:loopnumbers}
\ee
where $n_\alpha$ and $n_\beta$ are the numbers of loops with homotopy $\{0\}$ and $\{1,0\}$ that do {\em not} cross the upper and lower horizontal edges.

However, this is different from the weight in $Z_L$ of the corresponding configuration on the torus: 
Along with the factors of $\beta$ and $\alpha$ for the $\{0\}$ and $\{1,0\}$ loops that {\em do} touch the upper and lower horizontal edges, the factors $\alpha_{a,b}$ for loops with homotopy $\{a,b\}$ are also missing. 
(Here and in the following, we must exclude $(a,b)=(1,0)$ in order to avoid counting twice the loops of homotopy $\{1,0\}$.) 
To assign the correct weight to a configuration on the torus, we thus define the {\em linear functional} 
\be
\mathcal F: \ \mathcal EPTL_N(\alpha, \beta) \rightarrow \mathbb C,\qquad \mathcal F(c)=\alpha^{\bar n_\alpha} \beta^{\bar n_\beta}
\!\!\!\!\prod_{\substack{a\wedge b=1\\[.07cm] (a,b)\neq(1,0)}}\!\!\!\!\!\alpha_{a,b}^{\bar n_{a,b}}
\label{eq:operatorF}
\ee
where an $\bar n$ counts the loops crossing the horizontal edge along which the cylinder is glued into a torus. It is evident that every loop with homotopy $\{a,b\}\neq \{1,0\}$ crosses this edge, so $n_{a,b} = \bar n_{a,b}$. 

As an illustration of the evaluation of $\mathcal{F}$, let us first view the particular configuration in Figure~\ref{fig:loopconfig} as a product of connectivities in $\mathcal EPTL_N(\alpha, \beta)$. The loop numbers in \eqref{eq:loopnumbers} are then $n_\alpha=0$ and $n_\beta=2$. Removing the $n_\beta=2$ contractible loops from the configuration yields the connectivity
\begin{equation}
\psset{unit=0.6}
\begin{pspicture}(-0,2)(6,4.2)
\rput(-1,2){$c=$}
\psdots[linewidth=1pt](0.5,4)(1.5,4)(2.5,4)(3.5,4)(4.5,4)(5.5,4)(0.5,0)(1.5,0)(2.5,0)(3.5,0)(4.5,0)(5.5,0)
\psset{linewidth=1.7pt}
\psset{linecolor=myc}
\psarc{-}(5,4){0.5}{180}{0}
\psarc{-}(2,4){0.5}{180}{0}
\psarc{-}(2,0){0.5}{0}{180}
\psarc{-}(5,0){0.5}{0}{180}
\psarc{-}(0,4){0.5}{-90}{0}
\psbezier{-}(0.5,0)(0.5,2)(6.5,2)(6.5,4)
\psbezier{-}(3.5,0)(3.5,2)(9.5,2)(9.5,4)
\psbezier{-}(0,1.825)(1.5,2.40)(3.5,2.82)(3.5,4)
\psset{linecolor=black}\unlw
\psframe[fillstyle=solid,linecolor=white,linewidth=0pt](6,0)(10,4)
\psframe[fillstyle=solid,linecolor=white,linewidth=0pt](-0.4,0)(0,4)
\psline[linewidth=1.5pt](0,0)(0,4)(6,4)(6,0)(0,0)
\end{pspicture}
\label{eq:loopconnect}
\end{equation} 
\vspace{0.6cm}

\noindent and it readily follows that $\mathcal F(c) = \beta^2 \alpha_{1,1}^2$.

In (\ref{eq:operatorF}), the linear functional $\mathcal{F}$ is defined {\em algebraically} as an operator mapping connectivities $c\in\mathcal EPTL_N(\alpha, \beta)$ onto $\mathbb{C}$. By construction, $\mathcal{F}$ can be interpreted {\em geometrically} as gluing together the horizontal edges of a cylinder on which the loop configuration corresponding to a connectivity $c$ is realized. Indeed, $\mathcal{F}$ has been designed such that the value $\mathcal{F}(c)\in\mathbb{C}$ is {\em identical} to the weight of the loop configuration on the torus resulting from this gluing process. 

It is recalled that the product $cc'$ of two connectivities is itself a connectivity multiplied by a constant. Let us denote the resulting connectivity by $c''$ and the constant by $\kappa$. We thus have $cc'=\kappa c''$ and $\mathcal{F}(cc')=\kappa\mathcal{F}(c'')$ is the weight of the loop configuration on the torus corresponding to the product $cc'$.
The constant $\kappa$ is given by the product of the fugacities of the loops in this configuration that do not intersect the horizontal edges of the cylinder, while the fugacities of the loops that do intersect the edges are incorporated in $\mathcal{F}(c'')$. Since $c'c$ may differ from $cc'=\kappa c''$, the separation into the similar contributions coming from $c'c$ may be different.
On the other hand, because the two products $cc'$ and $c'c$ are equivalent as loop configurations on the torus,
it readily follows that the two products $cc'$ and $c'c$ are mapped to the same product of fugacities by $\mathcal{F}$.
We have thus established that $\mathcal{F}$ has the cyclicity property
\be
 \mathcal{F}(cc')=\mathcal{F}(c'c),\qquad c,c'\in\mathcal EPTL_N(\alpha, \beta).
\label{Fcycl}
\ee 
Being also linear, $\mathcal{F}$ is therefore a {\em trace operator} on $\mathcal EPTL_N(\alpha, \beta)$.

By construction, the partition function of the loop model on the torus is given by
\be
Z_L = \mathcal F\big(\Tb^M(u)\big).
\label{ZLF}
\ee
As we will see below, 
$Z_L$ can be expressed as a {\em modified matrix trace} in the twist representations $\omega_d(\Tb^M(u))$.

\subsection{Traces in the twist representations}
\label{sec:traces}

Our next goal is to write the linear functional $\mathcal F$ in terms of traces in the twist representations. The main idea is to use the winding parameter $v$ to decompose the traces of $\omega_d(c)$ into Fourier modes and to assign the correct weights to each mode to produce $\mathcal F(c)$. 
Because $\mathcal F$ is linear, it can be applied to linear combinations of connectivities in $\mathcal EPTL_N(\alpha,\beta)$ such as $\Tb^M(u)$. 
As written in (\ref{eq:Z}) below, the expression for $\mathcal{F}(c)$ applies to connectivities $c$ with a maximum of $M$ strands crossing the virtual boundary.
Here we write the Fourier decomposition in integral form and introduce the convenient notation
\be
 \mathrm{Tr}_d(c)=\mathrm{Tr}\,\omega_d(c).
\ee 
The final result is the following proposition which we prove in the remainder of this section.
\begin{Proposition} The partition function for the loop model on an $M \times N$ torus is given by
\be
Z_L = \mathcal F\big(\Tb^M(u)\big) \qquad \mathrm{where} \qquad
\left\{\begin{array}{l} \mathcal F(c) = \displaystyle{\frac1{2\pi}\int_0^{2 \pi} d\mu \sum_d \mathcal G_d(\mu, \boldsymbol \alpha) \mathrm{Tr}_{d}(c)} , \vspace{0.2cm}
\\   \mathcal G_0(\mu, \boldsymbol\alpha) = t_0 = 1, \vspace{0.2cm}
\\   \mathcal G_d(\mu, \boldsymbol\alpha) = \displaystyle{\sum_{k=-M}^{M} v^{-Nk} C_{k,d},} \quad (d>0) \vspace{0.2cm}
\\  C_{k,d} = t_{k\wedge d}(\alpha_{\frac{k}{k \wedge d},\frac{d}{k \wedge d}}), \vspace{0.22cm}
\\  t_n(\alpha) = (2- \delta_{n,0}) T_n(\frac{\alpha}{2}).  \end{array}\right.
\label{eq:Z}
\ee
The winding parameter is parameterized as $v = e^{i \mu}$ and $T_n(x)$ is the $n$-th Chebyshev polynomial of the first kind, given by $T_n(\cos \theta) = \cos(n \theta)$. The role of the integral is only to select out the constant term in the Laurent expansion in $v = e^{i \mu}$.
\label{sec:Traceprop}
\end{Proposition}

For a given connectivity $c$, the matrix elements of $\omega_d(c)$ are powers of $\alpha$ and $\beta$, but most of them are zero. In fact, in every column of $\omega_d(c)$, {\it at most one} element is nonzero. Some matrices, such as $\omega_N(e_i)$, are simply equal to the zero matrix. To calculate $\mathrm{Tr}_d(c)$, we search for the link states in $B_N^d$ that are eigenvectors of $c$ with eigenvalue different from $0$ (for generic $\alpha$ and $\beta$). 
We will refer to such eigenstates as {\em link eigenstates}.

\paragraph{An example:} 
We start by working out an instructive example in detail: the connectivity $c = \Omega^4 \in \mathcal E PTL_8(\alpha, \beta)$,
\begin{equation*} 
\psset{unit=0.5}\overbrace{
\begin{pspicture}(0.7,-1.9)(8.2,2.4)
\rput(-1.4,0){$_{|i| =4}$}
\rput(-0.2,0){ $\Bigg \{$}
\psline[linewidth=1pt]{-}(0.5,2)(8.5,2)
\psline[linewidth=1pt,linestyle=dotted]{-}(0.5,2)(0.5,-2)
\psline[linewidth=1pt,linestyle=dotted]{-}(8.5,2)(8.5,-2)
\psline[linewidth=1pt]{-}(8.5,-2)(0.5,-2)
\psset{linewidth=1pt}
\psdots(1,2)(2,2)(3,2)(4,2)(5,2)(6,2)(7,2)(8,2)(1,-2)(2,-2)(3,-2)(4,-2)(5,-2)(6,-2)(7,-2)(8,-2)
\psset{linecolor=myc}
\psbezier{-}(4,2)(4,0.7)(1,-0.2)(0.5,-1.2)
\psbezier{-}(3,2)(3,0.9)(0.75,0.1)(0.5,-0.4)
\psbezier{-}(2,2)(2,1.2)(0.75,0.6)(0.5,0.4)
\psbezier{-}(1,2)(1,1.6)(0.75,1.4)(0.5,1.2)
\psbezier{-}(1,-2)(1,-0.5)(5,0.5)(5,2)
\psbezier{-}(2,-2)(2,-0.5)(6,0.5)(6,2)
\psbezier{-}(3,-2)(3,-0.5)(7,0.5)(7,2)
\psbezier{-}(4,-2)(4,-0.5)(8,0.5)(8,2)
\psbezier{-}(5,-2)(5,-0.7)(8,0.2)(8.5,1.2)
\psbezier{-}(6,-2)(6,-0.9)(8.25,-0.1)(8.5,0.4)
\psbezier{-}(7,-2)(7,-1.2)(8.25,-0.6)(8.5,-0.4)
\psbezier{-}(8,-2)(8,-1.6)(8.25,-1.4)(8.5,-1.2)
\end{pspicture}}^{N=8}
\end{equation*}
As indicated in the illustration of this connectivity, four strands are crossing the virtual boundary. Thus, if $\Omega^4$ is realized on an $M\times N$ torus (with $N=8$), $M$ necessarily satisfies $M\ge 4$, while the ensuing loop configuration on the torus has four loops with homotopy $\{1,2\}$. It follows that
\be
 \mathcal F(\Omega^4)\big|_{N=8} = \alpha_{1,2}^4.
\ee 
We now proceed to verify that the prescription \eqref{eq:Z} yields the same result. 

The link pattern with $8$ defects is an eigenstate of $c$, with eigenvalue $v^{32}$. A simple way to calculate the winding $\Delta$ of such an eigenstate is to count the number of loop segments connected to defects that cross the virtual boundary and multiply this number by $N$. Examples of other link eigenstates of $c$ are
$
\psset{unit=0.3}
\begin{pspicture}(0.1,0.9)(8.5,1.8)
\psline[linewidth=1pt]{-}(0.5,1)(8.5,1)
\psset{linewidth=0.75pt}
\psdots(1,1)(2,1)(3,1)(4,1)(5,1)(6,1)(7,1)(8,1)
\psset{linecolor=myc2}
\psarc(0.5,1){0.5}{0}{90}
\psarc(4.5,1){0.5}{0}{180}
\psarc(8.5,1){0.5}{90}{180}
\psline(2,1)(2,2)
\psline(3,1)(3,2)
\psline(7,1)(7,2)
\psline(6,1)(6,2)
\end{pspicture}
$
\,and\!
$
\psset{unit=0.3}
\begin{pspicture}(0.1,0.9)(8.5,1.8)
\psline[linewidth=1pt]{-}(0.5,1)(8.5,1)
\psset{linewidth=0.75pt}
\psdots(1,1)(2,1)(3,1)(4,1)(5,1)(6,1)(7,1)(8,1)
\psset{linecolor=myc2}
\psbezier(1,1)(1,2.25)(4,2.25)(4,1)
\psbezier(5,1)(5,2.25)(8,2.25)(8,1)
\psarc(2.5,1){0.5}{0}{180}
\psarc(6.5,1){0.5}{0}{180}
\end{pspicture}$
with eigenvalues $v^{16}$ and $1$, respectively. Only a fraction of the four loop segments that cross the virtual boundary are connected to defects; 
$\frac{1}{2}$ and $0$ for the first and second state, respectively. The connectivity $c$ has no link eigenstates with $2$ or $6$ defects. For which sectors do these eigenstates exist and how many are there? If one takes any link state in $B_4^{d'}$ for $d' = 0, 2, 4$ and places two copies of it side by side, thereby forming a state in $B_8^{2d'}$, the result is an eigenstate of $c$ with eigenvalue $v^{8d'}$\!. 
Since this procedure to construct link eigenstates is exhaustive,
the number of link eigenstates of $\Omega^4$ in the sectors $d = 0, 4$ and $8$ are simply the number of states in $B_4^{0}$, $B_4^{2}$ and $B_4^{4}$, respectively. It therefore follows that
\be
\mathrm{Tr}_8(\Omega^4) = v^{32}, \quad \mathrm{Tr}_4(\Omega^4) = \begin{pmatrix} 4 \\ 1\end{pmatrix}v^{16}, \quad \mathrm{Tr}_0(\Omega^4) = \begin{pmatrix} 4 \\ 2\end{pmatrix}, \quad \mathrm{Tr}_6(\Omega^4) = \mathrm{Tr}_2(\Omega^4) = 0
\ee
and subsequently
\be
\left. \mathcal F(\Omega^4)\right|_{N=8} = \frac1{2\pi}\int_0^{2 \pi} d\mu \Big(\mathcal G_0(\mu, \boldsymbol \alpha) \mathrm{Tr}_{0}(\Omega^4) + \mathcal G_4(\mu, \boldsymbol \alpha) \mathrm{Tr}_{4}(\Omega^4) + \mathcal G_8(\mu, \boldsymbol \alpha) \mathrm{Tr}_{8}(\Omega^4) \Big).
\ee
The first term is easily evaluated to $\left(\!\begin{smallmatrix} 4 \\ 2\end{smallmatrix} \!\right)$. In the second term, only the contribution for $k=2$ in the sum in $\mathcal G_4(\mu, \boldsymbol \alpha)$ survives the integration, thereby yielding $\left(\!\begin{smallmatrix} 4 \\ 1\end{smallmatrix}\! \right)C_{2,4} = 4t_2( \alpha_{1,2}) = 4 (\alpha_{1,2}^2 - 2)$. In a similar fashion, we find the last term to be $C_{4,8} = t_4(\alpha_{1,2}) = \alpha_{1,2}^4 -4 \alpha_{1,2}^2 + 2$ and, overall,
\be\left. \mathcal F(\Omega^4)\right|_{N=8} = 6+4 (\alpha_{1,2}^2 - 2) + (\alpha_{1,2}^4 -4 \alpha_{1,2}^2 + 2) = \alpha_{1,2}^4 \ee
as required.

\paragraph{For \boldmath{$c$} a power of \boldmath{$\Omega$}:}
If $c=\Omega^i$ for some $i \in \mathbb Z$ (where $\Omega^0 = I$), the diagram of $c$ has $|i|$ loop segments that cross the virtual boundary. If $\Omega^i$ is a connectivity that contributes to $Z_L$ and $\Tb^M(u)$, then $M\ge |i|$. When the top and bottom of $c$ are connected into a torus, the number of loops is $n = N \wedge i$ and their homotopy is $\{a,b\} = \{i/n, N/n\}$, so that 
\be
 \mathcal F(\Omega^i) = \alpha_{i/n, N/n}^n,\qquad n = N \wedge i.
\label{FOi}
\ee 
The link state with only defects is an eigenstate with eigenvalue  $v^{Ni}$. In general, a link eigenstate of $\Omega^i$ is constructed by adjoining $\frac N n$ copies of a link state with $n$ nodes and $r$ half-arcs 
($0\le r \le \lfloor \frac{n}{2} \rfloor$). 
This gives $|B_n^{n-2r}|=\left(\!\begin{smallmatrix} n \\ r \end{smallmatrix} \!\right)$
link eigenstates in the sector with $d = \frac{N}{n}(n-2r)$ defects. Among the connections to the virtual boundary, only $\frac{n-2r}{n}\times |i|$ are connected to defects. The eigenvalue is $v^{Ni(n-2r)/n}$ and
\be
\mathrm{Tr}_{d} (\Omega^i) = \left\{\begin{array}{cl}
\!\!\begin{pmatrix} n \\ r \end{pmatrix} v^{Ni(n-2r)/n}, & \qquad d = \frac{N}{n}(n-2r), \quad r = 0, \ldots, \lfloor \frac{n}{2} \rfloor, \\ \!\!0, & \qquad\mathrm{otherwise,}\end{array} \right.
\ee
while 
\begin{align}
\mathcal F(\Omega^i) &=  \frac1{2\pi}\int_0^{2 \pi} d\mu \sum_{r=0}^{\lfloor\frac{n}{2}\rfloor} \mathcal G_{\frac{N}{n}(n-2r)}(\mu, \boldsymbol \alpha) \begin{pmatrix} n \\ r \end{pmatrix} v^{Ni(n-2r)/n} \nonumber \\ \\[-0.6cm] 
& = \sum_{r=0}^{\lfloor\frac{n}{2}\rfloor} \begin{pmatrix} n \\ r \end{pmatrix} C_{\frac{i}{n}(n-2r), \frac{N}{n}(n-2r)} = \sum_{r=0}^{\lfloor\frac{n}{2}\rfloor} \begin{pmatrix} n \\ r \end{pmatrix} t_{n-2r}(\alpha_{i/n, N/n}). \nonumber
\end{align}
For our prescription to work, this must reproduce (\ref{FOi}), which is easily verified as a relation between Chebyshev polynomials. In terms of trigonometric functions where $\alpha_{i/n, N/n} = 2 \cos \theta$, the corresponding equality reads
\be
(2 \cos \theta)^n = \left\{\begin{array}{ll} 
\!\displaystyle{ \sum_{j=0}^{\frac{n-1}2}} \begin{pmatrix} n \\ \frac{n-1}2 - j\end{pmatrix} 2\cos \big((2j+1)\theta \big), & \quad n \; \mathrm{odd}, \\ 
\!\!\begin{pmatrix} n \\ \frac{n}2\end{pmatrix} +
\displaystyle{ \sum_{j=1}^{\frac{n}2}} \begin{pmatrix} n \\ \frac{n}2 - j\end{pmatrix} 2\cos \big(2j\theta \big), & \quad n \; \mathrm{even}. \\
\end{array} \right. \label{eq:FourierCheby}
\ee
This can be verified by an explicit computation of the Fourier series of $(2 \cos \theta)^n$.

\paragraph{The general case:} 
We now verify that the expression for $\mathcal F(c)$ given in equation \eqref{eq:Z} is correct for any connectivity $c$ different from $\Omega^i$. For such a connectivity, the state with only defects is not an eigenstate. Every connectivity has at least one link eigenstate, and among all the link eigenstates of $c$, one is particular: It has more defects than any other link eigenstate of $c$, and we will denote by $j(c)$ the number of defects of this particular state. We now present an algorithm to determine $j(c)$ for any connectivity $c$. Take for example the connectivity

\be 
c = 
\psset{unit=0.5}
\begin{pspicture}(0,-0.2)(8.5,0.9)
\psline[linewidth=1pt]{-}(0.5,1)(8.5,1)
\psline[linewidth=1pt,linestyle=dotted]{-}(0.5,1)(0.5,-1)
\psline[linewidth=1pt,linestyle=dotted]{-}(8.5,1)(8.5,-1)
\psline[linewidth=1pt]{-}(8.5,-1)(0.5,-1)
\psset{linewidth=1pt}
\psdots(1,1)(2,1)(3,1)(4,1)(5,1)(6,1)(7,1)(8,1)(1,-1)(2,-1)(3,-1)(4,-1)(5,-1)(6,-1)(7,-1)(8,-1)
\psset{linecolor=myc}
\psline{-}(1,1)(1,-1)
\psline{-}(6,1)(6,-1)
\psarc(2.5,1){-0.5}{0}{180}
\psarc(4.5,-1){0.5}{0}{180}
\psarc(7.5,1){-0.5}{0}{180}
\psarc(7.5,-1){0.5}{0}{180}
\psbezier{-}(2,-1)(2,0)(4,0)(4,1)
\psbezier{-}(3,-1)(3,0)(5,0)(5,1)
\end{pspicture}
\vspace{0.3cm}
\label{eq:specialc}
\ee
Because the nodes $4$ and $5$ on the bottom edge are connected by a loop segment, any link state $w$ satisfying $cw = \lambda w$ for some $\lambda \in \mathbb C$ will have the nodes $4$ and $5$ connected. The same argument applies to the nodes $7$ and $8$, and $w = 
\psset{unit=0.3}
\begin{pspicture}(0.1,0.9)(8.6,1.8)
\psline[linewidth=1pt]{-}(0.5,1)(8.5,1)
\psset{linewidth=0.5pt} 
\psdots(1,1)(2,1)(3,1)(4,1)(5,1)(6,1)(7,1)(8,1) 
\psset{linecolor=myc2}
\psarc(4.5,1){0.5}{0}{180}
\psarc(7.5,1){0.5}{0}{180}
\rput(1.05,1.7){$_?$}
\rput(2.05,1.7){$_?$}
\rput(3.05,1.7){$_?$}
\rput(6.05,1.7){$_?$}
\end{pspicture}
$,
where we are using question marks ``?" to indicate the nodes whose connections are not yet determined.
When this $w$ is drawn atop $c$, a factor $\beta$ appears because a loop is closed at positions $7$ and $8$, and the nodes $2$ and $3$ of the bottom edge are connected, so $w = 
\psset{unit=0.3}
\begin{pspicture}(0.1,0.9)(8.6,1.8)
\psline[linewidth=1pt]{-}(0.5,1)(8.5,1)
\psset{linewidth=0.5pt}
\psdots(1,1)(2,1)(3,1)(4,1)(5,1)(6,1)(7,1)(8,1)
\psset{linecolor=myc2}
\psarc(2.5,1){0.5}{0}{180}
\psarc(4.5,1){0.5}{0}{180}
\psarc(7.5,1){0.5}{0}{180}
\rput(1.05,1.7){$_?$}
\rput(6.05,1.7){$_?$}
\end{pspicture}
$. Now, $cw$ is seen to produce a factor of $\beta^2$, and no further connections are made between nodes on the lower edge of $c$. The unfixed nodes that remain can be replaced by defects, and their number is $j(c) = 2$. 

The general procedure is as follows. If nodes on the lower edge of $c$ are connected, the corresponding nodes must also be connected in any eigenstate $w$ of $c$. Acting on the partially known $w$ may then close new connections between nodes on the lower edge, which are in turn added to $w$. This process is repeated until no new connection is formed between nodes on the lower edge. The nodes of $w$ which are left undetermined can be replaced by defects and their number is $j(c)$. The resulting link state is clearly unique.

A small miracle operates here. First, the link state obtained from this algorithm has eigenvalue $\alpha^{\bar n_\alpha}\beta^{\bar n_\beta}  v^\Delta$ for some $\bar n_\alpha$, $\bar n_\beta$ and $\Delta$, where $\Delta$ depends on the permutation of the unfixed nodes. Second, the factor $\alpha^{\bar n_\alpha}\beta^{\bar n_\beta}$ is precisely the weight of the loops with homotopy $\{0\}$ and $\{1,0\}$ that intersect the upper/lower edge of $c$ when the torus is formed by gluing together the horizontal edges of the cylinder. 

If the above procedure for the connectivity $c$ fixes every node of $w$, then $\mathrm{Tr}_d(c) = 0$ for $d>0$, $w$ is the {\em unique} link eigenstate of $c$ and $\mathrm{Tr}_0(c) =  \alpha^{\bar n_\alpha}\beta^{\bar n_\beta}$. It subsequently follows that $c$ has no loops winding around the torus vertically and from equation \eqref{eq:Z}, $\mathcal F(c) = \frac1{2\pi} \int_{0}^{2\pi} \mathrm{Tr}_0(c) =   \alpha^{\bar n_\alpha}\beta^{\bar n_\beta}$ as expected.

If a subset of the nodes of $w$ are undetermined, the positions of these nodes are those in $c$ where loops with homotopy $\{a,b\}$ cross the horizontal edges. Up to a deformation of the loop segments, the algorithm produces a diagram where the connectivity $c$ acts on the $j(c)$ nodes with question marks by operating a cyclic permutation, as $\Omega^i$ did in the previous example. For example, for a given connectivity (different from \eqref{eq:specialc}), this diagram could be
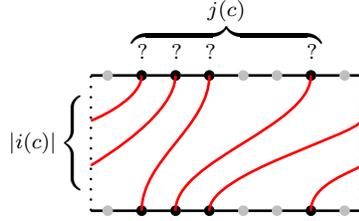
\begin{equation*} \hspace{1.1cm}
\psset{unit=0.5}\overbrace{
\begin{pspicture}(1.7,-2.9)(7.3,2.9)
\rput(-1.2,0){$_{|i(c)|}$}
\rput(-0.2,0){ $\Bigg \{$}
\psline[linewidth=1pt]{-}(0.5,2)(8.5,2)
\psline[linewidth=1pt,linestyle=dotted]{-}(0.5,2)(0.5,-2)
\psline[linewidth=1pt,linestyle=dotted]{-}(8.5,2)(8.5,-2)
\psline[linewidth=1pt]{-}(8.5,-2)(0.5,-2)
\psset{linewidth=1pt}
\psdots[linecolor=lightgray](1,2)(2,2)(3,2)(4,2)(5,2)(6,2)(7,2)(8,2)(1,-2)(2,-2)(3,-2)(4,-2)(5,-2)(6,-2)(7,-2)(8,-2)
\psdots(2,2)(3,2)(4,2)(7,2)(2,-2)(3,-2)(4,-2)(7,-2)
\rput(2.05,2.7){$_?$}
\rput(3.05,2.7){$_?$}
\rput(4.05,2.7){$_?$}
\rput(7.05,2.7){$_?$}
\psset{linecolor=myc}
\psbezier{-}(3,2)(3,0.9)(1,-0.36)(0.5,-0.66)
\psbezier{-}(2,2)(2,1.4)(1,0.86)(0.5,0.66)
\psbezier{-}(2,-2)(2,-0.5)(4,0.5)(4,2)
\psbezier{-}(3,-2)(3,-0.5)(7,0.5)(7,2)
\psbezier{-}(4,-2)(4,-0.9)(8.25,0.2)(8.5,0.66)
\psbezier{-}(7,-2)(7,-1.2)(8.25,-0.86)(8.5,-0.66)
\end{pspicture}}^{j(c)} 
\hspace{1cm}
\vspace{-0.5cm}
\end{equation*}
Replacing the question marks of $w$ by defects gives the unique link eigenstate with $j(c)$ defects. The shift of the $j(c)$ nodes is denoted by $i(c)$ and is measured by the number of loop segments that cross the left/right virtual boundary, $|i(c)|$. On the torus, the connections are divided into $n = i(c) \wedge j(c)$ loops of homotopy $\{i(c)/n,j(c)/n\}$. If $n>1$, $c$ will have other link eigenstates, where some of the question marks are replaced by half-arcs instead of defects. The picture is quite similar to what we encountered for $\Omega^i$: Separate the $j(c)$ nodes of $w$ into $j(c)/n$ sequences of $n$ nodes, choose a link state with $r$ half-arcs 
($r = 0, 1, \ldots, \lfloor\frac{n}{2}\rfloor$) and adjoin $j(c)/n$ copies of it side by side to form a link state with $d = \frac{j(c)}{n}(n-2r)$. The result is an eigenstate of $c$, and because only the fraction $\frac{n-2r}{n}$ of the $|i(c)|$ loop segments crossing the boundary are connected to defects, the associated eigenvalue is $\beta^{\bar n_\beta} v^{Ni(c)(n-2r)/n}$. The number of such link eigenstates in the sector with $d = \frac{j(c)}{n}(n-2r)$ defects is $\left(\!\begin{smallmatrix} n \\ r \end{smallmatrix}\!\right)$ and 
\begin{align} 
\mathcal F(c) &=  \beta^{\bar{n}_\beta}\times\frac{1}{2\pi}\int_0^{2 \pi} d\mu \sum_{r=0}^{\lfloor\frac{n}{2}\rfloor} \mathcal G_{\frac{j(c)}{n}(n-2r)}(\mu, \boldsymbol \alpha) \begin{pmatrix} n \\ r \end{pmatrix} v^{Ni(c)(n-2r)/n}\nonumber\\
& =\beta^{\bar{n}_\beta}\times \sum_{r=0}^{\lfloor\frac{n}{2}\rfloor} \begin{pmatrix} n \\ r \end{pmatrix} C_{\frac{i(c)}{n}(n-2r), \frac{j(c)}{n}(n-2r)} = \beta^{\bar{n}_\beta}\times\sum_{r=0}^{\lfloor\frac{n}{2}\rfloor} \begin{pmatrix} n \\ r \end{pmatrix} t_{n-2r}(\alpha_{i(c)/n, j(c)/n}) \\ 
&=\beta^{\bar{n}_\beta} (\alpha_{i(c)/n, j(c)/n})^n,\nonumber
\end{align}
where the last equality follows from \eqref{eq:FourierCheby}.
Because the expression for $\mathcal F(c)$ is linear in $c$, this completes the proof of Proposition~\ref{sec:Traceprop}.

%
\section{Critical dense polymers}
\label{sec:CDP}
%

An objective of this work is to express the partition function of the loop model on the torus in terms of the transfer matrix, and to test whether the result is modular invariant. From now on, we focus on $\beta = 0$, the value corresponding to critical dense polymers, while we initially keep the parities of $M$ and $N$ and the values of the non-contractible loop fugacities $\alpha_{i,j}$ free. 
In this general setup,  
the transfer matrix satisfies an inversion relation~\cite{PRV2010} which we solve for the twist representations $\omega_d$. 
As shown in Section~\ref{sec:traces}, the problem of computing the loop partition function boils down to calculating the eigenvalues of the transfer matrix in these representations.
As a result, we obtain the general expression for the partition function of the loop model for arbitrary $M$, $N$ and $\alpha_{i,j}$. Considerable simplifications are subsequently possible if we set all the non-contractible loop fugacities to $\alpha_{i,j}=2$. In this case, we prove that the partition function is modular invariant if both $M$ and $N$ are even.

The random cluster model, which is defined here only for $M$ and $N$ both even, is 
also expected to yield a modular invariant for some choice of the free parameters. 
It is recalled that, although both the loop model (Section~\ref{sec:loopmodel}) and the random cluster model (Section~\ref{sec:FKmodel}) are well-defined for $\beta = \sqrt Q = 0$ (and give nonzero partition functions), the passage from one to the other in Section~\ref{sec:FKmodel} is awkward. 
Consequently, the FK partition function we calculate
is of the no-cycle (``restricted") random cluster model with $Q=w_+=0$ discussed in Section~\ref{sec:FKcdp}. 

Of course, setting the weight of nontrivial clusters to zero gives a trivial modular invariant, $Z'_{FK} = 0$. 
This is obviously of very little interest, so one of our goals is to build another, nontrivial one from the transfer matrix approach to the loop model. 
With $\beta=0$, this method will allow us to compute the first term in \eqref{eq:FK0final}, but not the second one since this would require knowledge of the spectrum of the transfer matrix away from $\beta = 0$. 
Evidently, a direct way to eliminate the second term in (\ref{eq:FK0final}) is to set the weight of cross-topology clusters to zero, $w_+=0$, in which case the restricted FK partition function follows from that of the loop model. 

Let us comment on some of the terminology we will be using.
The number of defects $d$ classifying the representation $\omega_d$ is a quantum number that separates the theory into Ramond ($\frac{d}{2}$ even), Neveu-Schwarz ($\frac{d}{2}$ odd) and $\mathbb{Z}_4$ ($d$ odd) sectors. Following~\cite{SaleurSUSY,RS0106,PRV2010}, we are thus using terminology of supersymmetry even though we do not claim any superconformal symmetry in our model.

\subsection{Spectrum of the transfer matrix}
\label{sec:spectrumofT}

For $\beta = 0$, it was shown in~\cite{PRV2010} that the transfer matrix satisfies the inversion identity
\be
\Tb(u) \Tb(u+\textstyle{\frac{\pi}2}) = \left( \cos^{2N}\!u + (-1)^N \sin^{2N}\!u\right) \Ib + (\cos u \sin u)^N \Jb
\label{TTIJ}
\ee
where $\Ib$ is the vertical identity diagram while $\Jb \in \mathcal EPTL_N(\alpha, \beta=0)$ is given by 
\be
\Jb = \quad
\psset{unit=1}
\psset{linewidth=1pt}
\overbrace{
\begin{pspicture}(-0,-0.125)(5,1.2)
\pspolygon[fillstyle=solid,fillcolor=lightlightblue](0,1)(5,1)(5,-1)(0,-1)(0,1)
\psdots(0.5,-1)(1.5,-1)(2.5,-1)(4.5,-1)
\psdots(0.5,1)(1.5,1)(2.5,1)(4.5,1)
\lw
\psline{-}(-0.15,0.5)(0.0,0.5)
\psline{-}(5.15,0.5)(5.0,0.5)
\psline{-}(-0.15,-0.5)(0.0,-0.5)
\psline{-}(5.15,-0.5)(5.0,-0.5)
\unlw
\psline{-}(0,-1)(1,-1)(1,1)(0,1)(0,-1)
\psline{-}(0.1,-0.9)(0.9,-0.9)(0.9,0.9)(0.1,0.9)(0.1,-0.9)
\psline{-}(1,-1)(2,-1)(2,1)(1,1)(1,-1)
\psline{-}(1.1,-0.9)(1.9,-0.9)(1.9,0.9)(1.1,0.9)(1.1,-0.9)
\psline{-}(2,-1)(3,-1)(3,1)(2,1)(2,-1)
\psline{-}(2.1,-0.9)(2.9,-0.9)(2.9,0.9)(2.1,0.9)(2.1,-0.9)
\psline{-}(4,-1)(5,-1)(5,1)(4,1)(4,-1)
\psline{-}(4.1,-0.9)(4.9,-0.9)(4.9,0.9)(4.1,0.9)(4.1,-0.9)
\rput(3.5,0){$\dots$}
\psset{linecolor=myc}\unlw
\end{pspicture}}^N 
\qquad  \mathrm{with} \qquad
\begin{pspicture}(-0,-0.125)(1,1.2)
\pspolygon[fillstyle=solid,fillcolor=lightlightblue](0,-1)(1,-1)(1,1)(0,1)(0,-1)
\psdots(0.5,-1)(0.5,1)
\psline{-}(0.1,-0.9)(0.9,-0.9)(0.9,0.9)(0.1,0.9)(0.1,-0.9)
\end{pspicture} \, \, \,= \, \, \,
\begin{pspicture}(-0,-0.125)(1,1.2)
\pspolygon[fillstyle=solid,fillcolor=lightlightblue](0,-1)(1,-1)(1,1)(0,1)(0,-1)
\psdots(0.5,-1)(0.5,1)
\psline{-}(0,-1)(1,-1)(1,1)(0,1)(0,-1)
\psset{linecolor=blue,linewidth=1.5pt}
\psarc(1,1){0.5}{180}{270}
\psarc(0,0){0.5}{0}{90}
\psarc(1,0){0.5}{180}{270}
\psarc(0,-1){0.5}{0}{90}
\end{pspicture} 
\, \, \,- \, \, \,
\begin{pspicture}(-0,-0.125)(1,1.2)
\pspolygon[fillstyle=solid,fillcolor=lightlightblue](0,-1)(1,-1)(1,1)(0,1)(0,-1)
\psdots(0.5,-1)(0.5,1)
\psline{-}(0,-1)(1,-1)(1,1)(0,1)(0,-1)
\psset{linecolor=blue,linewidth=1.5pt}
\psarc(0,1){0.5}{-90}{0}
\psarc(1,0){0.5}{90}{180}
\psarc(0,0){0.5}{-90}{0}
\psarc(1,-1){0.5}{90}{180}
\end{pspicture}
\ee
\vspace{0.3cm}

\noindent Using the Drop-Down Lemma in~\cite{PRV2010}, we find
\be
\omega_d(\Jb) = \omega_d(\Ib) \times \left\{ \begin{array}{ll} 
\!\!(-1)^{\frac{N-d}2}\big((-1)^dv^{2N} + v^{-2N}\big), & \quad d>0, 
\\[.2cm] 
\!\!(-1)^{\frac{N}2}(\alpha^2-2), & \quad d=0.
\end{array}\right.
\ee
From the inversion relation, exact expressions for the eigenvalues of the transfer matrix can be found. Note that if $\alpha$ is replaced by $v^N + v^{-N}$, the expression for $\omega_d(\Jb)$ for $d>0$ also applies for $d=0$. Therefore, all the results for $d > 0$, where the eigenvalues 
of $\omega_d(\Tb(u))$, denoted by $T(u, \mu)$,
depend on the winding parameter $v = e^{i \mu}$, can be extended to $d = 0$ if we set 
\be
\alpha = 2 \cos \mu N.
\label{almu}
\ee 
Then, in the sector with $d$ defects, we have
\be 
\omega_d\big(\Tb(u) \Tb(u+ \textstyle{\frac{\pi}2})\big) = \omega_d(\Ib) \times F(u,v)
\ee
where 
\be
F(u,v)  = \Big(v^{N} \cos^N\!u +(-1)^{\frac{N-d}2} v^{-N} \sin^N\!u \Big) \Big(v^{-N} \cos^N\!u +(-1)^{\frac{N+d}2} v^{N} \sin^N\!u\Big).
\ee
The expression for $F(u,v)$ can be factorized, 
\be
 F(u, e^{i\mu}) = T(u, \mu) T(u+\tfrac{\pi}{2}, \mu),
\ee 
and the possible values for the eigenvalues are thus found by separating the zeros into contributions from $T(u, \mu)$ and $T(u+\tfrac{\pi}{2}, \mu)$. 
For $N$ odd, we find
\begin{align}
F(u,e^{i \mu})  = \frac {e^{-2 i u N} \sin^2(\frac{\pi}4 +(-1)^{\frac{d+1}2}\mu N)}{2^{2N-2}}\prod_{j=1}^N \Big( e^{4 i u} + \tan^2 \!\big(\tfrac{(2j-1)\pi}{4N}+(-1)^{j+\frac{d+1}2} \mu \big) \!\Big),
\end{align}
\vspace{-1.1cm}

\begin{align} 
T(u,\mu) &= \frac {\epsilon\, e^{-\frac{i\pi N}{4}}e^{- i u N} \sin(\frac{\pi}4 +(-1)^{\frac{d+1}2}\mu N)}{2^{N-1}}\prod_{j=1}^N \Big( e^{2 i u} + i \epsilon_j \tan\! \big(\tfrac{(2j-1)\pi}{4N}+(-1)^{j+\frac{d+1}2} \mu \big)\! \Big),
\label{Todd}
\end{align}
where $\epsilon,\epsilon_j\in \{+1, -1\}$ for all $j=1,\ldots,N$.
For $N$ even, we find
\be
F(u,e^{i \mu}) = \left\{ 
\begin{array}{l r l}
\!\!  \displaystyle{\frac{e^{-2 i u N} \sin^2( \mu N)}{2^{2N-2}}}\prod_{j=1}^N \Big( e^{4 i u} + \tan^2\! \big(\tfrac{j \pi}{N} - \mu \big) \!\Big), & & \,\,\frac{d}2 \; \mathrm{odd},\\ 
 \!\! \displaystyle{\frac {e^{-2 i u N} \cos^2(\mu N)}{2^{2N-2}}}\prod_{j=1}^N \Big( e^{4 i u} + \tan^2\! \big(\tfrac{(2j-1)\pi}{2N}- \mu \big)\! \Big), & &  \,\, \frac{d}2 \; \mathrm{even}, \end{array} \right.
\ee
\be
T(u,\mu) = \left\{ 
\begin{array}{l r l}
\!\!  \displaystyle{\frac{\epsilon (-i)^{\frac{N}2} e^{- i u N} \sin( \mu N)}{2^{N-1}}}\prod_{j=1}^N \Big( e^{2 i u} + i \epsilon_j \tan\! \big(\tfrac{j \pi}{N} - \mu \big)\! \Big), & & \frac{d}2 \; \mathrm{odd},\\ 
 \!\! \displaystyle{\frac {\epsilon (-i)^{\frac{N}2}e^{- i u N} \cos(\mu N)}{2^{N-1}}}\prod_{j=1}^N \Big(e^{2 i u} + i \epsilon_j \tan\! \big(\tfrac{(2j-1)\pi}{2N}- \mu \big) \!\Big), & & \frac{d}2 \; \mathrm{even}, \end{array} \right.
\label{Teven}
\ee
again with $\epsilon,\epsilon_j\in \{+1, -1\}$ for all $j=1,\ldots,N$.
For both parities of $N$, the eigenvalues are characterized by the choices of $\epsilon$ and $\epsilon_j$, $j = 1, \ldots, N$. This gives a total of $2^{N+1}$ possible eigenvalues. Because a given sector only has $\left(\!\!\begin{smallmatrix}N \\ \frac{N-d}2 \end{smallmatrix}\!\!\right)$ eigenvalues, only a subset of the possible eigenvalues appear in the sector $d$. This information is encoded in the selection rules of Section~\ref{sec:selection}.

The eigenvalues found in~\cite{PRV2010} are for $\mu = 0$, and extra care has to be taken in the Neveu-Schwarz sector ($N$ even and $\frac{d}{2}$ odd). Naively, the result for $T(u,\mu)$ in (\ref{Teven}) appears to be zero in that case because of the prefactor $\sin (\mu N)$, but the $j = \frac{N}{2}$ term of the product is singular in the limit $\mu \rightarrow 0$. The correct expression is obtained by applying l'H\^opital's rule. More generally, similar resolutions apply if $\mu$ is such that  $\sin(\frac{\pi}4 +(-1)^{\frac{d+1}2}\mu N )$, $\sin(\mu N)$ or $\cos(\mu N)$ is zero. 

To describe all three sets of eigenvalues in (\ref{Todd}) and (\ref{Teven}), we will henceforth use the short-hand notation
\be 
T(u,\mu) = K(\mu) \times \epsilon\times  \prod_{j=1}^N \Big(e^{i u} + i  e^{-i u} \epsilon_j \tan x_j \Big) ,
\label{Tumu}
\ee
where 
\vspace{-0.3cm}
\begin{align}   
 K(\mu) = \frac{e^{-\frac{i\pi N}{4}}}{2^{N-1}} \times \left\{ \begin{array}{ll} 
\!\! \sin(\frac{\pi}4 +(-1)^{\frac{d+1}2}\mu N), & \quad N \;\mathrm{odd,}\vspace{0.12cm}\\
\!\! \sin \mu N, & \quad N \;\mathrm{even},\; \frac{d}2 \; \mathrm{odd,}\vspace{0.12cm}\\
\!\! \cos \mu N, & \quad N \;\mathrm{even},\; \frac{d}2 \; \mathrm{even,}
 \end{array}\right.
\end{align}\vspace{-0.7cm} 

\begin{align}
x_j = \left\{ \begin{array}{l l} 
\!\! \frac{(2j-1)\pi}{4N}+(-1)^{j+\frac{d+1}2} \mu, & \quad N \;\mathrm{odd}, \vspace{0.12cm}\\ 
\!\!\frac{j \pi}{N}-\mu,  & \quad N \;\mathrm{even},\; \frac{d}2 \; \mathrm{odd},\vspace{0.12cm}\\ 
\!\!\frac{(2j-1)\pi}{2N} - \mu, & \quad N \;\mathrm{even},\; \frac{d}2 \;\mathrm{even}. \end{array} \right.
\label{xj}
\end{align}
It is noted that the eigenvalues (\ref{Tumu}) satisfy the {\em crossing symmetry}
\be
 \overline{T(\tfrac{\pi}{2}-\bar{u},\bar{\mu})}=T(u,\mu),
\ee
here indicated for complex $u$ and $\mu$.

\subsection{Selection rules}
\label{sec:selection}

In this section, we write down the selection rules dictating which of the solutions to the inversion identity appear as eigenvalues in the spectrum of $\omega_d(\Tb(u))$. The multiplicities of the eigenvalues also follow from the selection rules. In~\cite{PRV2010}, the corresponding selection rules for $\mu=0$ are expressed in terms of column configurations and excess parameters $\sigma$ and $\bar \sigma$. Here they are formulated exclusively in terms of conditions on the parameters $\epsilon_j$ and $\epsilon$, and their proof is given in Appendix~\ref{app:a}. That the selection rules reduce correctly to those of~\cite{PRV2010} for $\mu=0$ is also verified in the appendix.

Another component of the selection rules concerns the identification of the groundstates. Eigenvalues of $\omega_d(\Tb(u))$ are in general complex, as can be seen from $\Tb(u=0) = \Omega$ whose eigenvalues in the sector with $d$ defects are of the form $T(0,\mu) = v^d e^{\frac{2 i \pi m}N}$, for some $m \in \mathbb Z$. Hereafter, the groundstate eigenvalue of $\omega_d(\Tb(u))$ is defined as the one with maximal norm and is denoted by $T_d$. In Appendix \ref{app:a}, $T_d$ will be identified as the eigenvalue for which the linear term, $\textrm{Eig}(\mathcal H)$, is maximal.

\paragraph{\boldmath{$\mathbb Z_4$} sector (\boldmath{$N$} odd):} 

\begin{itemize}
\item[(i)]  
In the sector with $d$ defects, the parameters $\epsilon_j,\epsilon\in\{-1,+1\}$, $j=1,\ldots,N$, characterizing an eigenvalue of the transfer matrix satisfy
\be
\sum_{j=1}^N (-1)^j\epsilon_j  = d(-1)^{\frac{d+1}2},  \qquad \epsilon = (-1)^{\lfloor \frac{d+1}4 \rfloor}.
\label{eq:selrules1}
\ee
The number of different choices for the set of parameters $\{\epsilon_j\}$ satisfying (\ref{eq:selrules1}) is precisely $\left(\! \!\begin{smallmatrix} N \\ \frac{N-d}{2}\end{smallmatrix}\!\!\right)$, and each appears exactly once. 
\item[(ii)] Let $\bar \epsilon_j \equiv \epsilon_{N+1-j}$ for $j = 1, \ldots, \frac{N-1}2$. For $\mu\in[-\frac{\pi}{2N},\frac{\pi}{2N}]$, the groundstate in the sector with $d$ defects is characterized by
$\epsilon_{\frac{N+1}{2}}=+1$ and 
\be
\epsilon_j = \bar \epsilon_j = \left\{ \begin{array}{l l l} 
\!\!+1, & \quad &\ j=\frac{d+1}{2},\ldots,\frac{N-1}{2}, \\[.15cm]
\!\!(-1)^{j+\frac{d+1}{2}}, & \quad &\ j=1,\ldots,\frac{d-1}{2}. 
\end{array} \right.
\label{eq:gs}
\ee
\end{itemize}

If we set $\nu_k = (-1)^{k+\frac{d+1}2} \epsilon_k$, the selection rules simply state that exactly $\frac{N-d}{2}$ of these $N$ renormalized parameters satisfy $\nu_k=-1$. The full spectrum of $\omega_d(\Tb^M(u))$ can thus be written using a generating function as 
\be
\mathrm{Tr}_d \big(\Tb^M(u)\big) = \frac{1}{n!}\left( \frac{d^n G_d(z,v)}{d z^n}\right){\Big |}_{z= 0} \qquad \qquad  n = \frac{N-d}2,
\ee
where $v=e^{i\mu}$ and
\be
G_d(z,v) = (K(\mu) \times \epsilon)^M \prod_{j=1}^{N} \left(\! \left(e^{iu} + i e^{-iu} \tan x_j'  \right)^M 
+ z \left( e^{iu} - i e^{-iu} \tan x_j'  \right)^M  \right),
\label{eq:generatingodd}
\ee
\be
 x_j' = (-1)^{j + \frac{d+1}2} \frac{(2j-1)\pi}{4N} + \mu.
\ee

\paragraph{Ramond and Neveu-Schwarz sectors (\boldmath{$N$} even):}

\begin{itemize}
\item[(i)]  
In the sector with $d$ defects, the parameters $\epsilon_j,\epsilon\in\{-1,+1\}$, $j=1,\ldots,N$, characterizing an eigenvalue of the transfer matrix satisfy
\be
\sum_{j=1}^N \epsilon_j = -d,  \qquad \epsilon = (-1)^{\lfloor \frac{d+2} 4\rfloor}.
\label{eq:selrules2}
\ee
The number of different choices for the set of parameters $\{\epsilon_j\}$ satisfying (\ref{eq:selrules2}) is precisely $\left(\! \!\begin{smallmatrix} N \\ \frac{N-d}{2}\end{smallmatrix}\!\!\right)$, and each appears exactly once. 
\item[(ii)] For $\mu\in[-\frac{\pi}{2N},\frac{\pi}{2N}]$, the groundstate is characterized by
\be
\epsilon_j = \left\{ \begin{array}{ll} 
\!\!+1, & \quad  j=\lfloor \frac{d}4\rfloor+1,\ldots,\frac{N}2 - \lfloor \frac{d+2}4 \rfloor, \\[.15cm]
\!\!-1, & \quad \mathrm{otherwise}.
\end{array} \right.
\label{eq:gseven1}
\ee
\end{itemize}

The full spectrum of $\omega_d(\Tb^M(u))$ admits the expression
\be
\mathrm{Tr}_d \big(\Tb^M(u)\big) = \frac{1}{n!}\left( \frac{d^n G_d(z,v)}{d z^n}\right){\Big |}_{z= 0} \qquad  \qquad n = \frac{N-d}2  ,
\ee
where $v=e^{i\mu}$ and the generating functions are defined by
\be
G_d(z,v) = (K(\mu) \times \epsilon)^M \prod_{j=1}^{N} \left(\! \left(e^{iu} - i e^{-iu} \tan x_j  \right)^M 
+ z  \left(e^{iu} + i e^{-iu} \tan x_j  \right)^M  \right)
\label{eq:generatingeven}
\ee
with $x_j$ given in (\ref{xj}). Note that this expression differs from the similar one in (\ref{eq:generatingodd}).
For later convenience, it is also noted that
\be
 G_2(-1,1)=\lim_{\mu\to0}G_2(-1,e^{i\mu})=0.
\label{G2}
\ee

\subsection{Lattice partition function}
\label{sec:latticeZ}

We now proceed to find an exact expression for the partition function of the loop model $Z_L$ from the generating functions $G_d(z,v)$. 
First, we make two crucial observations:
\begin{itemize} 
\item[(i)] 
As a straightforward calculation shows, 
\be
 G_d(z,e^{i (\mu + \frac{\pi}N)}) = (-1)^M G_d(z, e^{i\mu}).
\ee 
Indeed, $K(\mu+\frac{\pi}{N})=-K(\mu)$ and the terms in the product \eqref{eq:generatingodd} (for $N$ odd) or \eqref{eq:generatingeven} (for $N$ even) are mapped onto one another when $\mu$ is shifted by $\frac{\pi}N$.
\item[(ii)] 
$G_d(z, v)$ admits a Laurent series expansion in the winding parameter $v$ with finitely many nonzero coefficients, with the terms with minimal and maximal powers of $v$ given by $v^{-NM}$ and $v^{NM}$, respectively. 

Indeed, for $N$ odd, for example, the generating function in \eqref{eq:generatingodd} can be expressed as
\be
G_d(z,v) = (e^{-\frac{i \pi N} 4} \epsilon)^M \prod_{j=1}^N \left((e^{i x_j'} \cos u + i \sin u\, e^{-i x_j'})^M + z (e^{-i x_j'} \cos u + i \sin u\,e^{i x_j'})^M\right),
\ee
where \eqref{eq:simplif1} has been used to simplify the factor $K^M(\mu)$.
Because 
\be 
 e^{i x_j'} = v \exp\!\Big(\frac{i\pi}{4N}(2j-1) (-1)^{j+\frac{d+1}2}\Big),
\ee 
the result readily follows. Using \eqref{eq:simplif2}, a similar argument carries through for $N$ even.
\end{itemize}
It follows from these two observations that $G_d(z,v)$ is $v^{-NM}$ times a Taylor series in $v^{2N}$. Because $G_d(z,v)$ is also polynomial in $z$, it can be written as
\be 
G_d(z, v) = \sum_{k = -M, -M+2, \ldots}^M\; \sum_{\ell = 0}^N\,v^{Nk}z^\ell  g_d(k,\ell).
\ee
In terms of the coefficients $g_d(k,\ell)$, the trace of $\omega_d(\Tb^M(u))$ is 
\be
\mathrm{Tr}_d\big(\Tb^M(u)\big) = \sum_{k = -M, -M+2, \ldots}^M v^{Nk} g_d(k,{\textstyle \frac{N-d}2}).
\label{eq:outofinspiration}
\ee
One can now use this last equation along with equation \eqref{eq:Z}  to write the partition function of the loop model as
\be
Z_L = \left\{ \begin{array}{ll}
\!\displaystyle \sum_{d=1,3, \ldots}^N \;\sum_{k = -M, -M+2, \ldots}^M C_{k,d} \, g_d(k,{\textstyle \frac{N-d}2}), & \quad N \; \mathrm{odd}, \\[.5cm] 
\!\mathrm{Tr}_0\big(\Tb^M(u)\big)  +
\displaystyle \sum_{d=2,4, \ldots}^N \; \sum_{k = -M, -M+2, \ldots}^M C_{k,d} \, g_d(k,{\textstyle \frac{N-d}2}), & \quad N \; \mathrm{even}. \end{array}\right. 
\label{eq:Zgeneral}
\ee
It is observed that the integers $k$ and $M$ always have the same parity. This is not a surprise since, in Section~\ref{sec:traces}, $k$ counts the number of times defects cross the virtual boundary, and this number always has the same parity as the number $M$ of layers of $\Tb(u)$. 

The special term $\mathrm{Tr}_0({\Tb}^M(u))$, which appears for $N$ even, can be rewritten as follows. First, it is recalled from (\ref{almu}) in Section~\ref{sec:spectrumofT} that the parameter $\alpha = \alpha_{1,0}$ is conveniently parameterized as $v^N + v^{-N}$. Because $\mathrm{Tr}_0(\Tb^M(u))$ is polynomial in $\alpha$, the coefficients in equation \eqref{eq:outofinspiration} for $d=0$ must satisfy the symmetry 
\be 
g_0(k, \tfrac N 2) = g_0(-k, \tfrac N 2).
\ee
(This also follows readily from equation \eqref{eq:syms} below.) We thus have 
\be
\mathrm{Tr}_0\big(\Tb^M(u)\big) = \tfrac12 \!\displaystyle{\sum_{\substack {0\le k \le M \\[.07cm] k = M \,\mathrm{mod} \, 2}}} (2-\delta_{k,0})\big(v^{Nk} + v^{-Nk}\big)\, g_0(k,\tfrac{N}2) =\displaystyle{\sum_{\substack {0\le k \le M \\[.07cm] k = M \,\mathrm{mod} \, 2}}} C_{k,0} \,g_0(k,\tfrac{N}2),
\label{eq:to}
\ee
where $C_{k,0}$, $0\leq k\leq M$, is defined as $C_{k,d}$ for $d>0$ in (\ref{eq:Z}):
\be
 C_{k,0}=\left\{\begin{array}{ll} 
   \!\! t_k(\alpha_{1,0})=t_k(\alpha),&k>0,\\[0.15cm]
  \!\!1,&k=0.
 \end{array}\right.
\ee
Equations \eqref{eq:Zgeneral} and  \eqref{eq:to}, combined with \eqref{eq:generatingodd} or \eqref{eq:generatingeven} for $N$ odd or even respectively, give the general expression for the partition function of the loop model. 

Considerable simplifications are possible 
if we set 
\be
\alpha_{i,j} = \alpha = 2
\ee
for all $i,j$. In this case, $t_n= 2-\delta_{n,0}$ and, from equations \eqref{eq:Zgeneral} and \eqref{eq:outofinspiration} at $v=1$, 
\be
Z_L\displaystyle\big|_{\alpha_{i,j} = 2}
= \sum_{\substack {0\le d \le N \\[.07cm] d = N \,\mathrm{mod} \, 2}} (2-\delta_{d,0}) \mathrm{Tr}_d\big(\Tb^M(u)\big)\displaystyle\Big|_{\substack{\alpha = 2\\ v=1}}\ .
\label{eq:ZT}
\ee
For each parity of the system size $N$, the total number of contributing link states is thus given by
\be
 \sum_{\substack {0\le d \le N \\[.07cm] d = N \,\mathrm{mod} \, 2}} (2-\delta_{d,0})\dim V_N^d=2^N
\label{2N}
\ee
where the dimensions are given by the binomial coefficients $\left(\!\!\begin{smallmatrix}N \\ \frac{N-d}2 \end{smallmatrix}\!\!\right)$ as in (\ref{Bdim}).
The explicit expression (\ref{eq:ZT}) giving the partition function as a sum of ordinary sector traces is key to the identification of the corresponding modular invariant discussed in Section~\ref{sec:modularZ}.

\subsection{Finitized partition functions}
\label{sec:finitizedZ}

The results of Section~\ref{sec:latticeZ} are expressed in terms of the $N+1$ functions $G_d(z,v)$ where $d = 0, \ldots, N$. 
In the following, we show how to extract the physical information from the three functions $G_0(z,v)$, $G_1(z,v)$ and $G_2(z,v)$ alone, and find that this allows the direct evaluation of the {\em finitized partition function} for $M$ even and $\alpha_{i,j} =2$ for all $i,j$. 
In contrast, similar results have not been obtained for $M$ odd.
It is noted that non-contractible loop fugacities are also set equal to $2$ in the work~\cite{SaleurSUSY}.

The expansion of the generating function $G_d(z,v)$ as a Taylor series in the variable $z$ is given by
\be
G_d(z,v) = \sum_{\ell=0}^NA^{(v)}_{d,\ell} z^\ell,\qquad\quad
 A^{(v)}_{d,\ell}=\sum_{k = -M, -M+2, \ldots}^M\! v^{Nk} g_d(k,\ell)
\label{Gdzv}
\ee 
and is such that 
\be
 \mathrm{Tr}_d\big(\Tb^M(u)\big) = A_{d,{\frac{N-d}2}}^{(v)}.
\ee 
Recalling that $T_d$ denotes the groundstate eigenvalue of $\omega_d(\Tb(u))$, we introduce {\em normalized} generating functions,
\be
 \hat{G}_d(z,v)=\frac{G_d(z,v)}{T_d(u,\mu)},\qquad v=e^{i\mu}.
\label{Ghat}
\ee

Now, to compute the finitized partition function, we first determine the appropriate finitizations $\hat{G}^{(N)}_d(u,v)$ of the normalized generating functions. This requires restricting $\mu$ to be of the order $1/N$, and we choose to parameterize it as 
\be   
 \mu = \frac{\pi a} N,\qquad a \in [-\tfrac{1}{2},\tfrac{1}{2}],
\label{mua}
\ee 
as in the description of the selection rules for the groundstates in Section~\ref{sec:selection}.
The finitizations $\hat{G}^{(N)}_d(u,v)$ are then obtained following the approach of~\cite{PRV2010} based on the evaluation of finite excitations. The results for $d=0,1,2$ are discussed below and are expressed in terms of the {\em aspect ratio}
\be
 \delta=\frac{M}{N}
\label{aspect}
\ee
and the {\em modular nome} $q$ and its complex conjugate $\bar{q}$, given by 
\be
q = \exp(-2\pi i \delta e^{-2 i u}),\qquad\bar q = \exp(2\pi i \delta e^{2 i u}).
\label{qqbar}
\ee 

Here it is recalled that the models we study are defined on a rectangular array of $M \times N$ square tiles
with periodic boundary conditions to cover the torus. 
In the continuum scaling limit, the periodicity of the torus is described by the {\em modular parameter} $\tau$ which lives in the upper half of the complex plane, i.e.~$\mathrm{Im}(\tau)>0$. The partition functions we compute are expressed in terms of a modular parameter $\tau$ through the associated nome $q$ and its complex conjugate $\bar q$ defined by
\be
q = e^{2 \pi i \tau}, \qquad \bar q = e^{-2 \pi i \bar \tau}.
\ee
Comparing this with (\ref{qqbar}), we see that
\be
 \tau=-\delta e^{-2iu}.
\label{tau1}
\ee
As discussed at the end of Appendix~\ref{app:Twist}, the spectral parameter $u$ (for $0<u<\frac{\pi}{2}$) can be interpreted~\cite{KimPearce87} geometrically as a measure of the spatial anisotropy of the lattice thus explaining its appearance in the definition of the modular parameter.
For isotropic interactions, $u=\frac{\lambda}{2}$ and 
\be
 \tau=-\delta e^{-\pi i/2}=i\,\frac{M}{N}
\ee
as expected.

\paragraph{\boldmath{$\mathbb Z_4$} sector:}

For $N$ odd, the generating functions have the symmetries
\be
\begin{array}{l}G_{4n+1}(z,v) = (-1)^{nM} G_1(z,v), \vspace{0.1cm}\\ G_{4n+3}(z,v) = (-1)^{nM} G_3(z,v),\end{array} \quad (n \in \mathbb N), \qquad G_3(z,v) = z^N (-1)^M G_1(z^{-1},v^{-1}).
\label{eq:oddsyms}
\ee
This allows us to write
\be
\mathrm{Tr}_d\big(\Tb^M(u)\big) =\left\{\begin{array}{ll}
\!\!(-1)^{\frac{d-1}4 M} A_{1,\frac{N-d}2}^{(v)}, \qquad \frac{d-1}2\; \mathrm{even}, \vspace{0.2cm}\\ 
\!\!(-1)^{\frac{d+1}4 M} A_{1,\frac{N+d}2}^{(1/v)}, \qquad \frac{d-1}2\; \mathrm{odd}.\end{array} \right.
\ee
The partition function for the loop model with $M$ even and $\alpha_{i,j}=2$ (see equation \eqref{eq:ZT}) then simplifies to
\begin{align}
Z_L \big|_{\alpha_{i,j}=2} &=2 \sum_{d = 1, 5, 9, \ldots} A^{(1)}_{1,\frac{N-d}2} + 2\sum_{d = 3, 7, 11, \ldots} A^{(1)}_{1,\frac{N+d}2} =  \sum_{\ell=0}^{N}\big(1+(-1)^{\ell+\frac{N-1}2}\big)A_{1,\ell}^{(1)}
\nonumber \\&= G_1(1,1) +(-1)^{\frac{N-1}2}G_1(-1,1).
\end{align}
As outlined above, with $\mu$ as in (\ref{mua}), we work out the finitized generating function
\begin{align}
 \hat{G}_1^{(N)} (z,e^{ \frac{i\pi a} N}) \, &= \prod_{j=1,3,\ldots}^{2\lfloor\frac{N+3}4 \rfloor-1}\!\big(1+z q^{\frac{2j-1}4+a}\big)
\prod_{j=2,4,\ldots}^{2\lfloor\frac{N+1}4 \rfloor}(q^{\frac{2j-1}4-a}+z)\nonumber \\
& 
 \times  \prod_{j=1,3,\ldots}^{2\lfloor\frac{N+1}4 \rfloor-1}\!\big(1+z (-1)^M\bar q^{\frac{2j-1}4-a}\big)\prod_{j=2,4,\ldots}^{2\lfloor\frac{N-1}4 \rfloor}\!\big((-1)^M\bar q^{\frac{2j-1}4+ a}+z \big)\label{eq:G1N}
\end{align}
where the modular nome is defined as in (\ref{qqbar}).
For $M$ even and $N$ odd, we then obtain the {\em finitized partition function}
\be Z_L^{(N)} \big|_{\alpha_{i,j}=2} 
=T_1^M|_{\mu = 0}\times \Big(\! \prod_{j=1}^{\frac{N+1}2} (1+q^{\frac{2j-1}4}) 
\prod_{j=1}^{\frac{N-1}2} (1+\bar q^{\frac{2j-1}4}) + \prod_{j=1}^{\frac{N+1}2} (1-q^{\frac{2j-1}4}) 
\prod_{j=1}^{\frac{N-1}2} (1-\bar q^{\frac{2j-1}4})\! \Big).
\label{ZLN}
\ee
The number of contributing link states is obtained by setting $q=\bar q=1$ and is readily found to be $2^N$ as in (\ref{2N}).

\paragraph{Ramond and Neveu-Schwarz sectors:}
For $N$ even, the generating functions have the symmetries
\be
\begin{array}{l}G_{4n}(z,v) = (-1)^{nM} G_0(z,v), \vspace{0.1cm}\\ G_{4n+2}(z,v) = (-1)^{nM} G_2(z,v),\end{array} \quad (n \in \mathbb N), \quad  \begin{array}{ll}\quad &G_0(z,v) = z^N G_0(z^{-1},v^{-1}), \\[.2cm] &
G_2(z,v) = z^N (-1)^M G_2(z^{-1},v^{-1})
\end{array}
\label{eq:syms}
\ee
from which it follows that 
\be
\mathrm{Tr}_d(\Tb^M(u)) =\left\{\begin{array}{ll}
\!\!(-1)^{\frac{d}4 M} A_{0,\frac{N-d}2}^{(v)}, \qquad\ \ \frac{d}2\ \mathrm{even}, \vspace{0.2cm}\\ 
\!\!(-1)^{\frac{d-2}4 M} A_{2,\frac{N-d}2}^{(v)}, \qquad \frac{d}2\ \mathrm{odd}.\end{array} \right.
\ee
Further simplifications are possible for $v=1$ and $M$ even as the second set of relations in \eqref{eq:syms} then translates into $A_{i,\ell}^{(1)}=A_{i,N-\ell}^{(1)}$ for $i = 0, 2$.
It subsequently follows that 
\begin{align}
G_i(1,1) &= \sum_{\ell = 0}^{\frac N 2} (2-\delta_{\ell,\frac N 2}) A_{i,\ell}^{(1)},  \qquad
G_i(-1,1) = \sum_{\ell = 0}^{\frac N 2} (2-\delta_{\ell,\frac N 2}) (-1)^\ell A_{i,\ell}^{(1)}
\end{align}
for $i = 0,2$, and the partition function for the loop model can be expressed as 
\begin{align}
Z_L \big|_{\alpha_{i,j}=2} &= \sum_{d = 0, 4, 8, \ldots}(2-\delta_{d,0}) A^{(1)}_{0,\frac{N-d}2} + 2\sum_{d = 2, 6, 10, \ldots} 
A^{(1)}_{2,\frac{N-d}2}\nonumber \\ 
&= \tfrac{1}{2}\Big(G_0(1,1) + (-1)^{\frac N 2} G_0(-1,1) + G_2(1,1) - (-1)^{\frac{N}2} \underbrace{G_2(-1,1)}_{=\,0}\Big)
\label{ZAA}
\end{align}
where the under-braced term vanishes due to (\ref{G2}).
With the parameterization (\ref{mua}), the finitizations of the (normalized versions (\ref{Ghat}) of the) generating functions appearing in (\ref{ZAA}) are given by
\begin{eqnarray}
\hat{G}_0^{(N)} (z,e^{ \frac{i\pi a} N}) &\!\!=\!\!& \prod_{j=1}^{\lfloor \frac{N+2}4\rfloor} \!\big(q^{\frac{2j-1}2 -a}+z\big)\big(1+zq^{\frac{2j-1}2+a}\big)\nonumber\\
 &&\qquad\times\,\prod_{j=1}^{\lfloor \frac{N}4\rfloor} \big((-1)^M \bar q^{\,\frac{2j-1}2 +a}+z\big)\big(1+(-1)^Mz\bar q^{\,\frac{2j-1}2-a}\big),\label{eq:G0N}
\end{eqnarray}
\vspace{-0.4cm}
\begin{align}
\hat{G}_2^{(N)} (z,e^{\frac{i\pi a}N}) &\,=\, \big(1+zq^{a}\big)\big(1+(-1)^Mz \bar{q}^{^{-a}}\big)\prod_{j=1}^{\lfloor \frac{N}4\rfloor} \big(q^{j -a}+z\big)\big(1+zq^{j+a}\big)\nonumber\\ 
& \hspace{3cm}\times \prod_{j=1}^{\lfloor \frac{N-2}4\rfloor} \!\big((-1)^M \bar q^{\,j +a}+z\big)\big(1+(-1)^Mz\bar q^{\,j-a}\big),\label{eq:G2N}
\end{align}
where the modular nome is defined as in (\ref{qqbar}).
Using the result 
\be
 {\displaystyle \lim_{\substack{M,N \gg 1 \\[0.05cm] M = \delta N}}}\;\frac{T_2^M\big|_{\mu=0}}{T_0^M\big|_{\mu=0}}=(q \bar q)^{\frac 18}
\label{qq18}
\ee
obtained in~\cite{PRV2010} from an Euler-Maclaurin analysis of the eigenvalues for $\mu=0$, we can now write the finitized partition function for $M$ and $N$ even as
\begin{align}
\!\!Z_L^{(N)} \big|_{\alpha_{i,j}=2}=&\; T_0^M\big|_{\mu = 0}\times \Big( 
\tfrac{1}{2}\!\!\prod_{j=1}^{\lfloor\frac{ N+2}4\rfloor} (1+q^{\frac{2j-1}2})^2 \prod_{j=1}^{\lfloor\frac{N}4\rfloor}(1+\bar q^{\frac{2j-1}2})^2 
\nonumber\\ 
&\hspace{-0.4cm}
 + \tfrac{1}{2}\!\!\prod_{j=1}^{\lfloor\frac{ N+2}4\rfloor} (1-q^{\frac{2j-1}2})^2 \prod_{j=1}^{\lfloor\frac{N}4\rfloor}(1-\bar q^{\frac{2j-1}2})^2
+ 2(q \bar q)^{\frac 1 8}  \prod_{j=1}^{\lfloor\frac{N}4\rfloor} (1+q^{j})^2 \prod_{j=1}^{\lfloor\frac{N-2}4\rfloor}(1+\bar q^{j})^2
\Big).
\label{Zqqbar}
\end{align}
The number of contributing link states is obtained by setting $q=\bar q=1$ and is readily found to be $2^N$ as in equation (\ref{2N}).

An interesting parallel can be drawn between the calculations of the finitized partition function of critical dense polymers and that of the Ising model carried out in~\cite{OPW}. The spin transfer matrix of the Ising model also satisfies an inversion identity from which its spectrum can be determined. 
The resulting finitized partition function of the Ising model is of a form very similar to the one of critical dense polymers (\ref{Zqqbar}) and also contains three terms. 
In the context of $Q$-state Potts models, the inversion identity satisfied by the spin transfer matrix is a feature unique to the Ising model ($Q=2$). Likewise in the context of the logarithmic minimal models $\mathcal{LM}(p,p')$, we believe that the loop transfer matrix satisfies a functional relation of degree $p'$, implying that the inversion identity (\ref{TTIJ}) is unique to critical dense polymers $\mathcal{LM}(1,2)$. We hope to discuss this elsewhere.

The analysis of the finitized partition functions above is based on a torus formed by gluing together a rectangle without twisting the boundaries. We generalize this to helical tori in Appendix~\ref{app:Twist} where we find that, for $\alpha_{i,j}=2$ and even helicity $t$, the finitized (twisted) partition function is given by the expression (\ref{ZLN}) or (\ref{Zqqbar}), depending on the parities of $M$ and $N$, but with $q$ replaced by the nome in (\ref{eq:seamedq}).

\subsection{Modular invariant partition function}
\label{sec:modularZ}

It is quite remarkable that the partition function (\ref{eq:ZT}) of the loop model on the torus for which all the non-contractible loop fugacities $\alpha_{i,j}$ are set to $2$ can be written exclusively in terms of traces in representations with $v=1$. In~\cite{PRV2010}, the behaviour of these traces was studied in the $M,N \rightarrow \infty$ limit. Here we demonstrate that the partition function (\ref{eq:ZT}) is {\em modular invariant} in the continuum scaling limit if $M$ and $N$ are both even.
It is recalled that the modular transformations are generated by
\be
 S:\ \tau\to-\frac{1}{\tau},\qquad T:\ \tau\to\tau+1
\label{ST}
\ee
acting on the modular parameter $\tau$ given in (\ref{tau1}).

Let us define $Z_d$ by
\be
\mathrm{Tr}_d\big(\Tb^M(u)\big)= T_{\mathrm{max}}^M \times \epsilon^M \times Z_d,
\vspace{0.05cm}
\ee
where $T_{\mathrm{max}}$ is the maximal eigenvalue of $\Tb(u)$, 
that is, $T_{\mathrm{max}}=T_0$ for $N$ even and $T_{\mathrm{max}}=T_1$ for $N$ odd. 
In~\cite{PRV2010}, it was found that these partition functions correspond to conformal partition functions associated with a logarithmic
CFT with central charge
\be
 c=-2
\ee
and conformal weights given in terms of the defect number $d$ by
\be
 \Delta_{t}=\frac{t^2-1}{8},\qquad t=\frac{d}{2}.
\ee
The corresponding Kac labels $r,s$, where $t=2r-s$, are integer in the Ramond and Neveu-Schwarz sectors ($N$ even) and half-integer in the $\mathbb{Z}_4$ sector ($N$ odd).

We recall the specialized Jacobi theta functions
\be 
 \vartheta_{s,p}(q)\;=\;\sum_{\lambda\,\in\,\mathbb{Z}+\frac{s}{2p}} q^{p\lambda^2}
\ee
and the Dedekind eta function
\be
 \eta(q)\;=\;q^{\frac{1}{24}}\prod_{n=1}^\infty (1-q^n),
\ee
where the modular nome is defined in (\ref{qqbar}). For the simplified expressions in Section~\ref{sec:finitizedZ} to apply, we let $M$ be even, in which case $\epsilon^M=1$ for all $d$.
From the analysis of the continuum scaling limit in~\cite{PRV2010}, and up to the divergent term $T_{\mathrm{max}}^M$, 
we then find that the conformal part of the partition function (\ref{eq:ZT}) is given by 
\be
 \sum_{d\in 2\mathbb{N}-1} Z_d(q)
   \;=\;\frac{|\vartheta_{\frac{1}{2},2}(q)|^2+|\vartheta_{\frac{3}{2},2}(q)|^2}{|\eta(q)|^2},
\label{ZZZodd}
\ee
for $N$ odd, while for $N$ even, it is given by
\be
 Z_0(q)+2\sum_{d\in 2\mathbb{N}} Z_d(q)=
  \frac{1}{|\eta(q)|^2} \sum_{s=0}^3|\vartheta_{s,2}(q)|^2.
\label{ZZZ}
\ee

As already observed in~\cite{PRV2010}, the partition function (\ref{ZZZodd}) for $N$ odd is not modular invariant. 
This is not a surprise since the sectors with $N$ odd are incompatible geometrically with the sector containing the vacuum.
The sesquilinear form in (\ref{ZZZ}), on the other hand, is a well-known modular invariant 
Coulombic partition function~\cite{FSZcoulomb1987,SaleurSUSY}. The conformal partition function of critical dense polymers on the torus with $M$ and $N$ both even is thus given by this modular invariant.
Writing it as 
\be
 Z_0(q)+2\sum_{d\in 2\mathbb{N}} Z_d(q)=|\hat\chi_{-\frac{1}{8}}(q)|^2+2|\hat\chi_0(q)\!+\!\hat\chi_1(q)|^2+|\hat\chi_{\frac{3}{8}}(q)|^2,
\label{ZZZW}
\ee
where the ${\cal W}$-irreducible characters~\cite{Kausch95} (see also~\cite{Flohr95}) are given by
\be
\begin{array}{rclrcl}
\hat\chi_{-\frac{1}{8}}(q)&\!\!=\!\!&\displaystyle{\frac{1}{\eta(q)}}\,\vartheta_{0,2}(q),\qquad\quad&
\hat\chi_0(q)&\!\!=\!\!& \displaystyle{\frac{1}{2\eta(q)}}\big(\vartheta_{1,2}(q)+\eta^3(q)\big),\\[14pt]
\hat\chi_{\frac{3}{8}}(q)&\!\!=\!\!& \displaystyle{\frac{1}{\eta(q)}}\,\vartheta_{2,2}(q),\qquad\quad&
\hat\chi_1(q)&\!\!=\!\!& \displaystyle{\frac{1}{2\eta(q)}}\big(\vartheta_{1,2}(q)-\eta^3(q)\big),
\end{array}
\ee
it is recognized as the modular invariant partition function of {\em symplectic fermions} on a $\mathbb{Z}_2$ orbifold~\cite{Kausch00}. 
From Section~\ref{sec:FKcdp}, this is also the partition function of the no-cycle FK cluster model at $Q=w_+=0$ in the continuum scaling limit.
 
The ${\cal W}$-characters incorporate the ${\cal W}$-extended symmetry of the triplet model associated with symplectic fermions. 
As discussed in~\cite{PRR2008}, the decomposition of the ${\cal W}$-irreducible characters in terms of irreducible Virasoro characters reflects that an arbitrary number of defects, compatible with the system size, is allowed in the corresponding loop model. Indeed, this is what we find in the expression (\ref{ZZZW}) for the modular invariant.
 
We stress that the modular invariant partition function in (\ref{ZZZ}) is {\em not} obtained by the usual matrix trace
summing over all even $N$ sectors since the gluing procedure governed by the linear functional $\mathcal{F}$ has changed the multiplicity of all of the $d\ne 0$ sectors by a factor of 2.
Indeed, the {\em standard} matrix trace {\em does not} yield a modular invariant. In stark contrast, the {\em modified} matrix trace in (\ref{ZZZ}) {\em does} yield a modular invariant partition function.

\section{Conclusion}
\label{sec:Conclusion}

Because of the nonlocal nature of the degrees of freedom of the logarithmic minimal models ${\cal LM}(p,p')$, the structure of their bulk logarithmic CFTs is much more complicated than in the case of rational theories. In particular, the bulk representations are no longer given as simple tensor products over the left- and right-chiral halves of the theory. Similar observations apply~\cite{GainutdinovEtAl} in the context of the $\mathfrak{gl}(1|1)$ super-spin chains. 
In the case of critical dense polymers ${\cal LM}(1,2)$, the bulk CFT of the associated 
$\mathcal{W}_{1,2}$
(or $c_{1,2}$) 
triplet model was constructed from the boundary CFT in~\cite{GaberdielRunkelBdyToBulk} and, for the case of critical percolation ${\cal LM}(2,3)$, the bulk CFT of the associated $\mathcal{W}_{2,3}$ triplet model was constructed from the boundary CFT in~\cite{GRW2010}. 
Generalizing the results of these papers, the modular invariant partition functions of the ${\cal LM}(p,p')$ models were conjectured in~\cite{PRcoset2011} from a lattice perspective. Although the representations do not factor, the modular invariant partition functions do factor in terms of characters and are given~\cite{PRcoset2011} as a combination of Coulombic and (rational) minimal modular invariant partition functions. To understand the bulk theory better, it is desirable to have a lattice derivation of these modular invariant partition functions. 

In this paper, we have taken the first step in this program by 
calculating, via an exact lattice derivation, the modular invariant partion function of 
critical dense polymers  ${\cal LM}(1,2)$. 
Starting with the work of Cardy, there are well established physical principles which assert that, with purely periodic boundary conditions (and no seams or mismatched size parities), the torus partition functions should be modular invariant. These arguments apply equally to rational and logarithmic CFTs. Indeed, the modular invariance of the torus partition function of critical dense polymers is confirmed by our exact lattice calculation.
In the lattice setting, the difficulty of closing the cylinder to the proper geometric torus is overcome by introducing a modified trace. In the continuum scaling limit, we obtain the expected modular invariant partition function of symplectic fermions on a $\mathbb{Z}_2$ orbifold. 
The modified trace acts to combine the 
three 
${\cal W}$-sectors from the cylinder with integer degeneracies $(1,2,1)$ as in (\ref{ZZZW}). 
These degeneracies coincide~\cite{PRcoset2011} with Coxeter exponents given by the right eigenvector of the Cartan matrix of the twisted affine Dynkin diagram $A_2^{(2)}$ with three nodes. The appearance of non-simply laced affine graphs and such degeneracies seems to be a feature of logarithmic CFTs. 
As is made clear by the lattice derivation, it is precisely the requirement of implementing a modified trace that leads to the correct degeneracies. From within CFT itself, and in the context of the general ${\cal LM}(p,p')$ models, it is still unclear precisely what principles select out the correct modular invariant partition functions.

Our modified trace (\ref{eq:Z}) holds for all of the logarithmic minimal models ${\cal LM}(p,p')$. The problem of calculating the modular invariant partition function of the general  ${\cal LM}(p,p')$ model is thus reduced to the problem of calculating the eigenvalues of the transfer matrix on the cylinder in sectors with $d$ defects. We hope to come back to this problem in later papers.

Helical boundary conditions can be applied to the construction of the torus by inserting a horizontal seam consisting of a power of the shift operator $\Omega$, as discussed in Appendix~\ref{app:Twist}.
Moving beyond the modular invariant torus partition function, it is possible to introduce vertical integrable (topological) seams yielding general toroidal conformal partition functions. For rational theories, the fusion of these vertical seams or their corresponding twist operators is described~\cite{PetkovaZuber} by the Ocneanu algebra~\cite{Ocneanu}. It is an interesting question as to what extent the Ocneanu algebra needs to be generalized in the logarithmic setting, but this question is beyond the scope of this paper.

\subsection*{Acknowledgments}

This work is supported by the Australian Research Council (ARC). JR is supported under the ARC Future Fellowship scheme, 
project number FT100100774. AMD is supported by the National Sciences and Engineering Research Council of Canada (NSERC) 
as a Postdoctoral Fellow. He is also grateful for support from the University of Queensland. The authors thank Yvan Saint-Aubin for helpful discussions.
Some of this work was done at Institut Henri Poincar\'e in Paris during the 2011 trimester on Advanced Conformal Field Theory and Applications. The authors thank the organizers and the institute for their generous hospitality.

%
%

\appendix
\section{Proof of the selection rules}
\label{app:a}

Here we prove the selection rules in Section~\ref{sec:selection}. For $N=1$, the transfer matrix is one-dimensional and its
single eigenvalue is given by
\be
 T(u,\mu)=v\cos u+v^{-1}\sin u.
\ee
This is readily seen to be reproduced by the selection rules which state that $\epsilon=\epsilon_1=1$.
To prove the selection rules for $N>1$, we will study the first terms in the Taylor series of $\Tb(u)$ around $u=0$  and compare their 
spectra with the similar objects in the periodic XX model. 

The expansion of $\Tb(u)$ around $u = 0$ is
\be
\Tb(u) = \Omega\, \big(I + u {\bf \mathcal H} + \mathcal O (u^2)\big), \qquad  \mathcal H = \sum_{i=1}^N e_i,
\ee
where we are using the sign convention for $\mathcal{H}$ of~\cite{AMDYSAinprep} for ease of comparison.
By expanding $T(u, \mu)$ around $u = 0$, one finds the eigenvalues of $\omega_d(\mathcal H)$ to be
\be
\mathrm{Eig}(\mathcal H) = \sum_{j=1}^N \epsilon_j \sin2x_j, \qquad
x_j = \left\{ \begin{array}{l l} 
\!\!\frac{(2j-1)\pi}{4N}+(-1)^{j+\frac{d+1}2} \mu, &\quad N \ \mathrm{odd}, \\[.15cm] 
\!\!\frac{j \pi}{N}-\mu,  &\quad N \ \mathrm{even},\; \frac{d}2 \ \mathrm{odd},\\[.15cm] 
\!\!\frac{(2j-1)\pi}{2N} - \mu, &\quad  N \ \mathrm{even}, \;\frac{d}2 \ \mathrm{even}. \end{array} \right.
\label{eq:eigenvaluesh}
\ee
Indeed,
\begin{align}
T(u,\mu)& = K(\mu)\times \epsilon\times \prod_{k=1}^N\frac{1}{\cos x_k} \prod_{j=1}^N \Big(e^{i u} \cos x_j+ i e^{-i u} \epsilon_j \sin x_j \Big) \label{eq:Tumu} \nonumber\\
& =  \underbrace{K(\mu) \times \epsilon\times \prod_{k=1}^N\frac{e^{i \epsilon_k x_k}}{\cos x_k}}_{\mathrm{Eig}(\Omega)}\prod_{j=1}^N\Big(\!\cos u + i \sin u e^{-2i \epsilon_j x_j}\Big),
\end{align}
where the under-braced prefactor is the corresponding eigenvalue of 
$\omega_d(\Omega)$. The eigenvalues of $\omega_d(\mathcal H)$ are then
\be
\mathrm{Eig}(\mathcal H) = i \sum_{j=1}^N ( \cos 2x_j - i \epsilon_j \sin 2 x_j) = \sum_{j=1}^N \epsilon_j \sin 2x_j,
\ee
because the first term sums to zero with the $x_j$ values given in (\ref{eq:eigenvaluesh}). As shown in~\cite{AMDYSAinprep}, the spectrum of $\omega_d(\mathcal H)$
 (for $\alpha=2\cos\mu N$ and $\beta=0$) is the same as the spectrum of the periodic XX Hamiltonian
\be
H = \sum_{j=1}^N \left( e^{2 i \mu} \sigma_j^+ \sigma_{j+1}^- + e^{-2 i \mu} \sigma_j^- \sigma_{j+1}^+ \right) 
\ee
in the sector with $S^z = \frac{d}2$, where
\be
 S^z=\tfrac{1}{2}\sum_{j=1}^N\sigma_j^z,\qquad \sigma_{N+1}^a\equiv\sigma_1^a,\qquad a\in\{+,-,z\}.
\ee 
To prove the selection rules, we proceed as in~\cite{AMD2011} by comparing the parameterizations of the spectra of the two hamiltonians $\mathcal{H}$ and $H$. First, we apply a Jordan-Wigner transformation to the XX Hamiltonian $H$.

\begin{Proposition} 
\label{PropA1}
In the sector with $S^z = \frac{d}2$, $H$ can be realized in terms of fermionic operators as
\be
H = \sum_q 2 \cos (q+2 \mu) \eta_q^{\dagger} \eta_q
\ee
where
\be
\quad q(m) = \left\{ \begin{array}{c l} \frac{(2m+1) \pi}{N}, &\quad \frac{N+d}2 \; \mathrm{even},  \vspace{0.12cm}\\  
\frac{2m \pi}{N}, &\quad \frac{N+d}2\; \mathrm{odd}, \end{array} \right. \qquad m = 0,\ldots, N-1.
\label{eq:diagHxxz}
\ee
The fermionic operators $\eta_q$ and $\eta_q^{\dagger}$ obey the usual
anti-commutation relations
\be
\{\eta_q^{\dagger}, \eta_{q'}\} = \delta_{q,q'}, \qquad \{\eta_q, \eta_{q'}\} = \{\eta^{\dagger}_q, \eta^{\dagger}_{q'}\} = 0,
\ee
and are given in terms of Pauli matrices by 
\begin{align}
\eta_q = \frac1{\sqrt{N}} \sum_{j=1}^N e^{-iqj} c_j,  \qquad  &\eta_q^\dagger = \frac1{\sqrt{N}} \sum_{j=1}^N e^{iqj} c_j^\dagger, \nonumber\\ \\[-0.55cm]
c_j  = \Big(\prod_{k=1}^{j-1}(-\sigma^z_k)\Big) \sigma_j^-,  \qquad & c_j^\dagger   = \Big(\prod_{k=1}^{j-1}(-\sigma^z_k)\Big) \sigma_j^+.\nonumber
\end{align}
\end{Proposition}
The proof for $\mu = 0$ appears in~\cite{Deguchi}. Its generalization to the present case is straightforward and therefore omitted.

Using $\sum_q \cos(q+2 \mu) = 0$, we immediately write 
\be
H = -\sum_q  2 \cos (q+2 \mu) \eta_q \eta_q^{\dagger}.
\ee
The state $| U \rangle = | +++ \,\ldots\, + \rangle$ is readily seen to be an eigenstate of $H$ with eigenvalue $0$. 
More generally, a basis of eigenvectors in the sector $S^z = \frac{d}{2}$ is generated by the states $\eta_{q_1} \ldots \eta_{q_n} | U \rangle$, $n = \frac{N-d}{2}$, with eigenvalues $\mathrm{Eig}(H)= -\sum_{i=1}^n 2 \cos (q_i+2 \mu)$.

Let us define 
\be
\Theta_m = 1-2\eta_{q} \eta_{q}^\dagger,\qquad m=0,1,\ldots,N-1,
\label{Th}
\ee 
where the relation between $q$ and $m$ is that of (\ref{eq:diagHxxz}). We define $\Theta_m$ for $m$ outside the interval in (\ref{Th}) by $\Theta_m \equiv \Theta_{m \,\mathrm{mod}\,N}$. All the eigenstates of $H$ 
indicated above are also eigenstates of $\Theta_m$ with eigenvalues $\pm 1$, and it follows that
\be
H = \sum_{m=0}^{N-1} \Theta_m \cos \big(q(m)+2 \mu \big). 
\label{eq:ThetaH}
\ee
The operators $\Theta_m$ play a role analogous to the parameters $\epsilon_j$ for the eigenvalues of $\Tb(u)$: They 
fix the eigenvalues of $H$ completely. Henceforth, we will consider them as constants, $\Theta_m\in \{+1, -1\}$, characterizing the eigenvalues of $H$ rather than as operators acting on spin states.

\subsection[$\mathbb Z_4$ sector ($N$ odd)]{\boldmath{$\mathbb Z_4$} sector (\boldmath{$N$} odd)}
\label{sec:selrulesodd}

By using the relation
\be
\prod_{j=1}^N \cos x_j  = 2^{-(N-1)} \sin \big(\tfrac{\pi}4 + (-1)^{\frac{d+1}2}\mu N \big),
\label{eq:simplif1}
\ee
the expression for $T(u,\mu)$ of equation \eqref{eq:Tumu} simplifies to  
\be 
T(u,\mu) = \epsilon\, e^{-\frac{i\pi N}{4}}  \exp\!\Big( i \sum_{k=1}^N\epsilon_k x_k\Big)\prod_{j=1}^N\Big(\! \cos u + i \sin u e^{-2i \epsilon_j x_j}\Big),
\label{eq:simplified1}
\ee
and the corresponding eigenvalue of $\omega_d(\Omega)$, found by setting $u= 0$, to   
\be
\mathrm{Eig}(\Omega) =T(0,\mu) =  \epsilon\, e^{-\frac{i\pi N}{4}}  \exp\! \Big( i \sum_{k=1}^N\epsilon_k x_k\Big).
\label{eq:eigomega1}
\ee
Because $\omega_d(\Omega^N) = v^{Nd} \omega_d(I)$, each eigenvalue of $\omega_d(\Omega)$ is $v^d$ times a $v$-independent phase. The 
dependence on $v$ in \eqref{eq:eigomega1} is hidden in the contribution $\exp(i \sum_{k=1}^N \epsilon_k x_k)$. By extracting from $x_k$ the $\mu$-dependent part and matching it with $v^d$, one finds
\be 
\sum_{k=1}^N (-1)^k\epsilon_k  = d(-1)^{\frac{d+1}2}
\label{sumk}
\ee
as announced in (\ref{eq:selrules1}). This constraint reduces the number of possible eigenvalues to $\left(\!\!\begin{smallmatrix} N \\ \frac{N-d}{2}\end{smallmatrix} \!\!\right)$, which is exactly the 
dimension of the representation $\omega_d$.

To prove that the eigenvalues of $\omega_d(\Tb(u))$ explore {\em all} the $\left(\!\!\begin{smallmatrix} N \\ \frac{N-d}{2}\end{smallmatrix}\!\! \right)$ possibilities (i.e.~that there are {\em no repetitions}), we consider the XX Hamiltonian and 
initially show it can be written as 
\be
H =  \left\{ \begin{array}{l l} 
\!\displaystyle{\sum_{j=1}^N} \Big( \!(-1)^{j+\frac{d+1}2}  \Theta_{ -\frac{N}{4} + \frac{2j-1}4(-1)^{j+\frac{d+1}2}} \!\Big) \sin 2x_j, & \quad \ \frac{N+d}2 \; \mathrm{odd},  \\[.15cm] 
 \!\displaystyle{\sum_{j=1}^N} \Big(\! (-1)^{j+\frac{d+1}2}  \Theta_{ -\frac{N+2}{4} + \frac{2j-1}4(-1)^{j+\frac{d+1}2}} \!\Big) \sin 2x_j, & \quad \ \frac{N+d}2 \; \mathrm{even}.
\end{array} \right.
\label{eq:Hrewritten}
\ee
This can be checked for all possible parities of $\frac{N-1}2$ and $\frac{d-1}2$. Here we only verify explicitly the case $\frac{N-1}2$ and $\frac{d-1}2$ both even. The other cases are very similar. Using the property $\Theta_m \equiv \Theta_{m \,\mathrm{mod}\,N}$, we thus compute

\begin{align}
H &= \sum_{m=0}^{N-1} \Theta_m \cos\! \big(\tfrac{2 m \pi}{N}+2 \mu \big) =  \sum_{m=0}^{N-1} \Theta_m \sin\! \big(\tfrac{(N + 4 m) \pi}{2N} + 2 \mu \big)  \nonumber\\ &= \sum_{n=0}^{N-1} \Theta_{n-\frac{N-1}4} \sin\!\big(\tfrac{(4n+1)\pi}{2N} + 2 \mu\big) = \Big( \sum_{n=0}^{\frac{N-1}2}  + \sum_{n=\frac{N+1}2}^{N-1} \Big)\Theta_{n-\frac{N-1}4} \sin\!\big(\tfrac{(4n+1)\pi}{2N} + 2 \mu\big)  \nonumber\\
& = \sum_{n=0}^{\frac{N-1}2} \Theta_{n-\frac{N-1}4} \sin\!\big(\tfrac{(4n+1)\pi}{2N} + 2 \mu\big) - \sum_{n'=1}^{\frac{N-1}2} \Theta_{- n' -\frac{N-1}4} \sin\!\big(\tfrac{(4n'-1)\pi}{2N} - 2 \mu\big)\nonumber \\
& = \sum_{r=1, 5, 9, \ldots}^{2N - 1} \Theta_{\frac{r-N}4 } \sin\!\big(\tfrac{r\pi}{2N} + 2 \mu\big) - \sum_{s=3, 7, 11, \ldots}^{2N - 3} \Theta_{-\frac{s+N}4 } \sin\!\big(\tfrac{s\pi}{2N} - 2 \mu\big) \nonumber\\
& = \sum_{j=1}^N (-1)^{j+1}  \Theta_{-\frac{N}{4}+\frac{2j-1}4(-1)^{j+1}} \sin\! \big( \underbrace{\tfrac{(2j-1)\pi}{2N} + (-1)^{j+1} 2 \mu}_{=\,2x_j} \big)
\end{align}
which is of the proposed form. 

Now, equation \eqref{eq:Hrewritten} can be directly related to the expression \eqref{eq:eigenvaluesh} for the eigenvalues of $\omega_d(\mathcal H)$, provided the identification
\be
\epsilon_j = \left\{ \begin{array}{ll}  
\!\!(-1)^{j+\frac{d+1}2}  \Theta_{-\frac{N}{4} + \frac{2j-1}4(-1)^{j+\frac{d+1}2}},  & \quad \ \frac{N+d}2 \; \mathrm{odd}, \\[.2cm]
\!\!(-1)^{j+\frac{d+1}2}  \Theta_{-\frac{N+2}{4} + \frac{2j-1}4(-1)^{j+\frac{d+1}2}},  & \quad \ \frac{N+d}2 \; \mathrm{even}
\end{array}\right. 
\label{eq:epsilonTheta1}
\ee
is made. As in Figure~\ref{fig:circle}, the eigenvalues of $H$ are non-degenerate (for generic $\mu$), making this identification unique. This induces a bijection between eigenvalues of $\omega_d(\mathcal H)$
 and of $H$. Because the $\left(\!\!\begin{smallmatrix} N \\ \frac{N-d}{2}\end{smallmatrix}\!\!\right)$ possible choices of $\Theta_m$ are all explored (exactly once) in the spectrum of $H$, the same is true for the $\epsilon_j$ parameters associated with $\omega_d(\mathcal{H})$. 
The proof of the selection rule for the overall sign $\epsilon$ in \eqref{eq:selrules1} is presented in Appendix~\ref{app:overallsign}.

\begin{figure}[ht]
\begin{center}
\psset{unit=4}
\begin{pspicture}(-1.7,0)(1.7,1.7)
\psline{->}(-1.1,0)(1.1,0)
\psline{->}(0,0)(0,1.1)
\psarc{-}(0,0){1}{0}{180}
\psline[linecolor=darkblue](0,0)(0.974928, 0.222521)
\psline[linecolor=darkblue](0,0)(0.781831, 0.62349)
\psline[linecolor=darkblue](0,0)(0.433884, 0.900969)
\psline[linecolor=darkblue](0,0)(0.,1)
\psline[linecolor=darkblue](0,0)(-0.433884, 0.900969)
\psline[linecolor=darkblue](0,0)(-0.781831, 0.62349)
\psline[linecolor=darkblue](0,0)(-0.974928, 
  0.222521)
\psline[linecolor=red, linestyle=dashed, dash=2pt 1pt](0,0)(0.983047, 0.183356)
\psline[linecolor=red, linestyle=dashed, dash=2pt 1pt](0,0)(0.756273, 0.654256)
\psline[linecolor=red, linestyle=dashed, dash=2pt 1pt](0,0)(0.469566, 0.882897)
\psline[linecolor=red, linestyle=dashed, dash=2pt 1pt](0,0)(-0.0399893, 0.9992)
\psline[linecolor=red, linestyle=dashed, dash=2pt 1pt](0,0)(-0.397508, 0.917599)
\psline[linecolor=red, linestyle=dashed, dash=2pt 1pt](0,0)(-0.806139, 0.591726)
\psline[linecolor=red, linestyle=dashed, dash=2pt 1pt](0,0)(-0.96525, 0.26133)
\rput(1.16991, 0.26702){$\epsilon_1$}
\rput(0.938198, 0.748188){$\epsilon_2$}
\rput(0.52066, 1.08116){$\epsilon_3$}
\rput(0., 1.2){$\epsilon_4$}
\rput(-0.52066, 1.08116){$\epsilon_5$}
\rput(-0.938198, 0.748188){$\epsilon_6$}
\rput(-1.16991, 0.267025){$\epsilon_7$}
\rput(1.41365, 0.322655){\darkgreen $\Theta_5$}
\rput(1.13366, 0.90406){\darkgreen$ -\Theta_4$}
\rput(0.629131, 1.3064){\darkgreen $\Theta_6$}
\rput(0., 1.45){\darkgreen $ -\Theta_3$}
\rput(-0.629131, 1.3064){\darkgreen$ \Theta_0$}
\rput(-1.13366, 0.90406){\darkgreen $ -\Theta_2$}
\rput(-1.41365, 0.322655){\darkgreen $ \Theta_1$}
\rput(1.65738, 0.378286){\orange $-\Theta_5$}
\rput(1.32911, 1.05993){\orange $ \Theta_6$}
\rput(0.737602, 1.53165){\orange $ -\Theta_4$}
\rput(0., 1.7){\orange $ \Theta_0$}
\rput(-0.737602, 1.53165){\orange $ -\Theta_3$}
\rput(-1.32911, 1.05993){\orange $ \Theta_1$}
\rput(-1.65738, 0.378286){\orange $ -\Theta_2$}
\end{pspicture}
\caption{$\mathbb Z_4$ sector ($N$ odd). 
For $N = 7$, $j = 1, \ldots, N$ and $\mu = 0.02$, the angles $2\hspace{0.02cm}x_j(\mu = 0)$ and $2\hspace{0.02cm}x_j(\mu)$, see \eqref{xj}, are indicated respectively in blue (full lines) and in red (dashed lines). 
Associated to a given angle labelled by $\epsilon_j$, $\Theta_m$ is indicated in green (innermost) for $\frac{d-1}2$ even and orange (outermost) for $\frac{d-1}2$ odd. The groundstate for $d=1$ is characterized by $\Theta_2, \Theta_3, \Theta_4 = -1$ and by $\Theta_3, \Theta_4 = -1$ for $d=3$. In both cases, the remaining $\Theta_m$ values are $+1$. For $\mu$ generic, there are no degeneracies.} 
\label{fig:circle}
\end{center}
\end{figure}
In a given sector $d$, the identification of an eigenvalue as a groundstate energy depends on $\mu$. To reproduce the selection rule given in~\cite{PRV2010},  we consider $\mu$ close to zero and study the largest eigenvalue of $H$. 
It is recalled that the eigenstates of $H$ in the sector $S^z = \frac{d}{2}$ are of the form $\eta_{q_1} \ldots \eta_{q_n} | U \rangle$ with $n = \frac{N-d}{2}$, i.e.~$|\{m;\, \Theta_m = -1\}| = \frac{N-d}{2}$ and $|\{m;\, \Theta_m = +1\}| = \frac{N+d}{2}$.
From equation (\ref{eq:ThetaH}), the groundstate is built by choosing to be $-1$ the 
$\frac{N-d}{2}$ parameters $\Theta_m$ with $\cos(q(m))$ negative and as large as possible in absolute value. 
It follows that the correct choice is 
\be
 \Theta_m=\left\{\begin{array}{lc}
 \!\!-1,\quad &m=\big\lfloor\frac{N+d+2}{4}\big\rfloor,\ldots,N-1-\big\lfloor\frac{N+d}{4}\big\rfloor,
 \\[.15cm]
 \!\!+1,\quad &\mathrm{otherwise}.
 \end{array}\right.
\label{Tm1}
\ee
(For $\mu$ outside the interval $[-\frac{\pi}{2N}, \frac{\pi}{2N}]$, this 
does not characterize the groundstate.) For $\frac{N-1}2$ and $\frac{d-1}2$ both even,
for example, equation (\ref{eq:epsilonTheta1}) is rewritten as
\begin{align}
\epsilon_{2i} = - \Theta_{\frac{3N+1}4-i}, \quad i = 1, \ldots, {\textstyle \frac{N-1}2}, \qquad 
\epsilon_{2i+1} = \left\{\begin{array}{ll}
\!\!\Theta_{\frac{3N+1}4+i}, & \quad i = 0, \ldots, \frac{N-5}4, \\[.2cm]
\!\!\Theta_{-\frac{N-1}4+i}, & \quad i = \frac{N-1}4, \ldots, \frac{N-1}2.
  \end{array}\right.
\label{eq:eps} 
\end{align}
The negative $\Theta_m$ parameters are thus identified with the $\epsilon_j$ parameters of the form $\epsilon_{2i}$ for $i = \frac{d+3}4, \ldots, \frac{N-1}2 - \frac{d-1}4$. It follows that $\epsilon_{2i+1}=1$ for every $i$, and $\epsilon_{2i} = 1$ for $i = \frac{d+3}4, \ldots, \frac{N-1}2 - \frac{d-1}4$. The only negative $\epsilon_j$ parameters are $\epsilon_{2i}=\bar{\epsilon}_{2i}=-1$ for $i=1,\ldots,\frac{d-1}{4}$, here indicated in terms of barred and unbarred epsilons. This is precisely the statement of equation (\ref{eq:gs}) for $\frac{d-1}2$ even. The other three cases (where $\frac{N-1}2$ or $\frac{d-1}2$ is odd) are examined in a similar fashion.

\paragraph{Comparison with the \boldmath{$\mathbb{Z}_4$} sector selection rules of~\cite{PRV2010}:}
We now compare our selection rules for $N$ odd with those of~\cite{PRV2010}. In that work, the selection rules pertain to the limit $\mu\to0$ and are expressed in terms of the excess parameters $\sigma$ and $\bar{\sigma}$ defined in terms of our parameters $\epsilon_j$ and $\bar{\epsilon}_j=\epsilon_{N+1-j}$ by
\begin{eqnarray}
 \sigma&\!\!=\!\!&\big| \{ i=1,\ldots,\big\lfloor\tfrac{N+1}{4}\big\rfloor;\,\epsilon_{2i} = -1 \} \big| - 
  \big| \{ i=0,\ldots,\big\lfloor\tfrac{N-1}{4}\big\rfloor;\,\epsilon_{2i+1} = -1 \} \big|,\nonumber \\[-.12cm] \label{eq:oddexcesses} \\[-.12cm]
 \bar{\sigma}&\!\!=\!\!&\big| \{ i=1,\ldots,\big\lfloor\tfrac{N-1}{4}\big\rfloor;\,\bar{\epsilon}_{2i} = -1 \} \big| - 
  \big| \{ i=0,\ldots,\big\lfloor\tfrac{N-3}{4}\big\rfloor;\,\bar{\epsilon}_{2i+1} = -1 \} \big|. \nonumber
\end{eqnarray}
Using (\ref{sumk}), we thus find
\begin{eqnarray}
 \sigma+\bar{\sigma}&\!\!=\!\!&\big| \{ i=1,\ldots,\tfrac{N-1}{2};\,\epsilon_{2i} = -1 \} \big|
  - \big| \{ i=0,\ldots,\tfrac{N-1}{2};\,\epsilon_{2i+1} = -1 \} \big| \nonumber\\
 &\!\!=\!\!&\tfrac{1}{2}\Big(\tfrac{N-1}{2}-\sum_{i=1}^{\frac{N-1}{2}}\epsilon_{2i}\Big)
  -\tfrac{1}{2}\Big(\tfrac{N+1}{2}-\sum_{i=0}^{\frac{N-1}{2}}\epsilon_{2i+1}\Big)
  =-\tfrac{1}{2}-\tfrac{1}{2}\sum_{j=1}^{N}(-1)^j\epsilon_{j}\nonumber\\
 &\!\!=\!\!&\left\{ \begin{array}{l l} 
\!\!\frac{d-1}2, & \quad \frac{d-1}2 \; \mathrm{even}, \\[.2cm] 
\!\!-\frac{d+1}2, & \quad \frac{d-1}2 \; \mathrm{odd}, \end{array} \right.
\label{ssd}
\end{eqnarray}
which is how the corresponding selection rule is stated in~\cite{PRV2010}. It follows readily that $\sigma-\bar{\sigma}$
is an even integer. 
For the groundstate, from \eqref{eq:oddexcesses} and \eqref{eq:gs}, we have
\be
 \sigma=\bar{\sigma}=\left\{ \begin{array}{l l} 
\!\!\frac{d-1}4, & \quad \frac{d-1}2 \; \mathrm{even}, \\[.2cm] 
\!\!-\frac{d+1}4, & \quad \frac{d-1}2 \; \mathrm{odd}, \end{array} \right.
\ee
again in accordance with~\cite{PRV2010}.

\subsection[Ramond and Neveu-Schwarz sectors ($N$ even)]{Ramond and Neveu-Schwarz sectors (\boldmath{$N$} even)}

First, we simplify the expressions for $T(u, \mu)$ by using the identities 
\begin{align}
\prod_{j=1}^N \cos x_j = (-1)^{\frac{N}{2}} 2^{-(N-1)} \times 
\left\{\begin{array}{ll} 
\!\!\sin \mu N, & \frac{d}2 \;\mathrm{odd}, \vspace{0.2cm}\\ 
\!\!\cos \mu N, & \frac{d}2 \;\mathrm{even}. 
\end{array}\right. 
\label{eq:simplif2}
\end{align} 
For both parities of $\frac d 2$, we can thus write
\be 
T(u,\mu) = \epsilon \, i^{\frac{N}2}  \exp\! \Big( i \sum_{k=1}^N\epsilon_k x_k\Big)\prod_{j=1}^N\Big(\! \cos u + i \sin u e^{-2i \epsilon_j x_j}\Big).
\label{eq:simplified2}
\ee
In the limit $u \rightarrow 0$, one finds the corresponding eigenvalue of $\omega_d(\Omega)$ to be given by
\be
 \mathrm{Eig}(\Omega) =T(0, \mu) =  \epsilon \, i^{\frac{N}2}  \exp\! \Big( i \sum_{k=1}^N\epsilon_k x_k\Big)
\label{eq:eigomega2}
\ee
and this must equal $v^d$ times a phase. The dependence on the winding parameter is $\exp(-i\mu \sum \epsilon_k)$, and hence $\sum_{j=1}^N\epsilon_j = -d$ as announced in (\ref{eq:selrules2}). 

To show that there are no repetitions between the $\left(\!\!\begin{smallmatrix} N \\ \frac{N-d}{2}\end{smallmatrix}\!\!\right)$ possible choices for the set of parameters $\{\epsilon_j\}$ satisfying $\sum_{j=1}^N\epsilon_j = -d$, we proceed as in the $\mathbb{Z}_4$ sector by establishing a map between the eigenvalues of $\omega_d(\mathcal H)$ and those of $H$.
First, we note that the XX Hamiltonian can be written as
\be
 H=\left\{ \begin{array}{l l} 
\!\!\displaystyle{-\sum_{j=1}^N}\,\Theta_{\frac{3N}{4}-j} \sin 2x_j, & \quad \ \frac{N}2 \; \mathrm{even},  \\[.15cm] 
\!\!\displaystyle{-\sum_{j=1}^N}\,\Theta_{\frac{3N+2}{4}-j} \sin 2x_j, & \quad \ \frac{N}2 \; \mathrm{odd},\ \frac{d}{2}\;\mathrm{even},  \\[.15cm] 
\!\!\displaystyle{-\sum_{j=1}^N}\,\Theta_{\frac{3N-2}{4}-j} \sin 2x_j, & \quad \ \frac{N}2 \; \mathrm{odd},\ \frac{d}{2}\;\mathrm{odd}.
\end{array} \right.
\label{HTeven}
\ee
For $\frac N 2$ and $\frac d 2$ both even, for example, this follows from
\begin{align} 
H &= \sum_{m=0}^{N-1} \Theta_m \cos\! \Big(\tfrac{(2 m +1) \pi}{N}+2 \mu \Big) = \sum_{m=0}^{N-1} \Theta_m \sin\! \Big(\!\big(\tfrac{N}2 - 2m -1 \big)\tfrac{\pi}{N} - 2 \mu \Big)  \nonumber\\ 
&= \sum_{i=1}^{N} \Theta_{\frac{N}4-i} \sin\! \Big(\tfrac{(2i-1)\pi}{N} - 2 \mu \Big)=-\sum_{j=1}^{N}\Theta_{\frac{3N}4-j}\sin \!\Big(\underbrace{\tfrac{(2j-1)\pi}{N} - 2 \mu}_{=\,2x_j} \Big).
\label{HT2}
\end{align}
The other cases are examined in a similar way.

\begin{figure}[ht] 
\begin{center}
\psset{unit=4}
\begin{pspicture}(-0.6,-1.15)(0.6,0.9)
\psline{->}(-0.6,0)(0.6,0)
\psline{->}(0,-0.6)(0,0.6)
\psarc{-}(0,0){0.5}{0}{360}
\psline[linecolor=darkblue](0,0)(0.46194, 0.191342)
\psline[linecolor=darkblue](0,0)(0.191342, 0.46194)
\psline[linecolor=darkblue](0,0)(-0.191342, 0.46194)
\psline[linecolor=darkblue](0,0)(-0.46194, 0.191342)
\psline[linecolor=darkblue](0,0)(-0.46194, -0.191342)
\psline[linecolor=darkblue](0,0)(-0.191342, -0.46194)
\psline[linecolor=darkblue](0,0)(0.191342, -0.46194)
\psline[linecolor=darkblue](0,0)(0.46194, -0.191342)
\psline[linecolor=red, linestyle=dashed, dash=2pt 2pt](0,0)(0.475753, 0.153814)
\psline[linecolor=red, linestyle=dashed, dash=2pt 2pt](0,0)(0.227646, 0.445171)
\psline[linecolor=red, linestyle=dashed, dash=2pt 2pt](0,0)(-0.153814, 0.475753)
\psline[linecolor=red, linestyle=dashed, dash=2pt 2pt](0,0)(-0.445171, 0.227646)
\psline[linecolor=red, linestyle=dashed, dash=2pt 2pt](0,0)(-0.475753, -0.153814)
\psline[linecolor=red, linestyle=dashed, dash=2pt 2pt](0,0)(-0.227646, -0.445171)
\psline[linecolor=red, linestyle=dashed, dash=2pt 2pt](0,0)(0.153814, -0.475753)
\psline[linecolor=red, linestyle=dashed, dash=2pt 2pt](0,0)(0.445171, -0.227646)
\rput(0.600522, 0.248744){$\epsilon_1$}
\rput(0.267878, 0.646716){$\epsilon_2$}
\rput(-0.267878, 0.646716){$\epsilon_3$}
\rput(-0.646716, 0.267878){$\epsilon_4$}
\rput(-0.646716, -0.267878){$\epsilon_5$}
\rput(-0.267878, -0.646716){$\epsilon_6$}
\rput(0.267878, -0.646716){$\epsilon_7$}
\rput(0.646716, -0.267878){$\epsilon_8$}
\rput(0.831492, 0.344415){$-\Theta_5$}
\rput(0.344415, 0.831492){$-\Theta_4$}
\rput(-0.344415, 0.831492){$-\Theta_3$}
\rput(-0.831492, 0.344415){$-\Theta_2$}
\rput(-0.831492, -0.344415){$-\Theta_1$}
\rput(-0.344415, -0.831492){$-\Theta_0$}
\rput(0.344415, -0.831492){$-\Theta_7$}
\rput(0.831492, -0.344415){$-\Theta_6$}
\rput(0,-1.1){(a)}
\end{pspicture}
\hspace{4cm}
\begin{pspicture}(-0.6,-1.15)(0.6,0.9)
\psline{->}(-0.6,0)(0.6,0)
\psline{->}(0,-0.6)(0,0.6)
\psarc{-}(0,0){0.5}{0}{360}
\psline[linecolor=darkblue](0,0)(0.353553, 0.353553)
\psline[linecolor=darkblue](0,0)(0, 0.5)
\psline[linecolor=darkblue](0,0)(-0.353553, 0.353553)
\psline[linecolor=darkblue](0,0)(-0.5, 0)
\psline[linecolor=darkblue](0,0)(-0.353553, -0.353553)
\psline[linecolor=darkblue](0,0)(0, -0.5)
\psline[linecolor=darkblue](0,0)(0.353553, -0.353553)
\psline[linecolor=darkblue](0,0)(0.5, 0)
\psline[linecolor=red, linestyle=dashed, dash=2pt 2pt](0,0)(0.380677, 0.324169)
\psline[linecolor=red, linestyle=dashed, dash=2pt 2pt](0,0)(0.0399573, 0.498401)
\psline[linecolor=red, linestyle=dashed, dash=2pt 2pt](0,0)(-0.324169, 0.380677)
\psline[linecolor=red, linestyle=dashed, dash=2pt 2pt](0,0)(-0.498401, 0.0399573)
\psline[linecolor=red, linestyle=dashed, dash=2pt 2pt](0,0)(-0.380677, -0.324169)
\psline[linecolor=red, linestyle=dashed, dash=2pt 2pt](0,0)(-0.0399573, -0.498401)
\psline[linecolor=red, linestyle=dashed, dash=2pt 2pt](0,0)(0.324169, -0.380677)
\psline[linecolor=red, linestyle=dashed, dash=2pt 2pt](0,0)(0.498401, -0.0399573)
\rput(0.459619, 0.459619){$\epsilon_1$}
\rput(0, 0.65){$\epsilon_2$}
\rput(-0.459619, 0.459619){$\epsilon_3$}
\rput(-0.67, 0){$\epsilon_4$}
\rput(-0.459619, -0.459619){$\epsilon_5$}
\rput(0, -0.65){$\epsilon_6$}
\rput(0.459619, -0.459619){$\epsilon_7$}
\rput(0.67, 0){$\epsilon_8$}
\rput(0.636396, 0.636396){$-\Theta_5$}
\rput(0, 0.87){$-\Theta_4$}
\rput(-0.636396, 0.636396){$-\Theta_3$}
\rput(-0.9, 0){$-\Theta_2$}
\rput(-0.636396, -0.636396){$-\Theta_1$}
\rput(0, -0.87){$-\Theta_0$}
\rput(0.636396, -0.636396){$-\Theta_7$}
\rput(0.9, 0){$-\Theta_6$}
\rput(0,-1.1){(b)}
\end{pspicture}
\caption{For $N = 8$, $j = 1, \ldots, N$ and $\mu = 0.04$, 
the angles $2\hspace{0.02cm}x_j(\mu = 0)$ and $2\hspace{0.02cm}x_j(\mu)$, see \eqref{xj}, are indicated respectively in blue (full lines) and in red (dashed lines), in (a) the Ramond sector ($N$ even, $\frac d 2$ even) and (b) the Neveu-Schwarz sector ($N$ even, $\frac d 2$ odd). The groundstate for $d=0$ is characterized by $\Theta_2, \Theta_3, \Theta_4, \Theta_5 = -1$ and the one for $d=2$ by $\Theta_3, \Theta_4, \Theta_5=-1$, with all other $\Theta_m$ values equal to $+1$. In both sectors, because $\sin 2x_j = - \sin 2x_{j+\frac{N}2}$, the eigenvalues of $H$ are degenerate and, a priori, this could lead to two possible identifications.}
\label{fig:semicircle}
\end{center}
\end{figure}
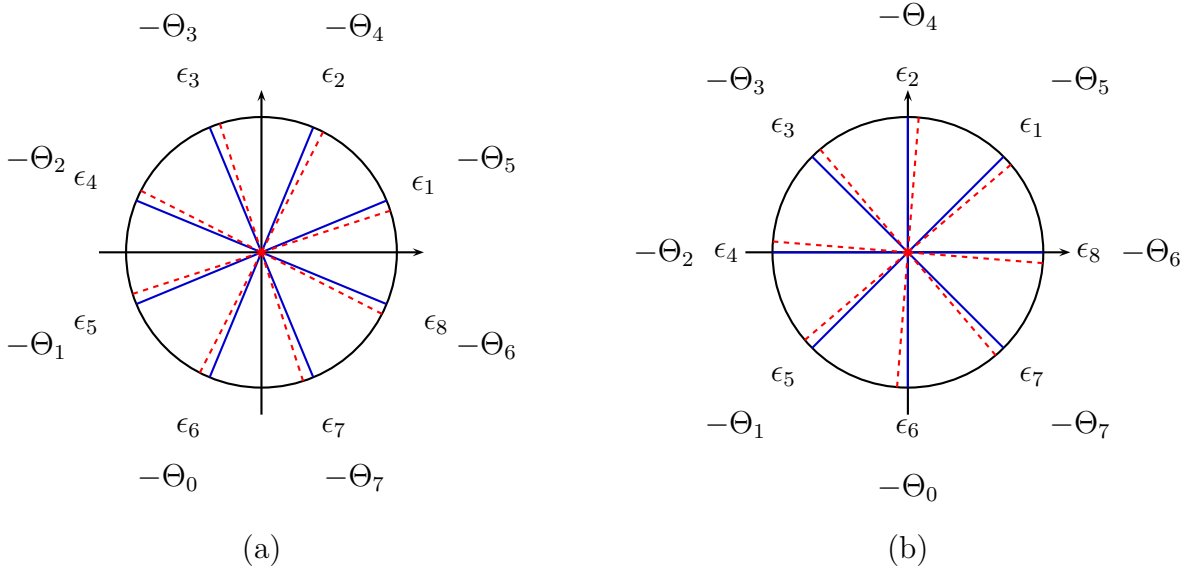

Now, the final expression in (\ref{HT2}) suggests the identification $\epsilon_j = - \Theta_{\frac{3N}4-j}$. However, from the expression just before it, one might be tempted to make the alternative identification $\epsilon_j = \Theta_{\frac{N}4-j}$. This extra possibility is caused by degeneracies in the spectrum of $\omega_d(\mathcal H)$ that are not transferred to the spectrum of $\omega_d(\Tb(u))$, see Figure~\ref{fig:semicircle}. 
To resolve this issue, we recall that $|\{m;\, \Theta_m = \pm 1\}| = \frac{N\pm d}{2}$ and note that with the identification $\epsilon_j = - \Theta_{\frac{3N}4-j}$, 
 \be
 \sum_{j=1}^N \epsilon_j = -\sum_{m=0}^{N-1}\Theta_m = \big|\{m;\, \Theta_m = - 1\}\big| - \big|\{m;\, \Theta_m = + 1\}\big| = -d,
 \ee
in accordance with \eqref{eq:selrules2}. The same calculation based on the alternative identification, $\epsilon_j = \Theta_{\frac{N}4-j}$, yields $\sum_{j} \epsilon_j =d$ and is therefore {\it incompatible} with \eqref{eq:selrules2} for $d>0$. As discussed in Appendix~\ref{app:overallsign}, even for $d=0$, only the first identification results in a fully consistent picture in general. 

As in Appendix~\ref{sec:selrulesodd}, the identification is a one-to-one map between the eigenvalues of the two models. Because all the different choices of $\Theta_m$ are explored in the XX model (with the constraint $|\{m;\, \Theta_m = -1\}| = \frac{N-d}2$), there are likewise no repetitions in the column configurations characterizing the eigenvalues in the loop model. In general, the identification depends on the parities of $\frac N 2$ and $\frac d 2$, as we have
\be
\epsilon_j = \left\{ \begin{array}{ll}
\!\! - \Theta_{\frac{3N}4 - j}, &\quad \frac{N}2 \ \mathrm{even}, \\[.15cm] 
\!\!- \Theta_{\frac{3N+2}4 - j}, &\quad \frac{N}2 \ \mathrm{odd},\; \frac{d}2 \ \mathrm{even}, \\[.15cm] 
\!\! - \Theta_{\frac{3N-2}4 - j}, &\quad \frac{N}2\ \mathrm{odd},\; \frac{d}2 \ \mathrm{odd},
\end{array}\right.
\label{eq:epsilonTheta2}
\ee
in accordance with (\ref{HTeven}).

The proof of the selection rule for the overall sign $\epsilon$ in (\ref{eq:selrules2}) is presented in Appendix~\ref{app:overallsign}. 
As in Appendix~\ref{sec:selrulesodd}, the groundstate of $H$ is obtained by fixing the parameters $\Theta_m$ as in \eqref{Tm1}.
Here this translates into the selection rule \eqref{eq:gseven1} when \eqref{eq:epsilonTheta2} is applied.

\paragraph{Comparison with the Ramond sector selection rules of~\cite{PRV2010}:}
We now compare our selection rules for $N$ and $\frac{d}{2}$ both even with those of~\cite{PRV2010}. In that work, the selection rules pertain to the limit $\mu\to0$ and are expressed in terms of the excess parameters $\sigma$ and $\bar{\sigma}$ defined in terms of our parameters $\epsilon_j$ by
\begin{eqnarray}
 \sigma&\!\!=\!\!&\big| \{ j=1,\ldots,\big\lfloor\tfrac{N+2}{4}\big\rfloor;\,\epsilon_{j} = -1 \} \big| - 
  \big| \{ j=1,\ldots,\big\lfloor\tfrac{N+2}{4}\big\rfloor;\,\epsilon_{N+1-j} = +1 \} \big|,\nonumber \\[-.12cm] \label{eq:Ramondexcesses} \\[-.12cm]
 \bar{\sigma}&\!\!=\!\!&\big| \{ j=1,\ldots,\big\lfloor\tfrac{N}{4}\big\rfloor;\,\epsilon_{\frac{N}{2}+1-j} = -1 \} \big| - 
  \big| \{ j=1,\ldots,\big\lfloor\tfrac{N}{4}\big\rfloor;\,\epsilon_{\frac{N}{2}+j} = +1 \} \big|.\nonumber
\end{eqnarray}
Using $\sum_{j=1}^N\epsilon_j = -d$, we thus find
\begin{eqnarray}
 \sigma+\bar{\sigma}
 &\!\!=\!\!&\tfrac{1}{2}\Big(\big\lfloor\tfrac{N+2}{4}\big\rfloor-\sum_{j=1}^{\lfloor\frac{N+2}{4}\rfloor}\epsilon_{j}\Big)
  -\tfrac{1}{2}\Big(\big\lfloor\tfrac{N+2}{4}\big\rfloor+\sum_{j=1}^{\lfloor\frac{N+2}{4}\rfloor}\epsilon_{N+1-j}\Big)\nonumber\\
 &\!\!+\!\!&\tfrac{1}{2}\Big(\big\lfloor\tfrac{N}{4}\big\rfloor-\sum_{j=1}^{\lfloor\frac{N}{4}\rfloor}\epsilon_{\frac{N}{2}+1-j}\Big)
  -\tfrac{1}{2}\Big(\big\lfloor\tfrac{N}{4}\big\rfloor+\sum_{j=1}^{\lfloor\frac{N}{4}\rfloor}\epsilon_{\frac{N}{2}+j}\Big)\nonumber\\
 &\!\!=\!\!&-\tfrac{1}{2}\sum_{j=1}^{N}\epsilon_{j}\nonumber\\
 &\!\!=\!\!&\frac{d}{2}
\end{eqnarray}
which is the corresponding selection rule of~\cite{PRV2010}. It follows readily that $\sigma-\bar{\sigma}$
is an even integer. 
For the groundstate, from \eqref{eq:Ramondexcesses} and \eqref{eq:gseven1}, we have 
\be
 \sigma=\bar{\sigma}=\frac{d}{4},
\ee
again in accordance with~\cite{PRV2010}.

\paragraph{Comparison with the Neveu-Schwarz sector selection rules of~\cite{PRV2010}:}
We now compare our selection rules for $N$ even and $\frac{d}{2}$ odd with those of~\cite{PRV2010}. In that work, the selection rules pertain to the limit $\mu\to0$ and are expressed in terms of the excess parameters $\sigma$ and $\bar{\sigma}$ defined in terms of our parameters $\epsilon_j$ by
\begin{eqnarray}
 \sigma&\!\!=\!\!&-\delta+\big| \{ j=1,\ldots,\big\lfloor\tfrac{N}{4}\big\rfloor;\,\epsilon_{j} = -1 \} \big| - 
  \big| \{ j=1,\ldots,\big\lfloor\tfrac{N}{4}\big\rfloor;\,\epsilon_{N-j} = +1 \} \big|,\nonumber \\[-.12cm] \\[-.12cm]
 \bar{\sigma}&\!\!=\!\!&-\bar{\delta}+\big| \{ j=1,\ldots,\big\lfloor\tfrac{N-2}{4}\big\rfloor;\,\epsilon_{\frac{N}{2}-j} = -1 \} \big| - 
  \big| \{ j=1,\ldots,\big\lfloor\tfrac{N-2}{4}\big\rfloor;\,\epsilon_{\frac{N}{2}+j} = +1 \} \big|.\ \nonumber
\end{eqnarray}
Here we have introduced the two additional parameters
\be
 \delta,\bar{\delta}\in\{0,1\}
\ee
to implement the freedom in the definition in~\cite{PRV2010} of the excess parameters.
It is also noted that $\epsilon_{\frac{N}{2}}$ and $\epsilon_N$ do not appear in the expressions for $\sigma$ and $\bar{\sigma}$. 
We now evaluate
\begin{eqnarray}
 \sigma+\bar{\sigma}
 &\!\!=\!\!&-\delta-\bar{\delta}+\tfrac{1}{2}\Big(\big\lfloor\tfrac{N}{4}\big\rfloor-\sum_{j=1}^{\lfloor\frac{N}{4}\rfloor}\epsilon_{j}\Big)
  -\tfrac{1}{2}\Big(\big\lfloor\tfrac{N}{4}\big\rfloor+\sum_{j=1}^{\lfloor\frac{N}{4}\rfloor}\epsilon_{N-j}\Big)\nonumber\\
 &\!\!+\!\!&\tfrac{1}{2}\Big(\big\lfloor\tfrac{N-2}{4}\big\rfloor-\sum_{j=1}^{\lfloor\frac{N-2}{4}\rfloor}\epsilon_{\frac{N}{2}-j}\Big)
  -\tfrac{1}{2}\Big(\big\lfloor\tfrac{N-2}{4}\big\rfloor+\sum_{j=1}^{\lfloor\frac{N-2}{4}\rfloor}\epsilon_{\frac{N}{2}+j}\Big)\nonumber\\
 &\!\!=\!\!&-\delta-\bar{\delta}-\tfrac{1}{2}\Big(\sum_{j=1}^{N}\epsilon_{j}-\epsilon_{\frac{N}{2}}-\epsilon_N\Big)\nonumber\\
 &\!\!=\!\!&\tfrac{d-2}{2}+\tfrac{1}{2}\big(2+\epsilon_{\frac{N}{2}}+\epsilon_N-2\delta-2\bar{\delta}\big)
\end{eqnarray}
where we have used $\sum_{j=1}^N\epsilon_j = -d$. We thus recover the selection rule 
\be
 \sigma+\bar{\sigma}=\frac{d-2}{2}
\label{ssd2}
\ee 
of~\cite{PRV2010} by setting
\be 
 \delta=\frac{1+\epsilon_N}{2},\qquad \bar{\delta}=\frac{1+\epsilon_{\frac{N}{2}}}{2},
\ee
in which case
\begin{eqnarray}
 \sigma&\!\!=\!\!&\big| \{ j=1,\ldots,\big\lfloor\tfrac{N}{4}\big\rfloor;\,\epsilon_{j} = -1 \} \big| - 
  \big| \{ j=0,\ldots,\big\lfloor\tfrac{N}{4}\big\rfloor;\,\epsilon_{N-j} = +1 \} \big|,\nonumber \\[-.12cm] \\[-.12cm] \label{eq:NSexcesses}
 \bar{\sigma}&\!\!=\!\!&\big| \{ j=1,\ldots,\big\lfloor\tfrac{N-2}{4}\big\rfloor;\,\epsilon_{\frac{N}{2}-j} = -1 \} \big| - 
  \big| \{ j=0,\ldots,\big\lfloor\tfrac{N-2}{4}\big\rfloor;\,\epsilon_{\frac{N}{2}+j} = +1 \} \big|.\ \nonumber
\end{eqnarray}
It follows readily from the selection rule (\ref{ssd2}) that $\sigma-\bar{\sigma}$ is an even integer. 
For the groundstate, from \eqref{eq:NSexcesses} and \eqref{eq:gseven1}, we have
\be
 \sigma=\bar{\sigma}=\frac{d-2}{4},
\ee
again in accordance with~\cite{PRV2010}.

\subsection[The overall sign $\epsilon$]{The overall sign \boldmath{$\epsilon$}}
\label{app:overallsign}

We now prove the selection rules for $\epsilon$ given in equations \eqref{eq:selrules1} and \eqref{eq:selrules2}, again by resorting to the connection with the XX model. 
As shown in~\cite{AMDYSAinprep}, the representative of $\Omega^{\pm1}$ in the XX representation of $\mathcal EPTL_N(\alpha, \beta=0)$ is given by the matrix
\be 
\bar \Omega^{\pm1} = v^{\pm 2S^z} t^{\pm 1}. 
\ee 
The operators $t$ and $t^{-1}$ introduced here translate the $N$ spins to the left and right by one position, implying that 
\be
 t^{\pm 1} \sigma_j^a t^{\mp 1}= \sigma^a_{j\mp 1}.
\ee 

For both parities of $N$, every eigenvalue of $\omega_d(\Tb(u))$ 
(and $\omega_d(\mathcal H$)) is associated to an eigenvalue of $\omega_d(\Omega)$, see equations \eqref{eq:eigomega1} and \eqref{eq:eigomega2}. It is shown in~\cite{AMDYSAinprep} that
every element of $\mathcal EPTL_N(\alpha, \beta=0)$ has the same spectrum in the $\omega_d$ representation as in the XX representation with $S^z = \frac d 2$. 
This is therefore true for $\Tb(u)$, $\mathcal H$ and $\Omega^{\pm1}$, in particular. To determine $\epsilon$, we use the identifications \eqref{eq:epsilonTheta1} and \eqref{eq:epsilonTheta2} along with the equality $\mathrm{Eig}(\Omega) = \mathrm{Eig}(\bar \Omega)$.

First, we seek the eigenvalue of $\bar \Omega$ associated to the state $\eta_{q_1} \ldots \eta_{q_n} |U\rangle$,  
where $n=\frac{N-d}{2}$. The operator $v^{2S^z}$ acts on 
this state as the multiplicative constant $v^d$. By applying the translation operators to the 
fermionic operators in Proposition~\ref{PropA1}, one finds 
\begin{align}
t^{-1}c_j t &= (-\sigma_1^z) c_{j+1} = c_{j+1} (-\sigma_1^z), \qquad\quad j = 1, \ldots, N-1, \nonumber\\[.2cm] 
t^{-1}c_N t &= (-\sigma_1^z) c_{1} (-1)^{\frac{N}2+S^z+1} = c_{1} (-\sigma_1^z)(-1)^{\frac{N}2+S^z}
\end{align}
and
\begin{align}
t^{-1}\eta_q t &=(-\sigma^z_1) \Big(e^{iq} \eta_q + \frac{c_1}{\sqrt N}\big(e^{-iqN}(-1)^{\frac{N}2 + S^z+1} - 1\big)\!\Big) \label{eq:teta1}\\
&=\Big(e^{iq} \eta_q + \frac{c_1}{\sqrt N}\big(e^{-iqN}(-1)^{\frac{N}2 + S^z} - 1\big)\!\Big) (-\sigma^z_1). \label{eq:teta2}
\end{align}
Applying $t^{-1}$ on eigenvectors of $H$ yields 
\be 
t^{-1} \eta_{q_1}\ldots \eta_{q_n} |U\rangle = \big(t^{-1}\eta_{q_1}t\big) \ldots \big( t^{-1}\eta_{q_n}t\big)t^{-1}|U\rangle = (-1)^n e^{i \sum_j q_j}  \eta_{q_1} \ldots\eta_{q_n} |U\rangle
\ee
where we have used the two expressions \eqref{eq:teta1} and \eqref{eq:teta2}
to rewrite $t^{-1}\eta_{q_j} t$ for $j$ even and odd, respectively.
Because 
\vspace{-0.5cm}
\begin{align}
\sum_{j} q_j &= \sum_{m|_{\Theta_m = -1}} q(m) = \frac12 \Big( \sum_{m|_{\Theta_m = -1}} q(m) -  \sum_{m|_{\Theta_m = 1}} q(m) +  \sum_{m=0}^{N-1} q(m) \!\Big) \nonumber\\[.15cm]
& = -\frac12  \sum_{m=0}^{N-1} \Theta_m\, q(m)  + \frac\pi2 \times \left\{\begin{array}{cl} \!\!N-1, &  \quad \frac{N+d}2 \; \mathrm{odd},
\\[.2cm] \!\!N, &  \quad \frac{N+d}2 \; \mathrm{even},\end{array} \right.
\end{align}
one readily finds 
\be 
\mathrm{Eig}(\bar\Omega) = v^d (-i)^{d-a} \exp\!\Big(\frac{i}2 \sum_{m=0}^{N-1} \Theta_m\, q(m)\!\Big),\qquad
a=\left\{\begin{array}{ll}
\!\!1,\quad &\frac{N+d}{2}\;\mathrm{odd},\\[.15cm]
\!\!0,\quad &\frac{N+d}{2}\;\mathrm{even}.
\end{array}\right.
\label{Eigbar}
\ee 
The next and final step
is based on the identifications (\ref{eq:epsilonTheta1}) and (\ref{eq:epsilonTheta2}) of the $\epsilon_j$ and $\Theta_m$ parameters,
and is performed separately for the two parities of $N$. 

\paragraph{\boldmath{$\mathbb Z_4$} sector (\boldmath{$N$} odd):} 
The selection rule for $\epsilon$ we wish to establish states that
\be
\epsilon = \left\{\begin{array}{ll}
\!\!(-1)^{\frac{d-1}4}, & \quad \frac{d-1}2 \; \mathrm{even}, \\[.15cm] 
\!\!(-1)^{\frac{d+1}4}, & \quad \frac{d-1}2 \; \mathrm{odd}.
\end{array}\right.
\label{epsodd}
\ee
Let us be specific and consider $\frac{N-1}2$ and $\frac{d-1}2$ both even. In this case, the identification (\ref{eq:epsilonTheta1}) specializes to \eqref{eq:eps} which is equivalent to
\be
\Theta_m = \left\{\begin{array}{lll}
\!\! \epsilon_{\frac{N+1}2 + 2m}, & \quad  m = 0, \ldots, \frac{N-1}4, \\[.15cm]
\!\!-\epsilon_{\frac{3N+1}2 - 2m}, &  \quad  m = \frac{N+3}4, \ldots, \frac{3N-3}4, \\[.15cm]
\!\!\epsilon_{2m -\frac{3N-1}2 }, &\quad m = \frac{3N+1}4, \ldots, N-1.
\end{array} \right.
\ee
To evaluate $\mathrm{Eig}(\bar\Omega)$ in (\ref{Eigbar}), we now compute
\begin{align}
\exp &\!\Bigg(\frac i 2 \sum_{m=0}^{N-1}\Theta_m\, q(m)\!\Bigg) = \exp \!\Bigg(\! \frac{i \pi  }{N} \Big( \sum_{m=0}^{\frac{N-1}4}m \epsilon_{\frac{N+1}2 + 2m} - \!\!\sum_{m=\frac{N+3}4}^{\frac{3N-3}4}m \epsilon_{\frac{3N+1}2 - 2m}
+\!\! \sum_{m=\frac{3N+1}4}^{N-1}m \epsilon_{2m - \frac{3N-1}2}\Big)\! \!\Bigg) \nonumber\\
&= \exp \!\Bigg(\! \frac{i\pi}{2N} \Big( \sum_{j_1=\frac{N-1}4}^{\frac{N-1}2}\!\!{\textstyle (2j_1-\frac{N-1}2)} \epsilon_{2j_1+1} +\sum_{j_2=1}^{\frac{N-1}2}{\textstyle (2j_2-\frac{3N+1}2) } \epsilon_{2j_2}
+\sum_{j_3=0}^{\frac{N-5}4} 
{\textstyle (2j_3+\frac{3N+1}2)} \epsilon_{2j_3+1}\Big)\!\!\Bigg)
 \nonumber\\
& = \exp \!\Bigg(\! \frac {i\pi} {4N} \sum_{j=1}^N (2j-1)\epsilon_j \!\Bigg) \exp \!\Bigg(\! \frac{i \pi}{4} \Big( -\!\!\sum_{j_1=\frac{N-1}4}^{\frac{N-1}2} \epsilon_{2j_1+1}- 3  \sum_{j_2=1}^{\frac{N-1}2}  \epsilon_{2j_2}  + 3 \sum_{j_3=0}^{\frac{N-5}4} 
\epsilon_{2j_3+1}\Big)\! \!\Bigg) 
\nonumber \\
& =  \exp \!\Bigg(\! \frac {i\pi} {4N} \sum_{j=1}^N (2j-1)\epsilon_j \!\Bigg) \exp \!\Bigg(\!\!-\frac{3i \pi}{4} \sum_{k=1}^N (-1)^k \epsilon_k\!\Bigg) \exp \!\Bigg(\!\! -i \pi\! \sum_{j_1=\frac{N-1}4}^{\frac{N-1}2} \epsilon_{2j_1+1}\!\Bigg) \nonumber\\
& = \exp \!\Bigg(\! \frac {i\pi} {4N} \sum_{j=1}^N (2j-1) \epsilon_j\!\Bigg) \exp \!\bigg(\frac{3i \pi d}{4}\bigg) (-1)^{\frac{N+3}4},
\end{align}
where we have used (\ref{sumk}).
This allows us to compare $\mathrm{Eig}(\bar\Omega)$ with
\be    
\mathrm{Eig}(\Omega) = \epsilon\, e^{-\frac{i\pi N}{4}} v^d \exp\! \Big( \frac {i\pi}{4N} \sum_{j=1}^N(2j-1)\epsilon_j \!\Big),
\ee
from which we deduce that $\epsilon = (-1)^{\frac{d-1}4}$. 
Similar arguments carry through for the other parities of $\frac{N-1}2$ and $\frac{d-1}2$, 
thereby completing the proof of the selection rule (\ref{epsodd}). 

\paragraph{Ramond and Neveu-Schwarz sectors (\boldmath{$N$} even):}  
The selection rule for $\epsilon$ we wish to show is given by 
\be
\epsilon = \left\{\begin{array}{ll}
\!\!(-1)^{\frac{d}4}, &\quad \frac d 2 \; \mathrm{even}, \\[.15cm] 
\!\!(-1)^{\frac{d+2}4}, &\quad \frac d 2 \; \mathrm{odd}.\end{array}\right.
\label{epsRNS}
\ee
For $\frac{N}2$ and $\frac{d}2$ both even,
for example, the identification (\ref{eq:epsilonTheta2}) is equivalently given by
\be
\Theta_m = \left\{ \begin{array}{ll} 
\!\!-\epsilon_{\frac{3N}4 - m}, \quad m = 0, \ldots, \frac{3N}4 - 1, \\[.17cm]
\!\!-\epsilon_{\frac{7N}4 - m}, \quad m =  \frac{3N}4, \ldots, N-1,
\end{array} \right. 
\ee
and one computes
\begin{align}
\exp &\Bigg(\frac i 2 \sum_{m=0}^{N-1}\Theta_m\, q(m)\!\Bigg) = \exp\!\Bigg(\! \frac {i\pi} {2N}\Big(- \sum_{m=0}^{\frac{3N}4-1} (2m+1)\epsilon_{\frac{3N}4-m} - \sum_{m=\frac{3N}4}^{N-1} (2m+1)\epsilon_{\frac{7N}4-m}\Big)\!\!\Bigg) \nonumber\\
& = \exp\!\Bigg(\! \frac {i\pi} {2N}\Big( \sum_{j=1}^{\frac{3N}4} (2j-1-{\textstyle\frac{3N}2})\epsilon_{j} + \sum_{j=\frac{3N}4+1}^{N} (2j-1-\textstyle{\frac{7N}2})\epsilon_{j}\Big)\!\!\Bigg) \nonumber\\
& = \exp \!\Bigg(\frac{i \pi}{2N}\sum_{j=1}^N (2j-1)\epsilon_j  - \frac{3i\pi}{4}\sum_{j=1}^N \epsilon_j  - i \pi \sum_{j=\frac{3N}4+1}^{N} \epsilon_j\!\Bigg)\nonumber\\
& = \exp \!\Bigg(\frac{i \pi}{2N}\sum_{j=1}^N(2j-1)\epsilon_j \!\Bigg)\exp \!\bigg(\frac {3i \pi d} 4\bigg)(-1)^{\frac{N}4}.
\end{align}
By comparing 
$\mathrm{Eig}(\bar \Omega)$ with 
\be
\mathrm{Eig}(\Omega) = \epsilon \, i^{\frac N 2} v^d \exp\!\Big( \frac{i \pi}{2N}  \sum_{j=1}^N (2j-1)\epsilon_j\! \Big),
\ee 
valid for $\frac{N}2$ and $\frac{d}2$ both even,
we find $\epsilon = (-1)^{\frac d 4}$. Similar arguments show that the selection rule (\ref{epsRNS}) also holds for the other parities of $\frac{N}2$ and $\frac{d}2$.

The calculations above and the ensuing comparison with $\mathrm{Eig}(\Omega)$ could have been carried out for the alternative identification $\epsilon_j = \Theta_{\frac N 4 -j}$ discussed following (\ref{HT2}). In that case, we would have obtained 
\be
\epsilon = (-1)^{\frac d 4} \exp\!\Big(\!\!-\frac{i \pi}N \sum_{j=1}^N (2j-1)\epsilon_j \!\Big),
\ee 
but this is in general {\em incompatible} with the requirement $\epsilon\!\in\!\{+1, -1\}$.
Thus, even in the $d = 0$ sector, only the first identification yields a fully consistent picture in general.
 Similar arguments apply for the other parities of $\frac{N}2$ and $\frac{d}2$, thus demonstrating that the identification (\ref{eq:epsilonTheta2}) 
is the correct one in the general case.

\section{Loop model on helical tori}
\label{app:Twist}

\begin{figure}[ht]
\begin{center}
\psset{unit=0.50}
\begin{pspicture}(0,-1.5)(9,9)
\psset{linewidth=1pt}
\rput(4,-2){(a)}
\psset{linecolor=lightgray}
\psline[linewidth=0.25pt]{-}(0,-1)(0,9)
\psline[linewidth=0.25pt]{-}(1,-1)(1,9)
\psline[linewidth=0.25pt]{-}(2,-1)(2,9)
\psline[linewidth=0.25pt]{-}(3,-1)(3,9)
\psline[linewidth=0.25pt]{-}(4,-1)(4,9)
\psline[linewidth=0.25pt]{-}(5,-1)(5,9)
\psline[linewidth=0.25pt]{-}(6,-1)(6,9)
\psline[linewidth=0.25pt]{-}(7,-1)(7,9)
\psline[linewidth=0.25pt]{-}(8,-1)(8,9)
\psline[linewidth=0.5pt]{-}(-1,1)(9,1)
\psline[linewidth=0.5pt]{-}(-1,2)(9,2)
\psline[linewidth=0.5pt]{-}(-1,3)(9,3)
\psline[linewidth=0.5pt]{-}(-1,4)(9,4)
\psline[linewidth=0.5pt]{-}(-1,5)(9,5)
\psline[linewidth=0.5pt]{-}(-1,6)(9,6)
\psline[linewidth=0.5pt]{-}(-1,7)(9,7)
\psline[linewidth=0.5pt]{-}(-1,8)(9,8)
\psset{linecolor=black}
\psline[linewidth=1.5pt]{-}(0,0)(0,4)(6,4)(6,0)(0,0)
\psline[linewidth=1.5pt]{-}(2,4)(2,8)(8,8)(8,4)(2,4)
\psline[linewidth=1.5pt]{-}(0,0)(-1,0)
\psline[linewidth=1.5pt]{-}(4,0)(4,-1)
\psline[linewidth=1.5pt]{-}(0,4)(-1,4)
\psline[linewidth=1.5pt]{-}(2,8)(-1,8)
\psline[linewidth=1.5pt]{-}(6,0)(9,0)
\psline[linewidth=1.5pt]{-}(8,4)(9,4)
\psline[linewidth=1.5pt]{-}(8,8)(9,8)
\psline[linewidth=1.5pt]{-}(4,8)(4,9)
\psset{linecolor=lightermyc2}
\psset{linewidth=1.7pt}
\psarc(0,1){0.5}{0}{90}\psarc(1,2){0.5}{180}{270}
\psarc(0,2){0.5}{0}{90}\psarc(1,3){0.5}{180}{270}
\psarc(1,3){0.5}{0}{90}\psarc(2,4){0.5}{180}{270}
\psarc(2,0){0.5}{0}{90}\psarc(3,1){0.5}{180}{270}
\psarc(3,1){0.5}{0}{90}\psarc(4,2){0.5}{180}{270}
\psarc(3,2){0.5}{0}{90}\psarc(4,3){0.5}{180}{270}
\psarc(2,2){0.5}{0}{90}\psarc(3,3){0.5}{180}{270}
\psarc(3,3){0.5}{0}{90}\psarc(4,4){0.5}{180}{270}
\psarc(5,2){0.5}{0}{90}\psarc(6,3){0.5}{180}{270}
\psarc(4,2){0.5}{0}{90}\psarc(5,3){0.5}{180}{270}
\psarc(5,0){0.5}{0}{90}\psarc(6,1){0.5}{180}{270}
\psset{linecolor=lightergreen}
\psarc(1,0){0.5}{90}{180}\psarc(0,1){0.5}{270}{0}
\psarc(2,0){0.5}{90}{180}\psarc(1,1){0.5}{270}{0}
\psarc(2,1){0.5}{90}{180}\psarc(1,2){0.5}{270}{0}
\psarc(2,2){0.5}{90}{180}\psarc(1,3){0.5}{270}{0}
\psarc(1,3){0.5}{90}{180}\psarc(0,4){0.5}{270}{0}
\psarc(3,3){0.5}{90}{180}\psarc(2,4){0.5}{270}{0}
\psarc(3,1){0.5}{90}{180}\psarc(2,2){0.5}{270}{0}
\psarc(4,0){0.5}{90}{180}\psarc(3,1){0.5}{270}{0}
\psarc(5,0){0.5}{90}{180}\psarc(4,1){0.5}{270}{0}
\psarc(5,1){0.5}{90}{180}\psarc(4,2){0.5}{270}{0}
\psarc(6,1){0.5}{90}{180}\psarc(5,2){0.5}{270}{0}
\psarc(6,3){0.5}{90}{180}\psarc(5,4){0.5}{270}{0}
\psarc(5,3){0.5}{90}{180}\psarc(4,4){0.5}{270}{0}
\psset{linecolor=lightermyc2}
\psarc(2,5){0.5}{0}{90}\psarc(3,6){0.5}{180}{270}
\psarc(2,6){0.5}{0}{90}\psarc(3,7){0.5}{180}{270}
\psarc(3,7){0.5}{0}{90}\psarc(4,8){0.5}{180}{270}
\psarc(4,4){0.5}{0}{90}\psarc(5,5){0.5}{180}{270}
\psarc(5,5){0.5}{0}{90}\psarc(6,6){0.5}{180}{270}
\psarc(5,6){0.5}{0}{90}\psarc(6,7){0.5}{180}{270}
\psarc(4,6){0.5}{0}{90}\psarc(5,7){0.5}{180}{270}
\psarc(5,7){0.5}{0}{90}\psarc(6,8){0.5}{180}{270}
\psarc(7,6){0.5}{0}{90}\psarc(8,7){0.5}{180}{270}
\psarc(6,6){0.5}{0}{90}\psarc(7,7){0.5}{180}{270}
\psarc(7,4){0.5}{0}{90}\psarc(8,5){0.5}{180}{270}
\psset{linecolor=lightergreen}
\psarc(3,4){0.5}{90}{180}\psarc(2,5){0.5}{270}{0}
\psarc(4,4){0.5}{90}{180}\psarc(3,5){0.5}{270}{0}
\psarc(4,5){0.5}{90}{180}\psarc(3,6){0.5}{270}{0}
\psarc(4,6){0.5}{90}{180}\psarc(3,7){0.5}{270}{0}
\psarc(3,7){0.5}{90}{180}\psarc(2,8){0.5}{270}{0}
\psarc(5,7){0.5}{90}{180}\psarc(4,8){0.5}{270}{0}
\psarc(5,5){0.5}{90}{180}\psarc(4,6){0.5}{270}{0}
\psarc(6,4){0.5}{90}{180}\psarc(5,5){0.5}{270}{0}
\psarc(7,4){0.5}{90}{180}\psarc(6,5){0.5}{270}{0}
\psarc(7,5){0.5}{90}{180}\psarc(6,6){0.5}{270}{0}
\psarc(8,5){0.5}{90}{180}\psarc(7,6){0.5}{270}{0}
\psarc(8,7){0.5}{90}{180}\psarc(7,8){0.5}{270}{0}
\psarc(7,7){0.5}{90}{180}\psarc(6,8){0.5}{270}{0}
\psset{linecolor=lightermyc2}
\psset{linewidth=1.7pt}
\psarc(6,1){0.5}{0}{90}\psarc(7,2){0.5}{180}{270}
\psarc(6,2){0.5}{0}{90}\psarc(7,3){0.5}{180}{270}
\psarc(7,3){0.5}{0}{90}\psarc(8,4){0.5}{180}{270}
\psarc(8,0){0.5}{0}{90}\psarc(9,1){0.5}{180}{270}
\psarc(8,2){0.5}{0}{90}\psarc(9,3){0.5}{180}{270}
\psarc(-1,2){0.5}{0}{90}\psarc(0,3){0.5}{180}{270}
\psarc(-1,0){0.5}{0}{90}\psarc(0,1){0.5}{180}{270}
\psset{linecolor=lightergreen}
\psarc(7,0){0.5}{90}{180}\psarc(6,1){0.5}{270}{0}
\psarc(8,0){0.5}{90}{180}\psarc(7,1){0.5}{270}{0}
\psarc(8,1){0.5}{90}{180}\psarc(7,2){0.5}{270}{0}
\psarc(8,2){0.5}{90}{180}\psarc(7,3){0.5}{270}{0}
\psarc(7,3){0.5}{90}{180}\psarc(6,4){0.5}{270}{0}
\psarc(9,3){0.5}{90}{180}\psarc(8,4){0.5}{270}{0}
\psarc(9,1){0.5}{90}{180}\psarc(8,2){0.5}{270}{0}
\psarc(0,1){0.5}{90}{180}\psarc(-1,2){0.5}{270}{0}
\psarc(0,3){0.5}{90}{180}\psarc(-1,4){0.5}{270}{0}
\psset{linecolor=lightermyc2}
\psarc(8,5){0.5}{0}{90}\psarc(9,6){0.5}{180}{270}
\psarc(8,6){0.5}{0}{90}\psarc(9,7){0.5}{180}{270}
\psarc(-1,5){0.5}{0}{90}\psarc(0,6){0.5}{180}{270}
\psarc(-1,6){0.5}{0}{90}\psarc(0,7){0.5}{180}{270}
\psarc(-1,7){0.5}{0}{90}\psarc(0,8){0.5}{180}{270}
\psarc(1,6){0.5}{0}{90}\psarc(2,7){0.5}{180}{270}
\psarc(0,6){0.5}{0}{90}\psarc(1,7){0.5}{180}{270}
\psarc(1,4){0.5}{0}{90}\psarc(2,5){0.5}{180}{270}
\psset{linecolor=lightergreen}
\psarc(9,4){0.5}{90}{180}\psarc(8,5){0.5}{270}{0}
\psarc(9,7){0.5}{90}{180}\psarc(8,8){0.5}{270}{0}
\psarc(0,4){0.5}{90}{180}\psarc(-1,5){0.5}{270}{0}
\psarc(1,4){0.5}{90}{180}\psarc(0,5){0.5}{270}{0}
\psarc(1,5){0.5}{90}{180}\psarc(0,6){0.5}{270}{0}
\psarc(2,5){0.5}{90}{180}\psarc(1,6){0.5}{270}{0}
\psarc(2,7){0.5}{90}{180}\psarc(1,8){0.5}{270}{0}
\psarc(1,7){0.5}{90}{180}\psarc(0,8){0.5}{270}{0}
\psset{linecolor=lightermyc2}
\psset{linewidth=1.7pt}
\psarc(-1,-1){0.5}{0}{90}\psarc(0,0){0.5}{180}{270}
\psarc(1,-1){0.5}{0}{90}\psarc(2,0){0.5}{180}{270}
\psarc(5,-1){0.5}{0}{90}\psarc(6,0){0.5}{180}{270}
\psarc(7,-1){0.5}{0}{90}\psarc(8,0){0.5}{180}{270}
\psarc(6,8){0.5}{0}{90}\psarc(7,9){0.5}{180}{270}
\psarc(0,8){0.5}{0}{90}\psarc(1,9){0.5}{180}{270}
\psarc(3,8){0.5}{0}{90}\psarc(4,9){0.5}{180}{270}
\psset{linecolor=lightergreen}
\psarc(5,8){0.5}{90}{180}\psarc(4,9){0.5}{270}{0}
\psarc(6,8){0.5}{90}{180}\psarc(5,9){0.5}{270}{0}
\psarc(8,8){0.5}{90}{180}\psarc(7,9){0.5}{270}{0}
\psarc(9,8){0.5}{90}{180}\psarc(8,9){0.5}{270}{0}
\psarc(0,8){0.5}{90}{180}\psarc(-1,9){0.5}{270}{0}
\psarc(2,8){0.5}{90}{180}\psarc(1,9){0.5}{270}{0}
\psarc(3,8){0.5}{90}{180}\psarc(2,9){0.5}{270}{0}
\psarc(5,-1){0.5}{90}{180}\psarc(4,0){0.5}{270}{0}
\psarc(1,-1){0.5}{90}{180}\psarc(0,0){0.5}{270}{0}
\psarc(4,-1){0.5}{90}{180}\psarc(3,0){0.5}{270}{0}
\psarc(3,-1){0.5}{90}{180}\psarc(2,0){0.5}{270}{0}
\psarc(7,-1){0.5}{90}{180}\psarc(6,0){0.5}{270}{0}
\psarc(9,-1){0.5}{90}{180}\psarc(8,0){0.5}{270}{0}
\psline[linewidth=1.5pt,linecolor=myc,arrowscale=2]{->}(0,0)(2,4)
\psline[linewidth=1.5pt,linecolor=myc,arrowscale=2]{->}(0,0)(6,0)
\rput(5,0.9){\myc $\nu_1$} 
\rput(1,3){\myc $\nu_2$}
\end{pspicture} \qquad \qquad
\begin{pspicture}(-2,-0.5)(9,9)
\psset{linecolor=lightgray}
\psline[linewidth=0.25pt]{-}(-1,0)(-1,4)
\psline[linewidth=0.25pt]{-}(1,0)(1,4)
\psline[linewidth=0.25pt]{-}(2,0)(2,4)
\psline[linewidth=0.25pt]{-}(3,0)(3,4)
\psline[linewidth=0.25pt]{-}(4,0)(4,4)
\psline[linewidth=0.25pt]{-}(5,0)(5,4)
\psline[linewidth=0.25pt]{-}(7,0)(7,4)
\psline[linewidth=0.5pt]{-}(-2,1)(8,1)
\psline[linewidth=0.5pt]{-}(-2,2)(8,2)
\psline[linewidth=0.5pt]{-}(-2,3)(8,3)
\psline[linewidth=0.25pt]{-}(-1,6)(-1,10)
\psline[linewidth=0.25pt]{-}(1,6)(1,10)
\psline[linewidth=0.25pt]{-}(2,6)(2,10)
\psline[linewidth=0.25pt]{-}(3,6)(3,10)
\psline[linewidth=0.25pt]{-}(4,6)(4,10)
\psline[linewidth=0.25pt]{-}(5,6)(5,10)
\psline[linewidth=0.25pt]{-}(7,6)(7,10)
\psline[linewidth=0.5pt]{-}(-2,7)(8,7)
\psline[linewidth=0.5pt]{-}(-2,8)(8,8)
\psline[linewidth=0.5pt]{-}(-2,9)(8,9)
\psset{linecolor=black}
\psline[linewidth=1.5pt]{-}(0,0)(0,4)(6,4)(6,0)(0,0)
\psline[linewidth=1.5pt]{-}(0,6)(0,10)(6,10)(6,6)(0,6)
\psline[linewidth=1.5pt]{-}(-2,0)(0,0)
\psline[linewidth=1.5pt]{-}(8,0)(6,0)
\psline[linewidth=1.5pt]{-}(-2,4)(0,4)
\psline[linewidth=1.5pt]{-}(8,4)(6,4)
\psline[linewidth=1.5pt]{-}(-2,6)(0,6)
\psline[linewidth=1.5pt]{-}(8,6)(6,6)
\psline[linewidth=1.5pt]{-}(-2,10)(0,10)
\psline[linewidth=1.5pt]{-}(8,10)(6,10)
\psset{linecolor=lightermyc2}
\psset{linewidth=1.7pt}
\psarc(0,1){0.5}{0}{90}\psarc(1,2){0.5}{180}{270}\psarc(6,1){0.5}{0}{90}\psarc(7,2){0.5}{180}{270}
\psarc(0,2){0.5}{0}{90}\psarc(1,3){0.5}{180}{270}\psarc(6,2){0.5}{0}{90}\psarc(7,3){0.5}{180}{270}
\psarc(1,3){0.5}{0}{90}\psarc(2,4){0.5}{180}{270}\psarc(7,3){0.5}{0}{90}\psarc(8,4){0.5}{180}{270}
\psarc(2,0){0.5}{0}{90}\psarc(3,1){0.5}{180}{270}
\psarc(3,1){0.5}{0}{90}\psarc(4,2){0.5}{180}{270}
\psarc(3,2){0.5}{0}{90}\psarc(4,3){0.5}{180}{270}
\psarc(2,2){0.5}{0}{90}\psarc(3,3){0.5}{180}{270}
\psarc(3,3){0.5}{0}{90}\psarc(4,4){0.5}{180}{270}
\psarc(5,2){0.5}{0}{90}\psarc(6,3){0.5}{180}{270}\psarc(-1,2){0.5}{0}{90}\psarc(0,3){0.5}{180}{270}
\psarc(4,2){0.5}{0}{90}\psarc(5,3){0.5}{180}{270}\psarc(-2,2){0.5}{0}{90}\psarc(-1,3){0.5}{180}{270}
\psarc(5,0){0.5}{0}{90}\psarc(6,1){0.5}{180}{270}\psarc(-1,0){0.5}{0}{90}\psarc(0,1){0.5}{180}{270}
\psset{linecolor=lightergreen}
\psarc(1,0){0.5}{90}{180}\psarc(0,1){0.5}{270}{0}\psarc(7,0){0.5}{90}{180}\psarc(6,1){0.5}{270}{0}
\psarc(2,0){0.5}{90}{180}\psarc(1,1){0.5}{270}{0}\psarc(8,0){0.5}{90}{180}\psarc(7,1){0.5}{270}{0}
\psarc(2,1){0.5}{90}{180}\psarc(1,2){0.5}{270}{0}\psarc(8,1){0.5}{90}{180}\psarc(7,2){0.5}{270}{0}
\psarc(2,2){0.5}{90}{180}\psarc(1,3){0.5}{270}{0}\psarc(8,2){0.5}{90}{180}\psarc(7,3){0.5}{270}{0}
\psarc(1,3){0.5}{90}{180}\psarc(0,4){0.5}{270}{0}\psarc(7,3){0.5}{90}{180}\psarc(6,4){0.5}{270}{0}
\psarc(3,3){0.5}{90}{180}\psarc(2,4){0.5}{270}{0}
\psarc(3,1){0.5}{90}{180}\psarc(2,2){0.5}{270}{0}
\psarc(4,0){0.5}{90}{180}\psarc(3,1){0.5}{270}{0}
\psarc(5,0){0.5}{90}{180}\psarc(4,1){0.5}{270}{0}\psarc(-1,0){0.5}{90}{180}\psarc(-2,1){0.5}{270}{0}
\psarc(5,1){0.5}{90}{180}\psarc(4,2){0.5}{270}{0}\psarc(-1,1){0.5}{90}{180}\psarc(-2,2){0.5}{270}{0}
\psarc(6,1){0.5}{90}{180}\psarc(5,2){0.5}{270}{0}\psarc(0,1){0.5}{90}{180}\psarc(-1,2){0.5}{270}{0}
\psarc(6,3){0.5}{90}{180}\psarc(5,4){0.5}{270}{0}\psarc(0,3){0.5}{90}{180}\psarc(-1,4){0.5}{270}{0}
\psarc(5,3){0.5}{90}{180}\psarc(4,4){0.5}{270}{0}\psarc(-1,3){0.5}{90}{180}\psarc(-2,4){0.5}{270}{0}
\psset{linecolor=lightermyc2}
\psset{linewidth=1.7pt}
\psarc(0,7){0.5}{0}{90}\psarc(1,8){0.5}{180}{270}\psarc(6,7){0.5}{0}{90}\psarc(7,8){0.5}{180}{270}
\psarc(0,8){0.5}{0}{90}\psarc(1,9){0.5}{180}{270}\psarc(6,8){0.5}{0}{90}\psarc(7,9){0.5}{180}{270}
\psarc(1,9){0.5}{0}{90}\psarc(2,10){0.5}{180}{270}\psarc(7,9){0.5}{0}{90}\psarc(8,10){0.5}{180}{270}
\psarc(2,6){0.5}{0}{90}\psarc(3,7){0.5}{180}{270}
\psarc(3,7){0.5}{0}{90}\psarc(4,8){0.5}{180}{270}
\psarc(3,8){0.5}{0}{90}\psarc(4,9){0.5}{180}{270}
\psarc(2,8){0.5}{0}{90}\psarc(3,9){0.5}{180}{270}
\psarc(3,9){0.5}{0}{90}\psarc(4,10){0.5}{180}{270}
\psarc(5,8){0.5}{0}{90}\psarc(6,9){0.5}{180}{270}\psarc(-1,8){0.5}{0}{90}\psarc(0,9){0.5}{180}{270}
\psarc(4,8){0.5}{0}{90}\psarc(5,9){0.5}{180}{270}\psarc(-2,8){0.5}{0}{90}\psarc(-1,9){0.5}{180}{270}
\psarc(5,6){0.5}{0}{90}\psarc(6,7){0.5}{180}{270}\psarc(-1,6){0.5}{0}{90}\psarc(0,7){0.5}{180}{270}
\psset{linecolor=lightergreen}
\psarc(1,6){0.5}{90}{180}\psarc(0,7){0.5}{270}{0}\psarc(7,6){0.5}{90}{180}\psarc(6,7){0.5}{270}{0}
\psarc(2,6){0.5}{90}{180}\psarc(1,7){0.5}{270}{0}\psarc(8,6){0.5}{90}{180}\psarc(7,7){0.5}{270}{0}
\psarc(2,7){0.5}{90}{180}\psarc(1,8){0.5}{270}{0}\psarc(8,7){0.5}{90}{180}\psarc(7,8){0.5}{270}{0}
\psarc(2,8){0.5}{90}{180}\psarc(1,9){0.5}{270}{0}\psarc(8,8){0.5}{90}{180}\psarc(7,9){0.5}{270}{0}
\psarc(1,9){0.5}{90}{180}\psarc(0,10){0.5}{270}{0}\psarc(7,9){0.5}{90}{180}\psarc(6,10){0.5}{270}{0}
\psarc(3,9){0.5}{90}{180}\psarc(2,10){0.5}{270}{0}
\psarc(3,7){0.5}{90}{180}\psarc(2,8){0.5}{270}{0}
\psarc(4,6){0.5}{90}{180}\psarc(3,7){0.5}{270}{0}
\psarc(5,6){0.5}{90}{180}\psarc(4,7){0.5}{270}{0}\psarc(-1,6){0.5}{90}{180}\psarc(-2,7){0.5}{270}{0}
\psarc(5,7){0.5}{90}{180}\psarc(4,8){0.5}{270}{0}\psarc(-1,7){0.5}{90}{180}\psarc(-2,8){0.5}{270}{0}
\psarc(6,7){0.5}{90}{180}\psarc(5,8){0.5}{270}{0}\psarc(0,7){0.5}{90}{180}\psarc(-1,8){0.5}{270}{0}
\psarc(6,9){0.5}{90}{180}\psarc(5,10){0.5}{270}{0}\psarc(0,9){0.5}{90}{180}\psarc(-1,10){0.5}{270}{0}
\psarc(5,9){0.5}{90}{180}\psarc(4,10){0.5}{270}{0}\psarc(-1,9){0.5}{90}{180}\psarc(-2,10){0.5}{270}{0}
\psset{linecolor=myc}
\psbezier{-}(-1.5,4)(-1.5,4.36)(-1.75,4.56)(-2,4.73)
\psbezier{-}(-0.5,4)(-0.5,4.695)(-1.5,4.93)(-2,5.2725)
\psbezier{-}(0.5,4)(0.5,5)(-1.5,5)(-1.5,6)
\psbezier{-}(1.5,4)(1.5,5)(-0.5,5)(-0.5,6)
\psbezier{-}(2.5,4)(2.5,5)(0.5,5)(0.5,6)
\psbezier{-}(3.5,4)(3.5,5)(1.5,5)(1.5,6)
\psbezier{-}(4.5,4)(4.5,5)(2.5,5)(2.5,6)
\psbezier{-}(5.5,4)(5.5,5)(3.5,5)(3.5,6)
\psbezier{-}(6.5,4)(6.5,5)(4.5,5)(4.5,6)
\psbezier{-}(7.5,4)(7.5,5)(5.5,5)(5.5,6)
\psbezier{-}(7.5,6)(7.5,5.64)(7.75,5.44)(8,5.27)
\psbezier{-}(6.5,6)(6.5,5.305)(7.5,5.07)(8,4.7275)
\rput(3,-1){(b)}
\end{pspicture}
\vspace{0.2cm}
\caption{(a) 
A configuration $c$ of the loop model on a $4\times 6$ torus with helicity $t=2$, with five contractible loops and two non-contractible loops of homotopy $\{0,1\}$. 
(b) The weight $\mathcal F_2(c) $ of the configuration $c$ is equal to $\mathcal F(c \cdot\Omega^{-2})$.
}\label{fig:seamconfig}
\end{center}
\end{figure}
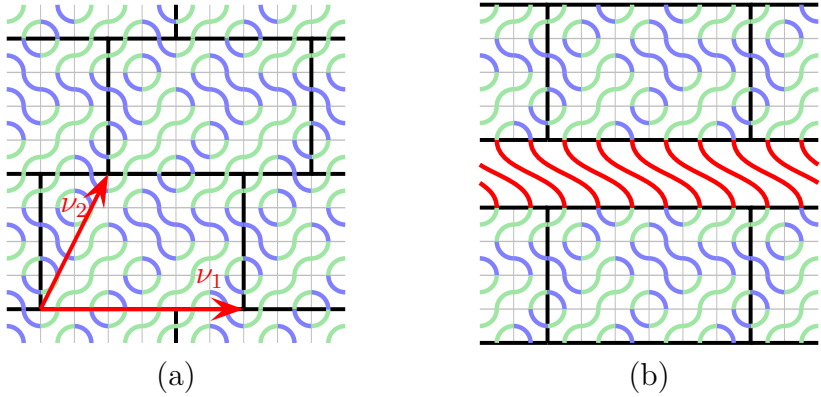

Here we generalize the loop model discussed in the bulk part of this paper by considering the model defined on helical tori~\cite{OkabeEtAl1999,LiawEtAl2006,IzmHu2007}.
Prior to forming the torus, the loop model is defined on a finite lattice, which we represent diagrammatically by a rectangular array of $M \times N$ square tiles. 
In this planar representation, the toroidal boundary conditions are encoded in the two periodicity vectors $\nu_1,\nu_2\in\mathbb Z^2$.
As indicated in Figure~\ref{fig:seamconfig}, we parameterize these vectors as
\be
 \nu_1=(N,0),\qquad \nu_2 =(t,M),
\ee 
where $t$ is the {\em helicity}. Its role is as follows. 
To construct the torus, one first identifies the right and left edges of the rectangle such that the tiles cover the outer surface of the resulting vertical cylinder. A torus is then formed by gluing together the upper and lower edges of the cylinder, but before doing that, one twists the
upper edge of the
cylinder $t$ tiles in the clockwise direction seen from above. The geometry of the helical lattice torus is thus defined by the triple $(M, N,t)$, where $t$ is negative if the twist is in the counterclockwise direction. 
In the bulk of this paper, we have exclusively considered $t=0$, but the results can be generalized in a straightforward manner as we will discuss in the following.

Loops (and similarly clusters) of homotopy $\{a,b\}$ now wind around the torus $a$ times in the $\nu_1$ direction and $b$ times in the $\nu_2$ direction, with the same convention for the sign of $a$ as in Section~\ref{sec:loopmodel}. Cross-topology clusters are those that wrap the torus along both $\nu_1$ and $\nu_2$. Of course, in the FK model, we restrict the helicity 
$t$ to {\em even} values only.

For a fixed connectivity $c$, the homotopy of non-contractible loops and the number of loops in general depend on $t$. The action of the abstract functional $\mathcal F(c)$ that assigns the correct weights to configurations when the cylinder is closed into a torus 
(defined in equation \eqref{eq:operatorF} for $t=0$) 
depends on the helicity and is here denoted by $\mathcal F_t(c)$ (with $\mathcal F_{0} \equiv \mathcal F$). However, as illustrated in Figure~\ref{fig:seamconfig} (b), studying a torus with helicity $t$ 
is equivalent to considering connectivities times $\Omega^{-t}$ on a torus with trivial helicity:
\be
 \mathcal F_t(c) =\mathcal F_0(c \cdot \Omega^{-t}).
\ee

To calculate the partition function $Z_L^t$, our prescription \eqref{eq:Z} for $\mathcal F(c)$ needs to be slightly modified: Connectivities 
contributing to $c = \Tb^M(u)\Omega^{-t}$ can have up to $M+t$ loops crossing the virtual boundary. This suggests changing the upper and lower bounds of the sum in $\mathcal G_d(\mu, \boldsymbol \alpha)$ to $\pm (M+t)$,
\be 
\mathcal G_d(\mu, \boldsymbol\alpha) = \displaystyle{\sum_{k=-(M+t)}^{M+t} v^{-Nk} C_{k,d}}, \qquad (d>0).
\label{eq:extraseam}
\ee
This simple prescription is not optimal, though, since it may include vanishing terms. Indeed, because the first $t$ layers of boxes are just $\Omega^{-t}$, some terms, the one corresponding to $k = M+t$ for instance, are simply zero. Nevertheless, with this slight modification of the definition of 
$\mathcal G_d(\mu, \boldsymbol\alpha)$, equation \eqref{eq:Z} remains valid.

The problem of computing the partition function $Z_L^t$ therefore boils down to calculating the eigenvalues of $\omega_d(\Tb^M(u)\Omega^{-t})$. The eigenvalues of $\omega_d(\Omega)$ corresponding to $T(u,\mu)$ are given in equations \eqref{eq:eigomega1} and \eqref{eq:eigomega2}, so in general
\be 
\mathrm{Eig}\big(\Tb^M(u)\Omega^{-t}\big) = \big(K(\mu)\big)^M \times \epsilon^{M-t}\times e^{\tfrac{i \pi N t}4(-1)^{N+1}} \prod_{j=1}^N e^{-it\epsilon_j x_j}\!\left(e^{iu}+i e^{-iu}\epsilon_j \tan x_j\right)^M
\ee
and the generating functions $G_d(z,v)$ should be modified in a similar way to depend on the helicity $t$. Below, we write $G_d^t(z,v)$ for these modified generating functions and $\hat{G}_d^t(z,v)$ for their normalized versions defined as in (\ref{Ghat}).

Finitizations of the generating functions $\hat{G}_d^t(z,v)$ are obtained by looking at finite excitations for $M, N,t \gg 1$, not only for fixed aspect ratio (\ref{aspect}), but also while keeping the ratio
\be
\gamma = \frac{t}N
\ee
fixed.
With the parameterization (\ref{mua}) for $\mu$, elementary excitations are given by 
\be
{\displaystyle \lim_{\substack{M,N,t \,\gg 1 \\[.04cm] M = \delta N,\,t=\gamma N}}}   e^{2it x_j}\Big(\frac{e^{iu}-i e^{-iu} \tan x_j}{e^{iu}+i e^{-iu} \tan x_j}\Big)^{\!\!M} \!= q_\gamma^{E_j(a)},
\ee
\be 
{\displaystyle \lim_{\substack{M,N,t \,\gg 1 \\[.04cm] M = \delta N,\,t=\gamma N}}}   e^{2it x_{f(j)}}\Big(\frac{e^{iu}-i e^{-iu} \tan x_{f(j)}}{e^{iu}+i e^{-iu} \tan x_{f(j)}}\Big)^{\!\!M} \!= (-1)^{M+t}\,\bar q_\gamma^{\,E_j(-a)},
\ee
where
\be
E_j(a) = \left\{\begin{array}{ll} 
\!\!\frac {2j-1} 4 + (-1)^{j+\frac{d+1}2}a, & N\ \mathrm{odd}, \\[.15cm]
 \!\!j -a, & N\ \mathrm{even},\; \frac{d}2 \, \mathrm{odd},\\[.15cm]
\!\!\frac{2j-1}2-a, \ & N\ \mathrm{even},\; \frac{d}2 \, \mathrm{even}, \
 \end{array} \right.\quad 
 f(j) = \left\{ \begin{array}{ll} 
 \!\!N+1-j & N\ \mathrm{odd},\\[.15cm] 
 \!\!\frac{N}2-j, & N\ \mathrm{even},\; \frac{d}2 \ \mathrm{odd},\\[.15cm] 
 \!\!\frac{N}2+1 -j, & 
 N\ \mathrm{even}, \;\frac{d}2 \ \mathrm{even},\end{array}\right.
\ee
and 
\be 
 q_\gamma = \exp\!\big(\!-2 \pi i (\delta e^{-2 i u} - \gamma)\big),\qquad \bar{q}_\gamma = \exp\!\big(2 \pi i (\delta e^{2 i u} - \gamma)\big).
\label{eq:seamedq}
\ee
The expressions for $\hat{G}_d^t(z,v)$ are now obtained from \eqref{eq:G1N}, \eqref{eq:G0N} and \eqref{eq:G2N} by replacing $q,\bar{q}$ by $q_\gamma,\bar{q}_\gamma$ and $(-1)^M$ by $(-1)^{M+t}$. For $N$ even, because the eigenvalues of $\omega_d(\Omega)$ corresponding to the groundstates in the $d=0$ and $d=2$ sectors are equal to $1$, the ratio (\ref{qq18}) of the maximal eigenvalues of $\omega_d(\Tb(u)\Omega^{-t})$ in these sectors is still given by $(q_\lambda \bar q_\lambda)^{\frac18}=(q \bar q)^{\frac18}$ in the continuum scaling limit. 

In this limit, for $M$ even, the partition function for critical dense polymers on the torus with helicity 
$t=\gamma N$ is then given by \eqref{ZZZodd} or \eqref{ZZZ} for $N$ odd and even, respectively, with the $\gamma$-dependent definition \eqref{eq:seamedq} for the nome. 
For $M$, $N$ and $t$ all even, this partition function is therefore modular invariant.

As discussed in~\cite{KimPearce87} on the related lattice spin models, the spectral parameter (for $0<u<\frac{\pi}{2}$) can be interpreted as measuring the spatial anisotropy of the lattice through the anisotropy angle
\be
 \theta=\frac{\pi u}{\lambda}=2u.
\ee
With this geometric interpretation, changing the spectral parameter away from the isotropic point $u=\frac{\lambda}{2}=\frac{\pi}{4}$ thus corresponds to distorting the elementary square faces (\ref{u}) on the lattice into rhombi whose bottom-right angle is given by $\theta$. It follows from the discussion above that a nontrivial helicity has the effect of changing the spatial anisotropy
of the lattice, in accordance with the situation for finite lattices as in Figure~\ref{fig:seamconfig} (a). Concretely, the helicity acts on the modular parameter
(see (\ref{tau1}))
\be
 \tau=\delta e^{i(\pi-\theta)}=-\delta e^{-2iu}
\label{tau}
\ee
as
\be
 \tau\ \to\ \tau'=\tau+\gamma.
\ee

%

\end{document}